\documentclass{aa}  
\usepackage{graphicx}
\usepackage{multicol}
\usepackage[hidelinks]{hyperref} 
\hypersetup{
    colorlinks,
    linkcolor={red},
    citecolor={blue},
}
\usepackage{caption}
\usepackage{subcaption}
\usepackage{txfonts}
\usepackage{physics}
\usepackage{amssymb}
\usepackage{pifont}
\usepackage[dvipsnames]{xcolor}

\usepackage[T1]{fontenc}
\usepackage[utf8]{inputenc}

\begin{document}

   \title{Combined X-ray and optical analysis to probe the origin of the plateau emission in $\gamma$-ray bursts afterglows}

   \author{S. Ronchini,
            \inst{1,2}
            G. Stratta,
            \inst{3,4,5,6}
            A. Rossi,
            \inst{6}
            D. A. Kann,
            \inst{7,8}\\
            G. Oganeysan,
            \inst{1,2}
            S. Dall'Osso,
            \inst{5}
            M. Branchesi,
            \inst{1,2}
            G. De Cesare
            \inst{6}
          }

   \institute{Gran Sasso Science Institute (GSSI), I-67100 L’Aquila, Italy
            \and INFN, Laboratori Nazionali Del Gran Sasso,  I-67100 Assergi, Italy
            \and Institut f\"ur Theoretische Physik, Goethe Universit\"at, Max-von-Laue-Str. 1, 60438 Frankfurt am Main, Germany
            \and INAF-IAPS, Via del Fosso del Cavaliere, 100, 00133, Roma, Italy
            \and INFN-La Sapienza, Piazzale Aldo Moro, 2, 00185 Rome, Italy
            \and INAF–Osservatorio di Astrofisica e Scienza dello Spazio, Via Piero Gobetti 93/3, 40129 Bologna, Italy
            \and  Instituto de Astrof\'isica de Andaluc\'ia (IAA-CSIC), Glorieta de la Astronom\'ia s/n, 18008 Granada, Spain,
            \and Hessian Research Cluster ELEMENTS, Giersch Science Center, Max-von-Laue-Strasse 12, Goethe University Frankfurt, Campus Riedberg, 60438 Frankfurt am Main}

   \date{}
 
  \abstract{ 
  A large fraction of $\gamma$-ray bursts (GRBs) shows a plateau phase during the X-ray afterglow emission, whose physical origin is still debated. In this work we select a 
  sample of 30 GRBs with simultaneous X-ray and optical data during and after the plateau phase. 
  Through a time-resolved spectral analysis of the X-ray plateaus, we test the consistency of the unabsorbed optical fluxes with those obtained via X-ray-to-optical spectral extrapolation by assuming a synchrotron spectrum.
  Combining X-ray with optical data, we find that 63\% (19/30) GRBs are compatible with a single synchrotron spectrum,
  thus suggesting that both the optical and X-ray radiations are produced from a single emitting region. For these GRBs
  we derive the temporal evolution of the break 
  frequency and we compare it with the expectations predicted by several models. For 11/30 GRBs the optical emission is above the predicted range of values extrapolated from the X-rays in at least one temporal bin of the light curve. These GRBs may not be explained with a single-zone emission, indicating the necessity of invoking two cooperating processes in order to explain the broad band spectral behaviour during X-ray plateaus. We discuss our findings in the framework of different scenarios invoked to explain the plateau feature, including the energy injection from a spinning-down magnetar and the high latitude emission from a structured jet.
  }

\titlerunning{X-ray and optical analysis of the GRB plateau}
\authorrunning{Ronchini et al.}
   \maketitle
%

\section{Introduction}
$\gamma$-ray bursts (GRBs) are among the most fascinating transients in astrophysics and the recent
association of a gravitational wave source GW170817 with the short GRB 170817A \citep{Abbott2017PRL,Abbott2017ApJLIGOGBM,Abbott2017ApJMMA,Goldstein2017ApJ,Savchenko2017ApJ} has
further increased the interest in these objects, also in the rapidly developing field of multi-messenger
astrophysics.
Although enormous advances have been made over the past two decades in understanding their emission mechanisms, many open questions still need to be answered.\\
The fast repointing capabilities (on  time scales of $\sim$minutes) of the X-Ray Telescope (XRT) onboard the \emph{Neil
Gehrels Swift Observatory} \citep[\emph{Swift} hereafter,][]{Gehrels2004ApJ} enabled us to discover several features not predicted by the standard afterglow
model (e.g. \citealt{mes1993ApJ...405..278M,mes1993ApJ...418L..59M,Sari1998ApJ}) in the early GRB afterglow light curve stages ($<0.5$ day). Before Swift, GRB afterglows were known as sources of non-thermal radiation quickly fading with time as power laws ($F\propto t^{-a}$ with $a>0.7-1$). 
Today we know that, soon after the end of the prompt emission,
a large fraction of GRB X-ray afterglow light curves show a peculiar shallow flux decay phase (``plateau'', e.g. \citealt{Nousek2006ApJ,zha2006ApJ...642..354Z,obr2006ApJ...647.1213O}) during which the temporal decay index reaches values in the range $a \in [0,\sim 0.7]$ 
for a temporal interval that typically lasts for few hours. Post-plateau fluxes decay according to standard afterglow predictions and the end of the plateau phase is typically not accompanied by any spectral evolution. These properties are not compatible with the standard afterglow model based on synchrotron emission from a shock-accelerated electron population (e.g. \citealt{Sari1998ApJ,Gra2002}). So far no firm conclusion has been reached on the origin of plateaus, but it is commonly believed that
they encode crucial information on the GRB central engine.

The leading interpretation proposed for the origin of plateaus invokes the presence of a central engine that can provide continuous energy injection, such as, for example, the fallback onto a newly formed black hole (BH) or dipole radiation from magnetar spin-down. Both these scenarios can fairly well describe the observed light curve morphologies (e.g. \citealt{Rowlinson2013MNRAS,Li2018ApJS}). An important step forward was recently made by \cite{Stratta2018ApJ}. They successfully compared the magnetar model with a data set of 51 GRBs, finding evidence of consistency
of the GRB distribution in the diagram of magnetic field versus spin period (B-P) with the well known
spin-up line for accreting Galactic X-ray pulsars (e.g. \citealt{Bhattacharya1991PhR,Pan2013ApSS}). In particular, the normalization of the observed relation perfectly matches the spin-up line
predictions for typical neutron star masses ($\sim1\textrm{M}_{\odot}$) and radii ($\sim10$ km), and for mass accretion rates expected in GRBs ($10^{-4}\,\textrm{M}_{\odot}/\textrm{s}<\dot{M}<0.1\,\textrm{M}_{\odot}$/s). This result was independently confirmed two years later \citep{Lin2020ApJ} supporting the presence of a magnetar as central engine. 

Another possible interpretation invokes a completely different scenario where the plateau is created by the high latitude emission (HLE, \citealt{fen1996ApJ...473..998F}) from a structured jet \citep{dai2001ApJ...552...72D,lip2001ARep...45..236L,ros2002MNRAS.332..945R}, associated with the prompt phase of the GRB  \citep{Oganesyan2020ApJ}. Before GW170817/GRB 170817A, the GRB jet structure was typically approximated with a uniform energy and velocity distribution within the cone aperture. The off-axis view of GRB 170817 enabled us for the first time to observe evidence of the structured nature of the jet, where energy and velocity distribution decrease gradually towards the jet edges. Independent of the jet structure, the HLE, namely the radiation observed from larger angles relative to the line of sight after the prompt emission from a curved surface is switched off \citep{Kumar2000ApJ}, is typically invoked to explain the initial steep decay phase of X-ray light curves (e.g. \citealt{Liang2006ApJ,Tag2005,Bar2005}) or X-ray Flares (e.g. \citealt{Rossi2011AA}).
In the case of a structured jet (SJ) the expected HLE lasts longer. Indeed, compared to a top-hat jet, photons departing from a given angular distance from the jet axis ($\theta$) are less Doppler boosted, and therefore the radiation is less beamed. This effect gives an extra contribution to the observed flux at late times, explaining the flattening and the longer duration of the HLE. 
This results in a shallow segment of the light curve for a portion of time consistent with the duration of the observed plateau.
The HLE from a SJ has the ability to reproduce the temporal evolution of the plateau and also its spectral properties.\\
\cite{ben2022} recently proposed a different interpretation where X-ray plateaus can form during the afterglow phase if the observer is slighly off-axis from the jet, though no direct comparison with observed light curves are performed. They show how this interpretation can explain most of the main plateau properties, including a correlation between the prompt energy, the plateau luminosity and duration, which is specific to their model and not expected in energy injection scenarios.
Since all these
interpretations can fairly well reproduce the plateau features observed in X-rays, it is challenging to identify the actual mechanism. 

This work is the continuation of a previous work (Stratta et al. 2022, in press) where 
we explore the behaviour of the plateaus over a larger energy band, in particular adding optical observations, to investigate if such a
degeneracy can be broken.
A number of events show a plateau feature also at optical wavelengths (e.g. \citealt{Si2018ApJ,Dainotti2020ApJ}), but a comprehensive consistency check of X-ray and optical light curves within the afterglow model is still confined to a few cases (e.g. \citealt{Gompertz2015MNRAS,Zhang2018}). 
In this work, we study a sample of 30 GRBs with evidence of an X-ray plateau and simultaneous optical data during and after the plateau
phase, to test different scenarios. 
In Section \ref{sample} we present the sample while in Section \ref{dataanalysis} we present our analysis of the X-rays and optical plateaus. We discuss our results in Section \ref{discussion} and draw our conclusions in Section \ref{conclusion}.  Throughout this work, the flux density of the afterglow is described as F$_\nu(t)\propto t^{-\alpha} \nu^{-\beta}$. We assumed a $\Lambda$CDM cosmological model for calculations. All errors are at $1\sigma$ unless otherwise specified.

\section{GRB sample}
\label{sample}
\subsection{X-ray plateau sample selection}

We consider all the GRBs detected with Swift in about 13 years (from 2005 up to mid 2018) with X-ray afterglow. In order to identify GRBs with evidence of a plateau feature in their afterglow, we use the publicly available Swift XRT Repository\footnote{\url{https://www.swift.ac.uk/xrt_live_cat/}}
\citep{Eva2007,Eva2009} and the provided data analysis tools to fit multiple power laws along the light curve of each GRB. 
We select all the GRBs that present in the X-ray light curve at least one segment with a temporal slope $-0.8 \leq \alpha \leq 0.8$ within errors. This phenomenology is indeed
challenging for the standard afterglow model assuming a constant density environment as typically observed in GRBs (e.g. \citealt{Sari1998ApJ}). A wind environment can give rise to a flat temporal evolution, but it requires a rising spectral slope in the $F_{\nu}-\nu$ plane, which is a condition never observed in X-ray afterglows. Defining $t_i$ and $t_f$ the initial and final time of the plateau, we also discard all GRBs where $t_f/t_i<2$. 
This requirement is necessary in order to select a plateau long enough to have a sufficient number of photons to perform a time-resolved spectral analysis, as explained in Sec.~\ref{trsa}.

\subsection{X-ray plateau optical counterpart sample}

\label{sec_xsam}
In order to investigate on the broad-band properties of the plateau, 
in this work we focus on
a sample subset with multi-band optical/NIR follow-up during the X-ray plateau phase. Specifically, we select GRBs for which multi-band data during the X-ray plateau and post-plateau phase are enough
to perform a time resolved spectral  analysis (see next section).

The optical sample is based on criteria and methods originally presented in \cite{Kann2006ApJ}. The sample includes GRBs with afterglow, redshift and good coverage in the UV/optical/NIR domain and is based on the updated sample and analysis of works in preparation (Kann et al. 2022a,b,c, in prep.). For this work in particular, we have re-analyzed GRB 110213A, while the analysis of GRB 180728A is taken from Rossi et al. (in prep.). 
Hereby, we assume achromaticity and that all bands follow the same temporal evolution.  
Any contribution to the afterglow flux from the presence of the GRB host galaxy and/or an associated supernova has been subtracted. 
The SEDs are fitted with local dust extinction laws \citep{Pei1992ApJ}, determining the intrinsic spectral slope $\beta_0$ and the extinction in the rest-frame $V$ band $A_V$ (in mag). Using the method of \cite{Kann2006ApJ}, the light curves are then corrected for line-of-sight extinction. 
The achromatic nature of the light curves allows us to shift data from other bands to the common $R_C$ band, creating a maximally dense light curve for each analyzed GRB afterglow. Please note that we use results based only on the optical/NIR SED.
We report the $A_V$ value of each GRB of our sample in Table \ref{tab_opt}.
Most of these measurements are found in the work presented in \cite{Kann2010ApJ,Kann2011ApJ} and Kann et al. 2022a,b,c, in prep. There are a few cases (indicated in the table) where we refer to works dedicated to single events.
Our final sample of GRBs with X-ray plateau and optical counterparts consists of 30 events. 
General information of the full GRB sample, including burst duration, redshift and energetics, are quoted in Tab.~\ref{tab_gen}.  

\section{Data analysis}
\label{dataanalysis}
\begin{figure}
    \centering
    \includegraphics[width=1.0\columnwidth]{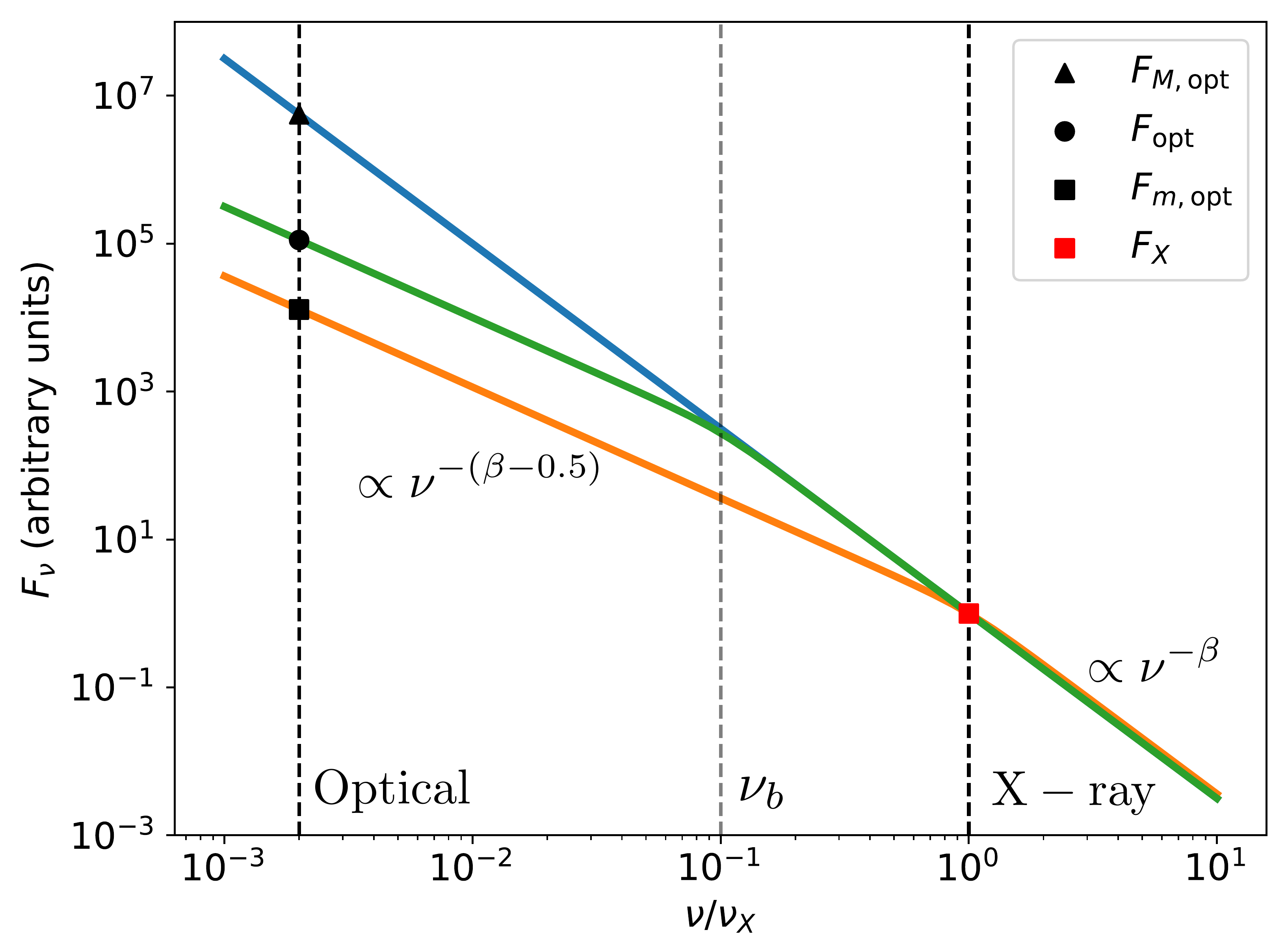}
    \caption{Schematic representation of the method used to compare X-ray and optical data. $F_{m,\rm opt}$ and $F_{M,\rm opt}$ are the minimum and maximum optical fluxes allowed by the standard afterglow model assuming that there is a single break $\nu_{\rm opt}<\nu_b<\nu_X$. $F_X$, $F_{\rm opt}$ and the X-ray spectral slope are derived from spectral analysis. Therefore, if $F_{m,\rm opt}<F_{\rm opt}<F_{M,\rm opt}$, we then consider that the optical counterpart is consistent with a single synchrotron spectrum and we can derive $\nu_b$. The blue line has no break, the orange line has $\nu_b=\nu_X$, while the green line has a break between optical and X-ray.} 
    \label{draw}
\end{figure}

Our goal is to verify that both the optical and X-ray emission during the X-ray plateau are consistent at each time with a single synchrotron component. To achieve this goal we perform a time resolved spectral analysis in the XRT band, fitting the  X-ray data with a single absorbed power law. From this analysis we derive the X-ray flux and the photon index. We then extrapolate the expected flux in the optical regime at the X-ray binning times (see below and Fig.~\ref{draw}), and we check whether the extrapolation is compatible with the optical observations shifted to the $R_C$ band (see sec.~\ref{sec_xsam}).

\subsection{X-ray time resolved spectral analysis}
\label{trsa}
The X-ray light curves have been re-binned imposing that the number of counts per bin in the band (0.5-10) keV is above a certain threshold $N_0$ and we choose $N_0>500$ counts. For bins with $N_0<500$ the spectral analysis gives too large errors on the parameters or the fit does not converge at all, due to the noisiness of the spectrum.
Since the X-ray light curves present some observational gaps, the temporal length is not the same for all the bins. Moreover, given the criteria described above, the bin length tend to be larger in the post-plateau phase, where the count rate decreases.\\
We perform for each bin a spectral analysis using XSPEC, version 12.10.1 and PyXspec. We consider only photons in the band $E=0.5-10$ keV.
Each spectrum is modeled by adopting an absorbed power law and for the absorption we use the T\"ubingen-Boulder model \citep{Wilms2000}. Since for all the GRBs of our sample the redshift is known, we distinguish a Galactic absorber and the host galaxy absorber. The Galactic absorption is taken from \cite{Kab2005}. The specific syntax in XSPEC is \textit{tbabs*ztbabs*po}. 
The estimation of the host equivalent hydrogen column density $N_H$ is performed on the time-integrated spectrum
of the post-plateau phase, where we do not expect strong spectral evolution, as verified by \cite{But2007} and \cite{Mu2016}. Once the host $N_H$ is determined, it is fixed to be the same for all the bins of the light curve. The only free parameters are photon index and normalization. Such a procedure is preferred to the case of leaving the host $N_H$ as a free parameter, because of the degeneracy between photon index and column density.

\subsection{Comparison of optical and X-ray data}
\begin{figure*}[t]
     \centering
     \begin{subfigure}{0.47\textwidth}
         \centering
         \includegraphics[width=\textwidth]{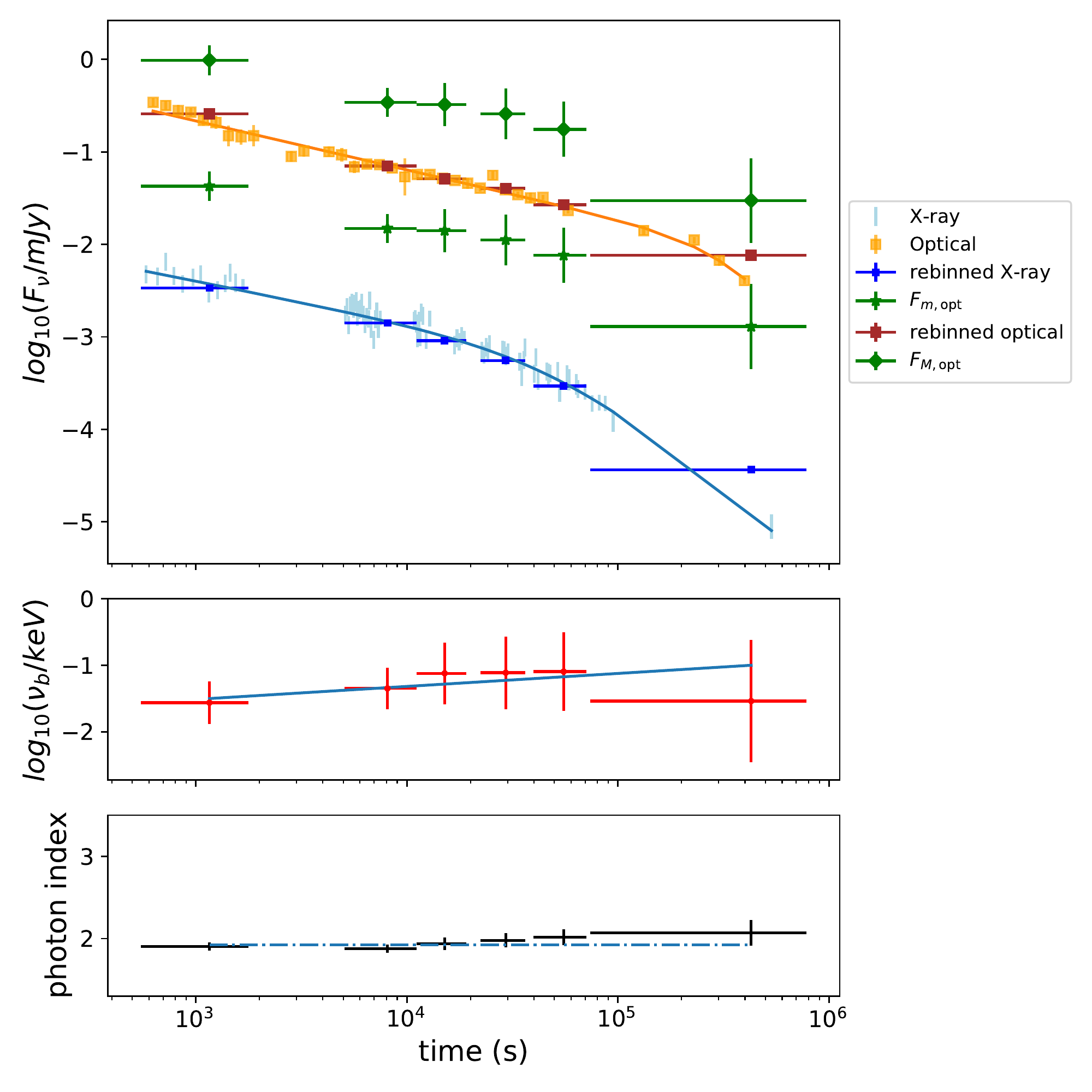}
         \caption{050319}
         \label{fig:y equals x}
     \end{subfigure}
     \hfill
     \begin{subfigure}{0.47\textwidth}
         \centering
         \includegraphics[width=\textwidth]{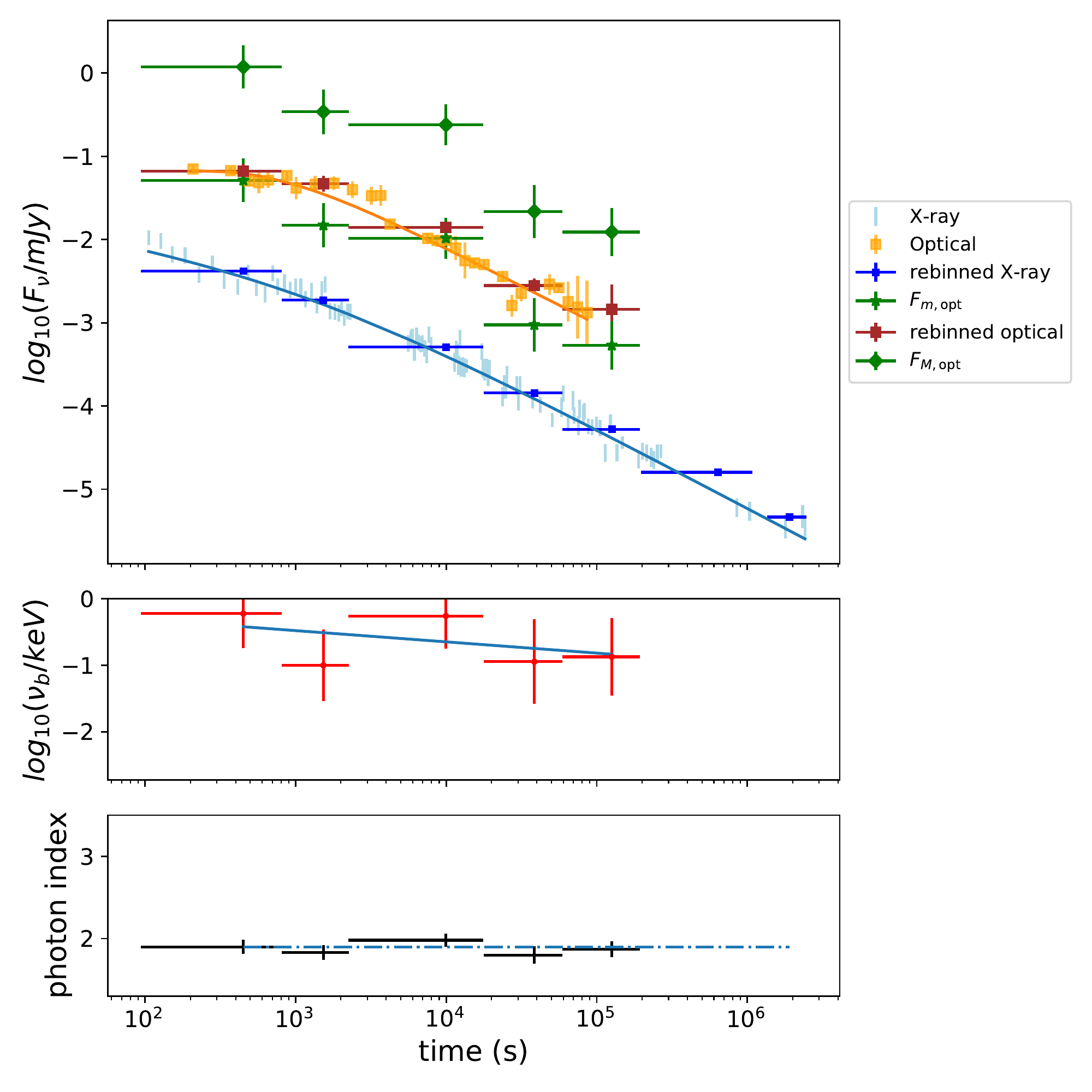}
         \caption{050416A}
         \label{fig:three sin x}
     \end{subfigure}
     \\
     \begin{subfigure}{0.47\textwidth}
         \centering
         \includegraphics[width=\textwidth]{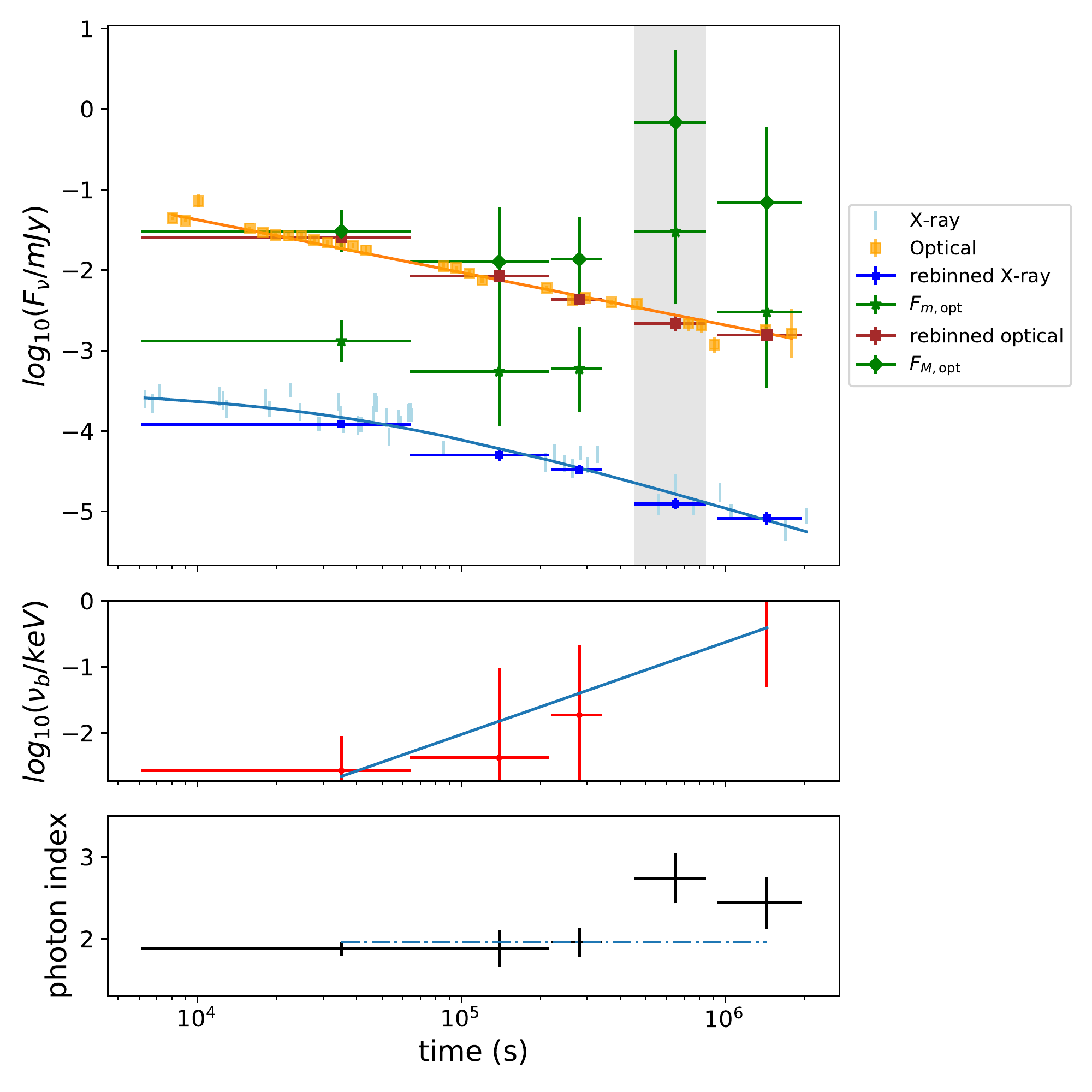}
         \caption{050824}
         \label{fig:five over x}
     \end{subfigure}
     \hfill
     \begin{subfigure}{0.47\textwidth}
         \centering
         \includegraphics[width=\textwidth]{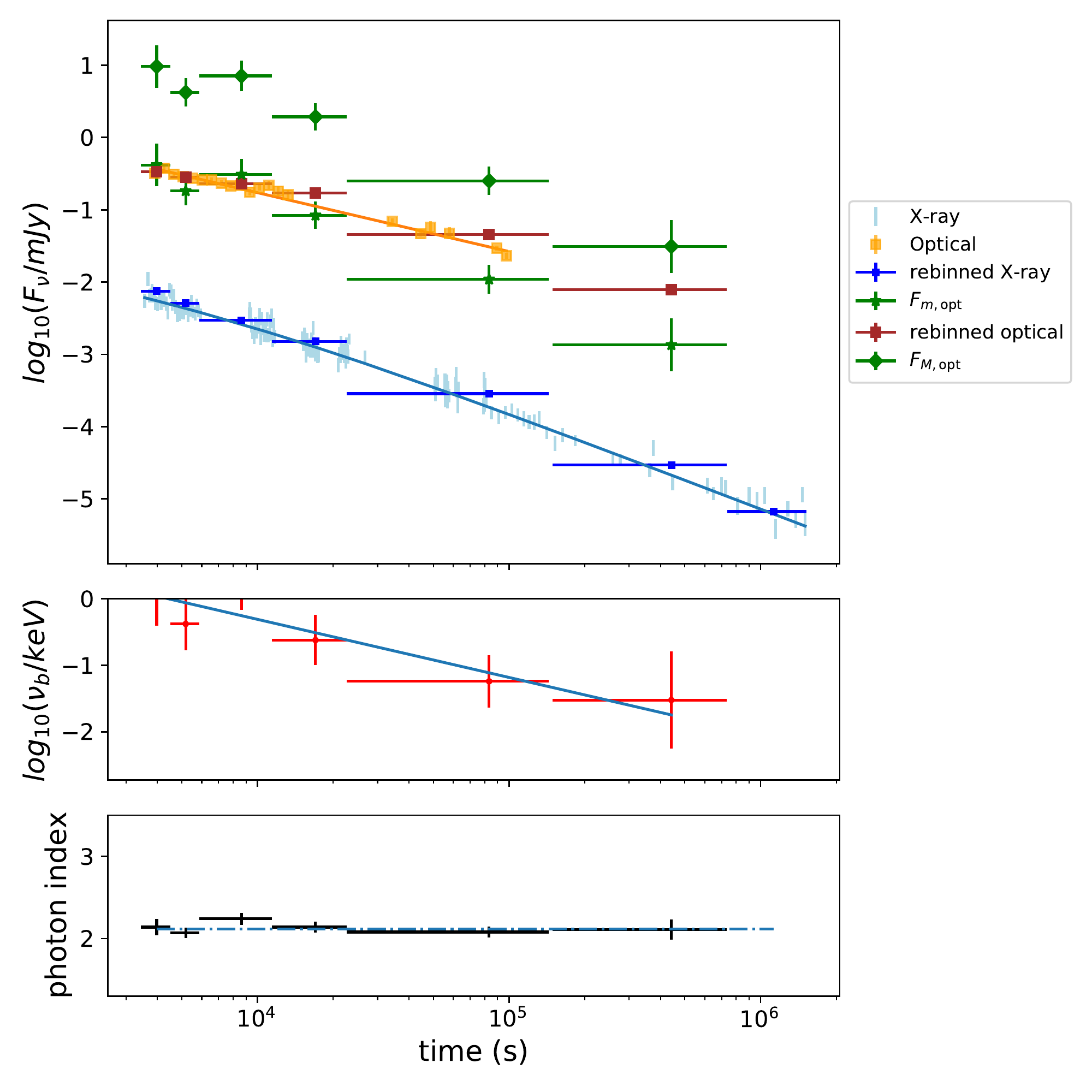}
         \caption{051109A}
         \label{fig:five over x}
     \end{subfigure}
        \caption{Summary plots of the simultaneous optical/X-ray spectral analysis for the first sample ({\it Sample 1}), where optical and X-ray are compatible with a single synchrotron spectrum. The top panel shows the X-ray (blue points) and the optical (orange points) light curves. The solid blue and orange lines are the best fit curves that interpolate the X-ray and optical light curves, respectively. $f_m$ is the optical flux density extrapolated from X-ray flux assuming a spectral break $\nu_b=\nu_X$, while $f_M$ is the optical flux density extrapolated from X-ray flux assuming no spectral break. Gray vertical band indicates when $f_{opt}<f_m$. In the middle panel we report the evolution of the break frequency $\nu_b$ as a function of time (red error bars), while the blue line indicates the best fit with a power law ($n_b\propto t^s$). In the bottom panel we show the value of the X-ray photon index. }
        \label{s1}
\end{figure*}

\begin{figure*}[t]
     \centering
     \begin{subfigure}{0.47\textwidth}
         \centering
         \includegraphics[width=\textwidth]{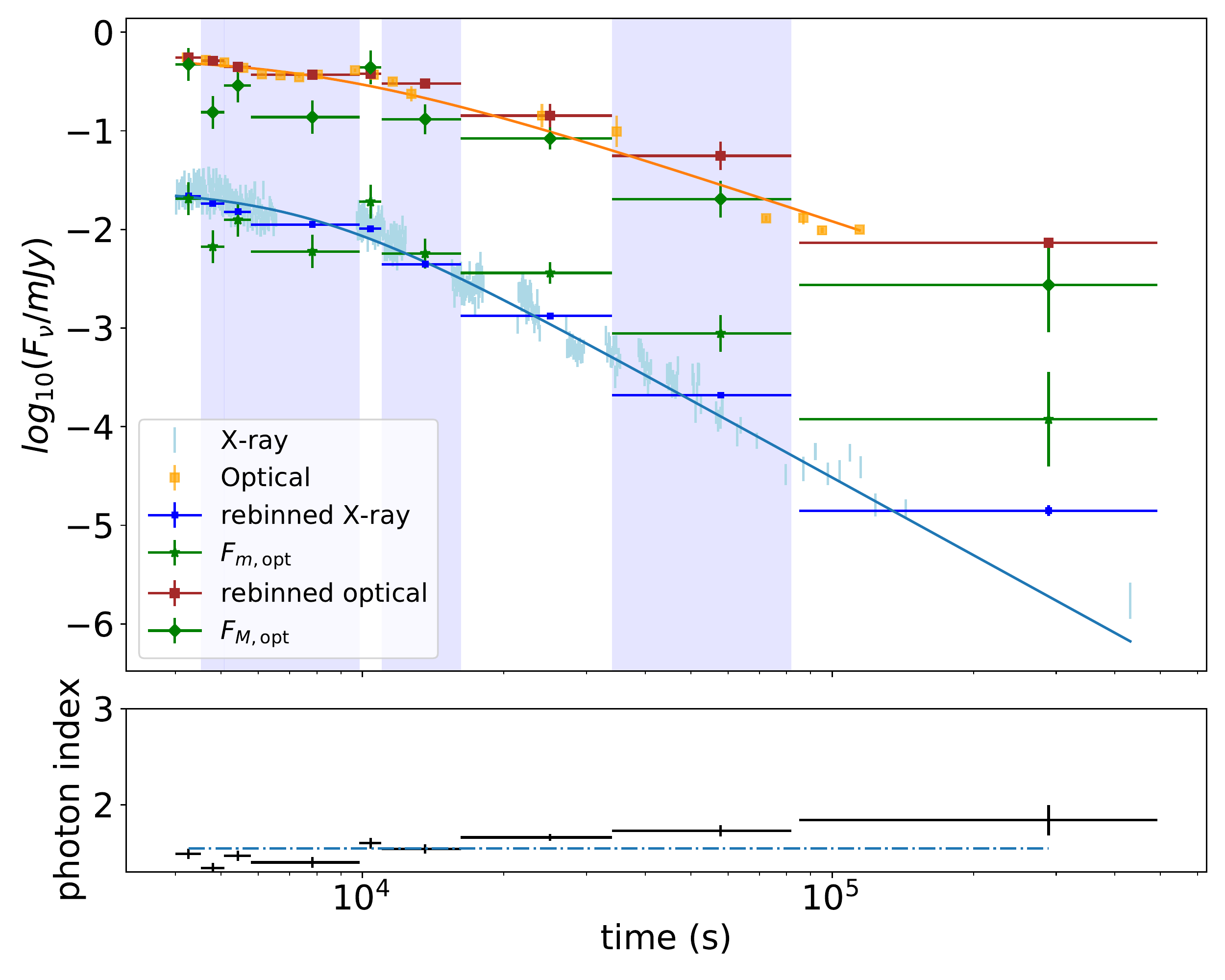}
         \caption{050730}
         \label{fig:y equals x}
     \end{subfigure}
     \hfill
     \begin{subfigure}{0.47\textwidth}
         \centering
         \includegraphics[width=\textwidth]{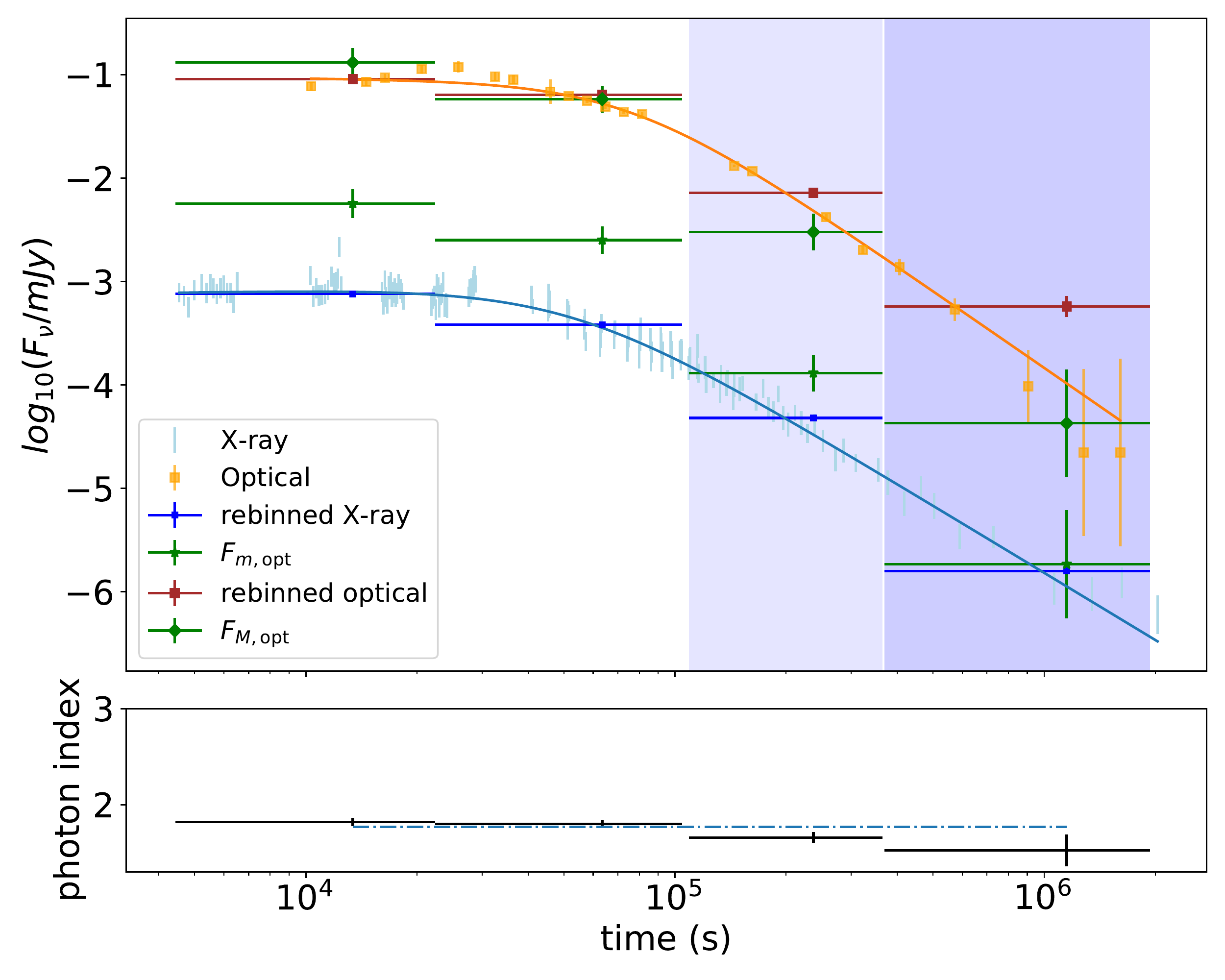}
         \caption{060614}
         \label{fig:three sin x}
     \end{subfigure}
     \\
     \begin{subfigure}{0.47\textwidth}
         \centering
         \includegraphics[width=\textwidth]{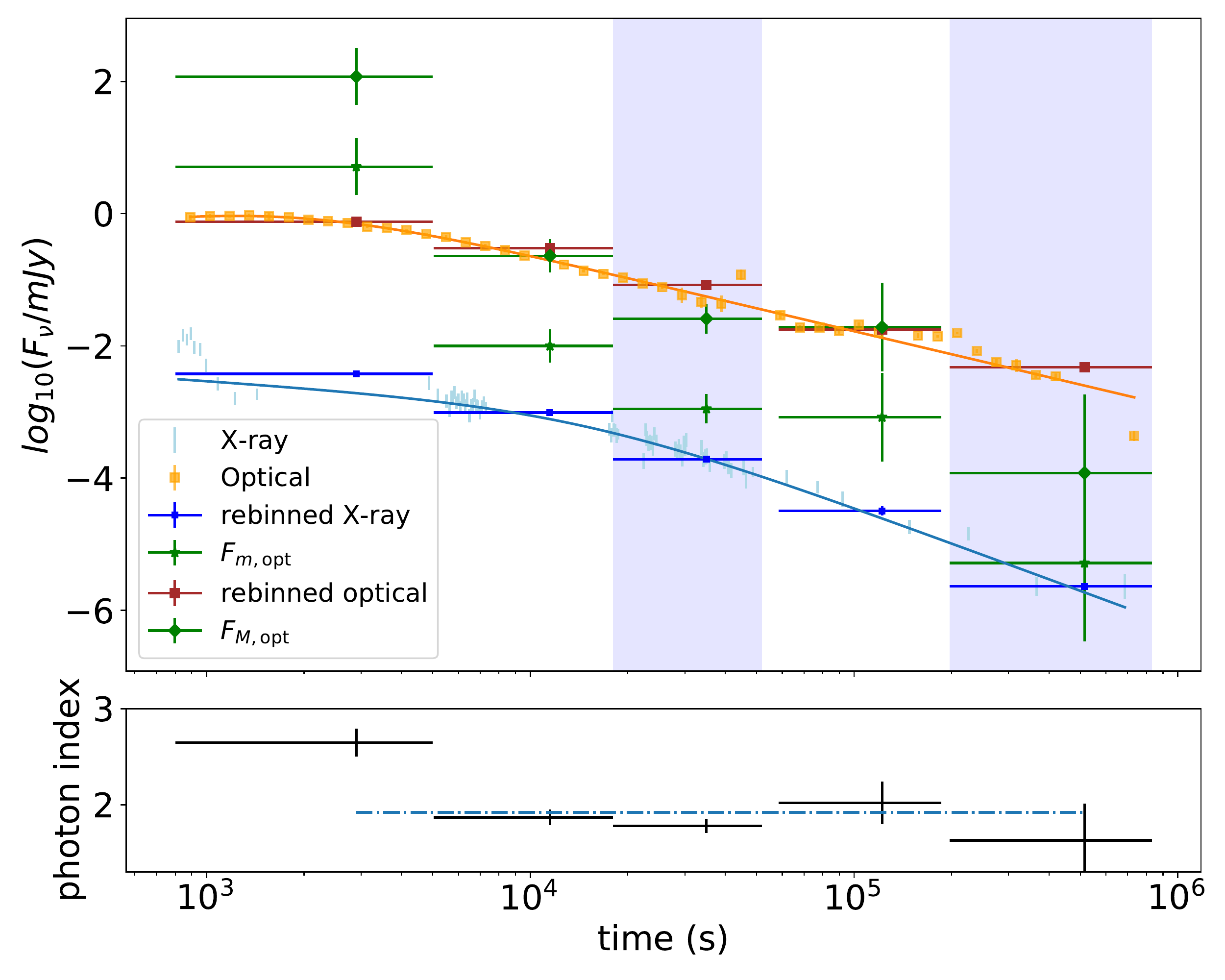}
         \caption{080310}
         \label{fig:five over x}
     \end{subfigure}
     \hfill
     \begin{subfigure}{0.47\textwidth}
         \centering
         \includegraphics[width=\textwidth]{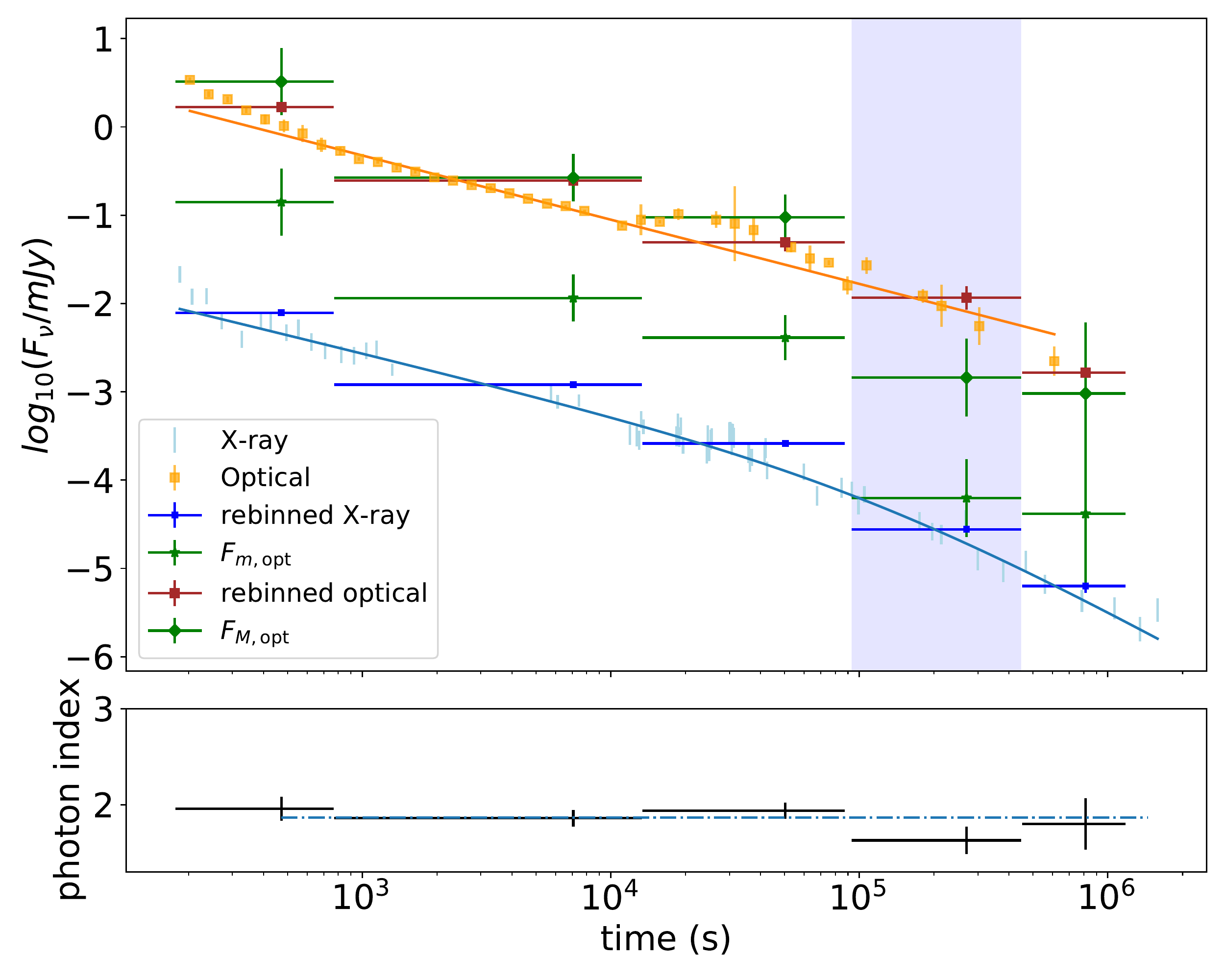}
         \caption{081007}
         \label{fig:five over x}
     \end{subfigure}
        \caption{Summary plots of the second sample of GRBs (\emph{Sample 2}) where for at least one temporal bin the optical flux is above $f_M$, indicating that optical and X-ray data are not compatible with a single synchrotron spectrum. The symbols are the same as in Fig.~\ref{s1} and we indicate with a vertical light-blue band the temporal bin where $f_{opt}>f_M$. The lower panel shows the temporal evolution of the X-ray photon index.}
        \label{s2}
        
\end{figure*}

\begin{table}[]
    \centering
    \begin{tabular}{l c c c}
    \hline
GRB 		& A & s & class					\\ \hline
050824		&	$	-8.99	\pm 4.45	$			&	$	-1.39	\pm 0.91	$	&   	(1)\\	\hline
050319		&	$	-2.16	\pm 1.07	$			&	$	-0.21	\pm 0.27	$	&   	(2)\\	\hline
050416A		&	$	0.03	\pm 1.09	$			&	$	0.17	\pm 0.28	$	&   	(2)\\	\hline
051109A		&	$	3.19	\pm 1.35	$			&	$	0.87	\pm 0.31	$	&   	(3)\\	\hline
060526		&	$	-3.84	\pm 2.04	$			&	$	-0.41	\pm 0.51	$	&   	(2)\\	\hline
060605		&	$	-4.26	\pm 2.37	$			&	$	-0.55	\pm 0.60	$	&   	(2)\\	\hline
060729		&	$	-0.71	\pm 0.64	$			&	$	0.24	\pm 0.13	$	&   	(3)\\	\hline
061121		&	$	2.07	\pm 0.61	$			&	$	0.76	\pm 0.16	$	&   	(3)\\	\hline
080413B		&	$	1.68	\pm 0.60	$			&	$	0.88	\pm 0.17	$	&   	(3)\\	\hline
080605		&	$	0.55	\pm 0.51	$			&	$	0.60	\pm 0.19	$	&   	(3)\\	\hline
090618		&	$	4.24	\pm 0.38	$			&	$	1.32	\pm 0.10	$	&   	(3)\\	\hline
091018		&	$	-2.81	\pm 0.76	$			&	$	-0.27	\pm 0.20	$	&   	(1)\\	\hline
091029		&	$	0.50	\pm 1.92	$			&	$	0.54	\pm 0.45	$	&   	(3)\\	\hline
100621A		&	$	-2.50	\pm 1.86	$			&	$	-0.37	\pm 0.48	$	&   	(2)\\	\hline
110213A		&	$	-0.34	\pm 0.69	$			&	$	0.37	\pm 0.17	$	&   	(3)\\	\hline
111228A		&	$	2.83	\pm 1.08	$			&	$	0.99	\pm 0.24	$	&   	(3)\\	\hline
130702A		&	$	0.59	\pm 2.17	$			&	$	0.48	\pm 0.38	$	&   	(3)\\	\hline
140419A	 & $ -3.74\pm1.23 $ 	 & $ -0.55\pm0.30 $ & (1) \\ \hline
180728A		&	$	0.68	\pm 0.88	$			&	$	0.61	\pm 0.19	$	&   	(3)\\	\hline

\end{tabular}
    \caption{Results of the fit of the temporal behaviour of $\nu_c$. The fitting function is $\log_{10}(\nu_c/\rm keV)=A-s\log_{10}(t/\rm s)$. Errors are given at the 1 sigma level of confidence. The last column identifies three classes, considering the value of s within the errors: (1) if $s<0$, (2) if $s=0$, (3) if $s>0$.}
    \label{fit_nuc}
\end{table}

In this section we show a method to derive the overall spectral properties of the plateau and post-plateau phases, exploiting the knowledge of the X-ray flux, the optical flux and the X-ray photon index. In the synchrotron scenario, two characteristic frequencies are defined: 1) the cooling frequency $\nu_c$, which is associated with the cooling Lorentz factor $\gamma_{c}=\frac{6 \pi m_{e} c}{\sigma_{\mathrm{T}} t B^2}$, where the time $t$ and the magnetic field $B$ are defined in the comoving frame; 2) the frequency $\nu_m$, associated with the minimum Lorentz factor $\gamma_m$ of the electron energy distribution. In the following, we work in the assumption of a slow cooling regime, namely $\nu_m<\nu_c$, as commonly found in GRB afterglows. Indeed, computing the ratio between the cooling frequency $\nu_c$ and $\nu_m$ \citep{Sari1998ApJ}, we have:
$$
\frac{\nu_c}{\nu_m} \sim 10^5 \varepsilon_{B,-3}^{-2} E_{52}^{-1} n^{-1} \varepsilon_{e,-1}^{-2} t_d
$$
where $\epsilon_B$ and $\epsilon_e$ are the fraction of energy that goes to magnetic field and electrons, respectively, E is isotropic kinetic energy, $n$ the circum-burst density and $t_d$ the time measured in days (we adopt the notation $Q_X=Q/10^X$). Therefore, for typical values we expect $\nu_c\gg \nu_m$.
In the slow cooling regime, we expect that the flux density goes like:
$$
F_{\nu}\propto \nu^{-\beta}
$$
for $\nu>\nu_c$ and 
$$
F_{\nu}\propto \nu^{-(\beta-1/2)}
$$
for $\nu<\nu_c$, where $\beta=p/2$ and $p$ is the power law index of the particle distribution. 
If $\nu_c<\nu_{\rm opt}$ then optical and X-ray flux can be connected with a single power law with spectral index $\beta$. In this case we define\footnote{As reference, throughout the paper we consider $\nu_X=$1 keV and $\nu_{\rm opt}=1.88\times 10^{-3}$ keV, corresponding to the central frequency of the R band.}:
$$
F_{M,\rm opt}=F_X(\nu_{\rm opt}/\nu_X)^{-\beta}
$$
If, instead, $\nu_{\rm opt}<\nu_c<\nu_X$ the optical flux is below the spectral extrapolation from X-rays assuming a single power law, i.e. $F_{\rm opt}<F_{M,\rm opt}$. The minimum expected optical flux corresponds to the case in which $\nu_c\sim\nu_X$ or $\nu_c>\nu_X$ and in this case we define
$$
F_{m,\rm opt}=F_X(\nu_{\rm opt}/\nu_X)^{-(\beta-1/2)}
$$
Therefore, whenever $F_{m,\rm opt}<F_{\rm opt}<F_{M,\rm opt}$, optical and X-ray emission are compatible with a synchrotron spectrum with $\nu_{\rm opt}<\nu_c<\nu_X$. In the case  $\nu_{\rm opt}<\nu_c<\nu_X$, we can write
$$
F_{\rm opt}=F_X\left(\frac{\nu_c}{\nu_X}\right)^{-\beta}\left(\frac{\nu_{\rm opt}}{\nu_c}\right)^{-(\beta-1/2)}.
$$ 
Two other possibilities are:
\begin{enumerate}
    \item $F_{\rm opt}>F_{M,\rm opt}$
    \item $F_{\rm opt}<F_{m,\rm opt}$
\end{enumerate}
In case 1) there is no way to justify optical and X-ray emission with a single synchrotron spectrum. Case 2) can be explained if we assume that $\nu_X<\nu_c$ and $\nu_{\rm opt}<\nu_m<\nu_X$. We recall that $F_{\nu}\propto \nu^{1/3}$ for $\nu<\nu_m$, but none of the analysed GRBs has an optical spectral index $\beta_{\rm opt}<0$. On the other hand, we have to take into account that the spectral breaks in synchrotron are not sharp, namely the spectral slope $dF_{\nu}/d\nu$ has a smooth transition across the break. Therefore, if $\nu_m\sim \nu_{\rm opt}$ and no other break is present between $\nu_{\rm opt}$ and $\nu_X$, it is still possible to measure $\beta_{\rm opt}>0$ and $F_{\rm opt}<F_{m,\rm opt}$. Hence, we conservatively consider case 2) as a case compatible with a single synchrotron spectrum.\\
In the case of $F_{m,\rm opt}<F_{\rm opt}<F_{M,\rm opt}$, since $F_{\rm opt}$, $F_X$ and $\beta$ are known, we can estimate $\nu_c$ as follows:
\begin{equation}
\label{nuc_sc}
    \nu_c=\left( \frac{F_X}{F_{\rm opt}}\right)^2 \left(\frac{\nu_{\rm opt}}{\rm keV}\right)^{1-p}  \rm keV
\end{equation}
The uncertainty on $\nu_c$ is derived through error propagation using the same formula.
Based on the comparison of optical and X-ray data, we define two sub-samples:
\begin{itemize}
    \item \emph{Sample 1}: for all the temporal bins the optical flux satisfies the condition $F_{m,\rm opt}<F_{\rm opt}<F_{M,\rm opt}$  or $F_{\rm opt}<F_{m,\rm opt}$. In this case optical and X-ray emissions are compatible with a single synchrotron spectrum. 19 GRBs satisfy this condition.
    \item \emph{Sample 2}: for at least one temporal bin $F_{\rm opt}>F_{M,\rm opt}$, which indicates an incompatibility with single component synchrotron origin. 11 GRBs fall within this case.
\end{itemize}
In the case of \emph{Sample 1} we derive the temporal evolution of $\nu_c$ during and after the X-ray plateau. If the number of temporal bins where both X-ray and optical data are available is larger than 4, we fit the temporal evolution of $\nu_c$ with a power law $\nu_c\propto t^{-s}$. The adopted fitting formula is:
\begin{equation}
\label{rel_nuc}
    \log _{10}\left(v_{c} / \mathrm{keV}\right)=\mathrm{A}-\mathrm{s} \log _{10}(\mathrm{t} /\mathrm{~sec}).
\end{equation}
The results of the fit are reported in Tab.~\ref{fit_nuc}. In our sample there are 5 GRBs with s consistent with 0 within the errors, 11 with $s>0$ and 3 with $s<0$.\\
The same analysis about the temporal evolution of the spectral break can be repeated relaxing the assumption that we are in slow cooling regime. All the considerations reported above still hold, but in fast cooling we have:
$$
F_{m,\rm opt}=F_X(\nu_{\rm opt}/\nu_X)^{-1/2}
$$
and the break frequency would be $\nu_m$, which can be analogously derived and corresponds to 
\begin{equation}
\label{nuc_fc}
    \nu_m=\left(\frac{F_X}{F_{\mathrm{opt}}}\right)^{2 /(p-1)}\left(\frac{\nu_{\mathrm{opt}}}{\mathrm{keV}}\right)^{1 /(1-p)} \mathrm{keV}
\end{equation}

The expression of $F_{M,\rm opt}$ is unchanged, since the spectral slope above the break is the same in both slow and fast cooling regime. This means that the selection condition $F_{\rm opt}\lessgtr F_{M,\rm opt}$ would produce the same classification of \textit{Sample 1} and \textit{Sample 2} as in slow cooling. Comparing eqs.~\ref{nuc_sc} and \ref{nuc_fc}, we notice that the break frequency derived in the slow cooling regime ($\nu_{c,SC}$) and the one derived for the fast cooling regime ($\nu_{m,FC}$) are connected by the following relation:
$$
\nu_{m,FC}\propto \left(\nu_{c,SC}\right)^{\frac{1}{p-1}}
$$
Therefore, having fitted the temporal behaviour of $\nu_{c,SC}$ as $\propto t^{-s}$, in the assumption of a fast cooling regime we would have obtained 
$$
\nu_{m,FC}\propto t^{\frac{s}{1-p}}
$$
This implies that, for typical values of $p>2$, if $\nu_{c,SC}$ decreases in time, $\nu_{m,FC}$ would decrease as well.

\subsection{Temporal fit of X-ray and optical light curves}

\begin{figure}
    \centering
    \includegraphics[width=1.0\columnwidth]{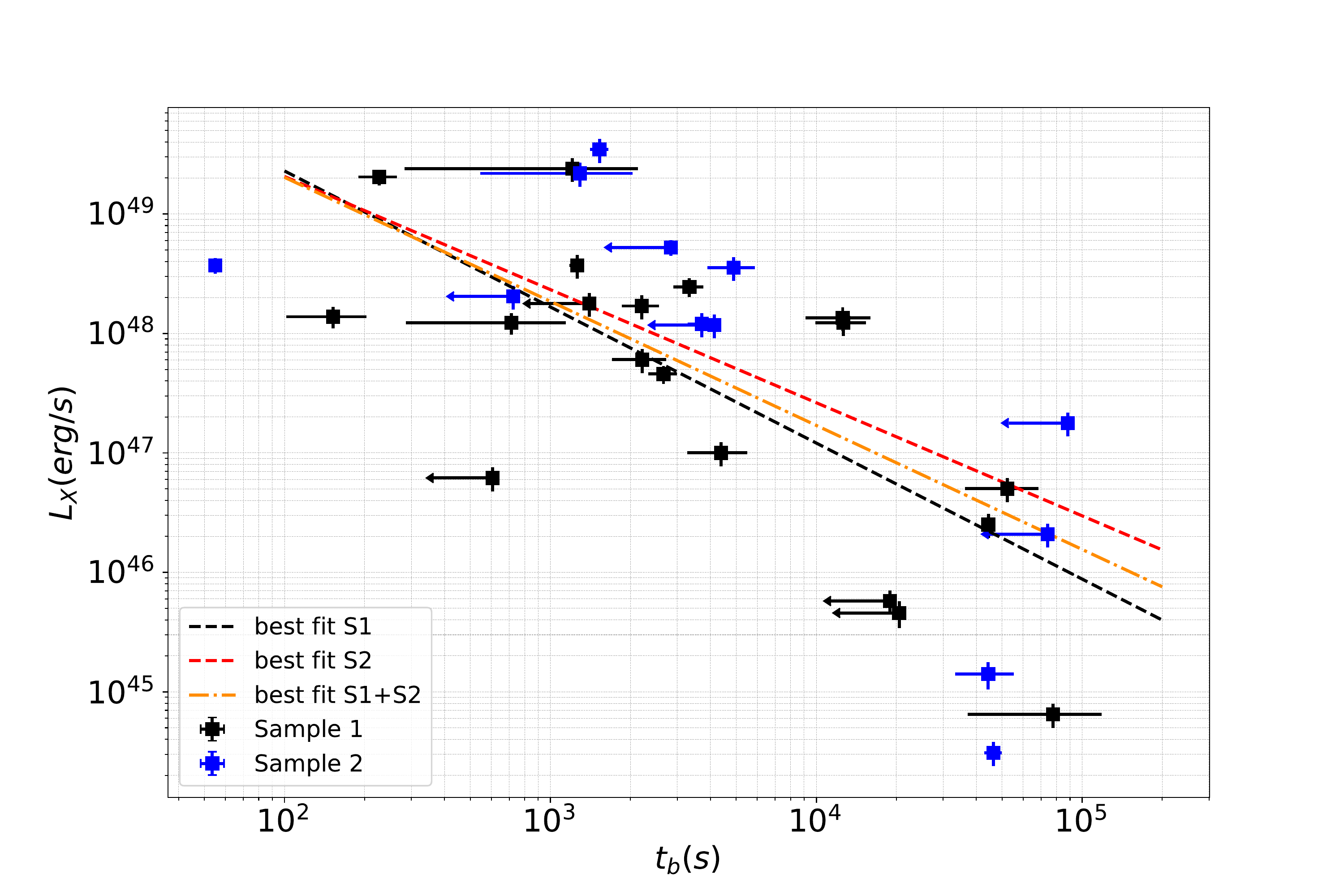}
    \caption{Relation between the X-ray luminosity and duration of the plateau. We distinguish \emph{Sample 1} and \emph{Sample 2} with black and blue points, respectively. Arrows indicate upper limits. Time is reported in the rest frame of the source. The black (red) dashed line is the best fit line for the \emph{Sample 1} (\emph{Sample 2}), where the fitting function is $\log_{10}(L_X)=N+\lambda \log_{10}(t_b)$. The orange dot-dashed line is the corresponding best fit for the union of \emph{Sample 1} and \emph{Sample 2}.}
    \label{Lx-tp}
\end{figure}

\begin{figure}
    \centering
    \includegraphics[width=1.0\columnwidth]{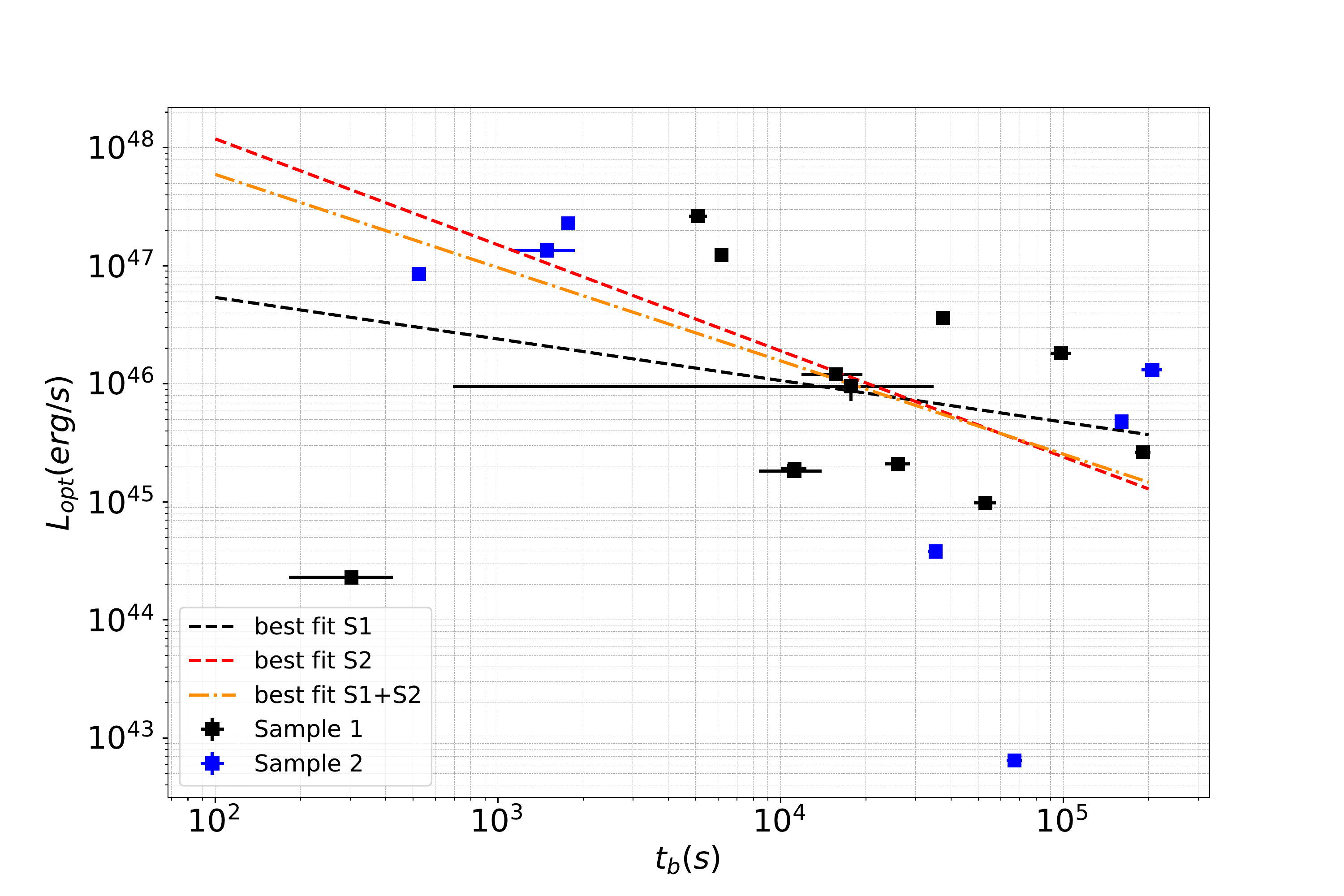}
    \caption{Relation between the optical luminosity and duration of the plateau. We distinguish \emph{Sample 1} and \emph{Sample 2} with black and blue points, respectively. Time is reported in the rest frame of the source. The black (red) dashed line is the best fit line for the \emph{Sample 1} (\emph{Sample 2}), where the fitting function is $\log_{10}(L_{\rm opt})=N+\lambda \log_{10}(t_b)$. The orange dot-dashed line is the corresponding best fit for the union of \emph{Sample 1} and \emph{Sample 2}.}
    \label{Lo-tp}
\end{figure}

\begin{figure}
    \centering
    \includegraphics[width=1.0\columnwidth]{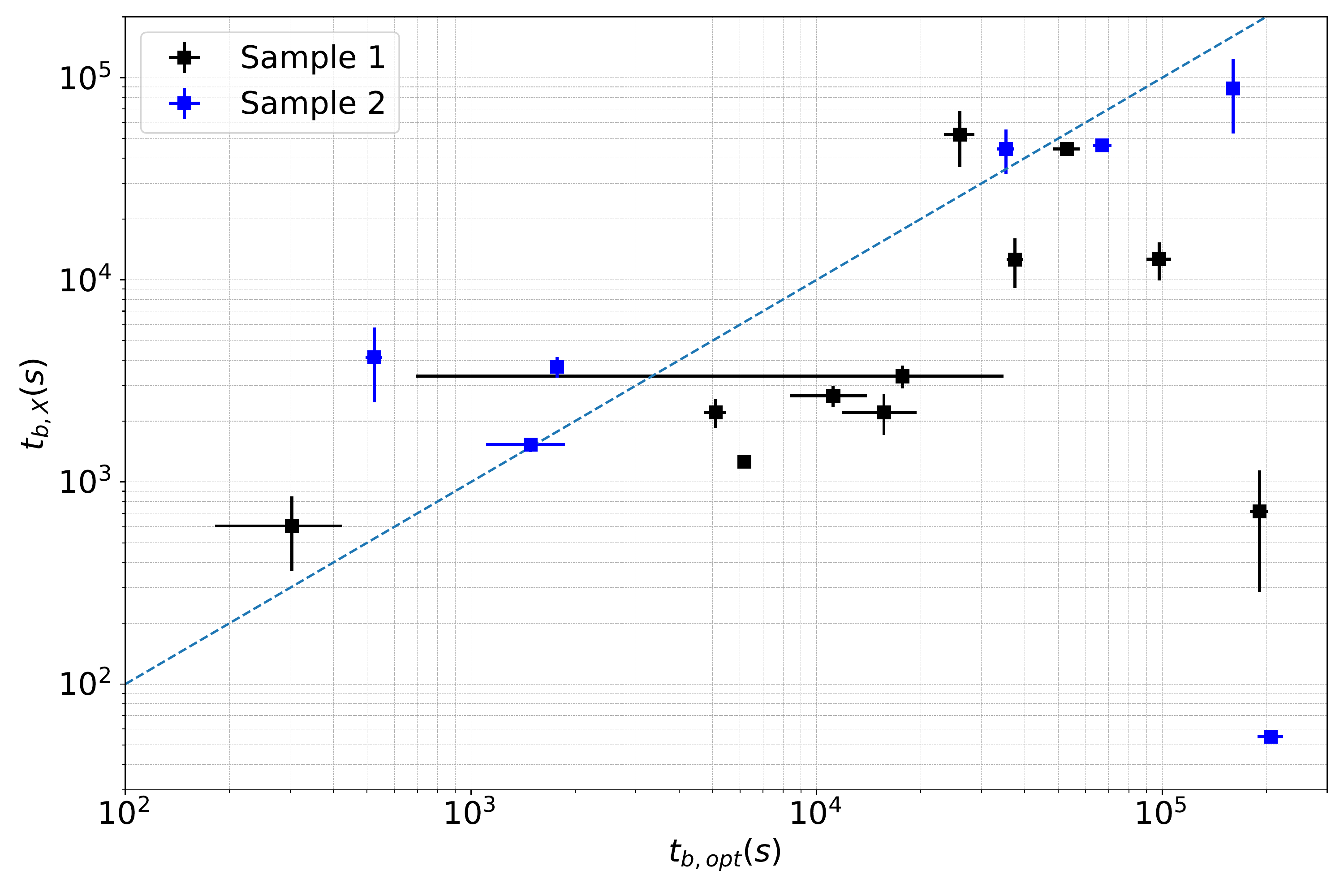}
    \caption{Relation between the duration of the X-ray and optical plateau. We distinguish \emph{Sample 1} and \emph{Sample 2} with black and blue points, respectively. Time is reported in the rest frame of the source. The blue dashed line is corresponds to $t_{b,\rm opt}=t_{b,X}$.
    }
    \label{tp-tp}
\end{figure}

In order to check if the X-ray plateau  has a corresponding plateau phase in the optical, we fit both X-ray and optical light curves of GRBs in \textit{Sample} 1 and \textit{Sample} 2 with an empirical broken power law, in the form:
$$
F(t)=\frac{N}{(t/t_b)^{a}+(t/t_b)^{b}}
$$
This functional form is valid if the temporal decay before the break is shallower than the one after the break. This condition is always satisfied for all the X-ray and optical light curves, with the exception of the optical light curve of GRB 140419A. In this case we adopt the following functional form:
$$
F(t)=N((t/t_b)^{a}+(t/t_b)^{b})
$$
In the case of optical light curves we rebin the data points as follows: if $t_i$ and $t_f$ are the initial and final times of the optical data, we rebin using a grid of 50 bins spaced logarthmically in the interval $[t_i-t_f]$.
The fit is performed using the python function \emph{curvefit}, which is based on least squares minimisation. If the fit does not converge, i.e. if one of the parameters of the model cannot be constrained, we fit
with a single power law. The results are reported in Tab.~\ref{fit_x} and \ref{fit_opt} for X-ray and optical light curves, respectively. The best fit curves are reported as solid blue (X-ray) and orange (optical) lines in Fig.~\ref{s1} and \ref{s2}. In all the cases, the fit of X-ray light curves gives a reasonably good value of $\chi^2$. Instead, for some optical light curves $\chi^2\gg1$, indicating the presence of more complex temporal structures that cannot be approximated by a simple broken power law. The cases showing the worst agreement are GRB 100621A \citep{gre2013}, GRB 110213A \citep{Cuc2011} and GRB 100814A \citep{nar2014}, which clearly exhibit the superposition of a bump over the power law decay.\\
Looking at Tab.~\ref{fit_opt}, 8 GRBs have an optical light curve compatible with a single power law with $a\in [-0.8,0.8]$. Therefore, the light curves of these GRBs significantly show the presence of a plateau phase. All the other GRBs have an optical light curve compatible with a broken power law where the flatter segment has a shallow temporal slope in the interval $a\in [-0.8,0.8]$. The only exceptions are GRB 091018 (with $a>0.8$), GRB 081029 and GRB 150910A (with $a<-0.8$). Therefore 19/30 GRBs show a plateau both in X-ray and optical.\\ 
For all the GRBs that show a plateau in optical and/or in X-ray, we compute the X-ray/optical luminosity during the plateau as:
$$
L_{X,opt}=\expval{F_{X,\rm opt }}\times 4\pi D_L(z)^2
$$
where $\expval{F_{X,\rm opt }}$ is the average flux in the specific band. The duration of the plateau is approximated with the break time. Such an approximation is valid in the limit in which the initial time of the plateau satisfies the condition $t_i\ll t_b$, which is usually the case.\\
\begin{table}[]
    \centering
    \begin{tabular}{c c c }
    & \multicolumn{2}{c}{X-ray}\\ \hline
    \hline
    &$\lambda$ & N \\ \hline

   \emph{Sample 1} & $-1.14\pm 0.03$ & $0.08\pm 0.03$ \\ \hline
   \emph{Sample 2} & $-0.95\pm 0.03$ & $0.42\pm 0.03$ \\ \hline 
   \emph{Sample 1}+\emph{Sample 2} & $-1.04\pm 0.02$ & $0.23\pm 0.02$ \\ \hline  
   \hline
    & \multicolumn{2}{c}{Optical}\\ \hline
    &$\lambda$ & N \\ \hline

    \emph{Sample 1} & $-0.35\pm 0.03$ & $-0.97\pm 0.01$ \\ \hline
    \emph{Sample 2} & $-0.90\pm 0.01$ & $-0.72\pm 0.01$ \\ \hline 
   \emph{Sample 1}+\emph{Sample 2} & $-0.79\pm 0.01$ & $-0.81\pm 0.01$ \\ \hline  
   \hline

    \end{tabular}
    \caption{Best fit values of the relation between the plateau luminosity and the plateau duration, in X-ray and optical. The values of $\lambda$ and $N$ are specified in the text.}
    \label{bf_daino}
\end{table}
The relation between X-ray plateau luminosity and duration is shown in Fig.~\ref{Lx-tp}. The points, even if quite scattered, seem to follow the Dainotti relation \citep{Dai2010}. We fitted the $L_X-t_b$ relation with the function:
\begin{equation}
\label{daino}
   \log_{10} \frac{L_x}{10^{47} \rm erg s^{-1}}= N + \lambda \log_{10} \frac{t_b}{10^4 \, \rm s}
\end{equation}
We show in Fig.~\ref{Lo-tp} the analogous relation between plateau luminosity and duration in the optical. Also in this case we fit the same power law relation of eq.~\ref{daino}. The best fit parameters of the $L_X-t_p$ and $L_{\rm opt}-t_p$ relations are reported in Tab.~\ref{bf_daino}. The value of the slope of the $L_{\rm X}-t_b$ relation found by \cite{Dai2013} is $\lambda=-1.32\pm0.28$, steeper but still compatible with our values. For the $L_{\rm opt}-t_b$ relation, instead, \cite{Dainotti2020ApJ} report $\lambda=-1.12\pm 0.26$, significantly steeper than our values. We note however that in \cite{Dainotti2020ApJ} $L_{\rm X}$ is defined slightly different as the luminosity at the end of the plateau.\\
Finally, for the sub-sample of GRBs with a plateau in both optical and X-ray bands, we show in Fig.~\ref{tp-tp} the relation between the duration of the plateau in optical and X-rays. There is a clear indication that optical plateaus tend to be longer than X-ray plateaus, which is in line with our findings on $\nu_{opt}$ being smaller than $\nu_c$ is most of the cases. Among all the the analysed GRBs with both an X-ray and optical plateau, only a few have $t_{b,X}\simeq t_{b,opt}$, implying that the majority shows a chromatic plateau. There are two particular cases, GRBs 110715A and 080413B, which are in the bottom right corner of the plot, indicating an optical break of the plateau at much later times with respect to the X-ray break. Such a behavior seems to significantly deviate from the average trend, but looking at Figs.~\ref{mmm} and \ref{fig_080413B} it is possible that a simple broken power law does not describe sufficiently well the temporal structure of the optical light curve, therefore giving a possibly biased estimation of the optical break. Apart these two exceptions, $t_{b,X}$ and $t_{b,opt}$ show a correlation. Excluding GRBs 110715A and 080413B, we derive a Pearson correlation coefficient $C_P=0.78$ and a 2-tailed p-value of $3.4 \times 10^{-4}$.

\section{Discussion}
\label{discussion}
\begin{figure}
    \centering
    \includegraphics[width=1.0\columnwidth]{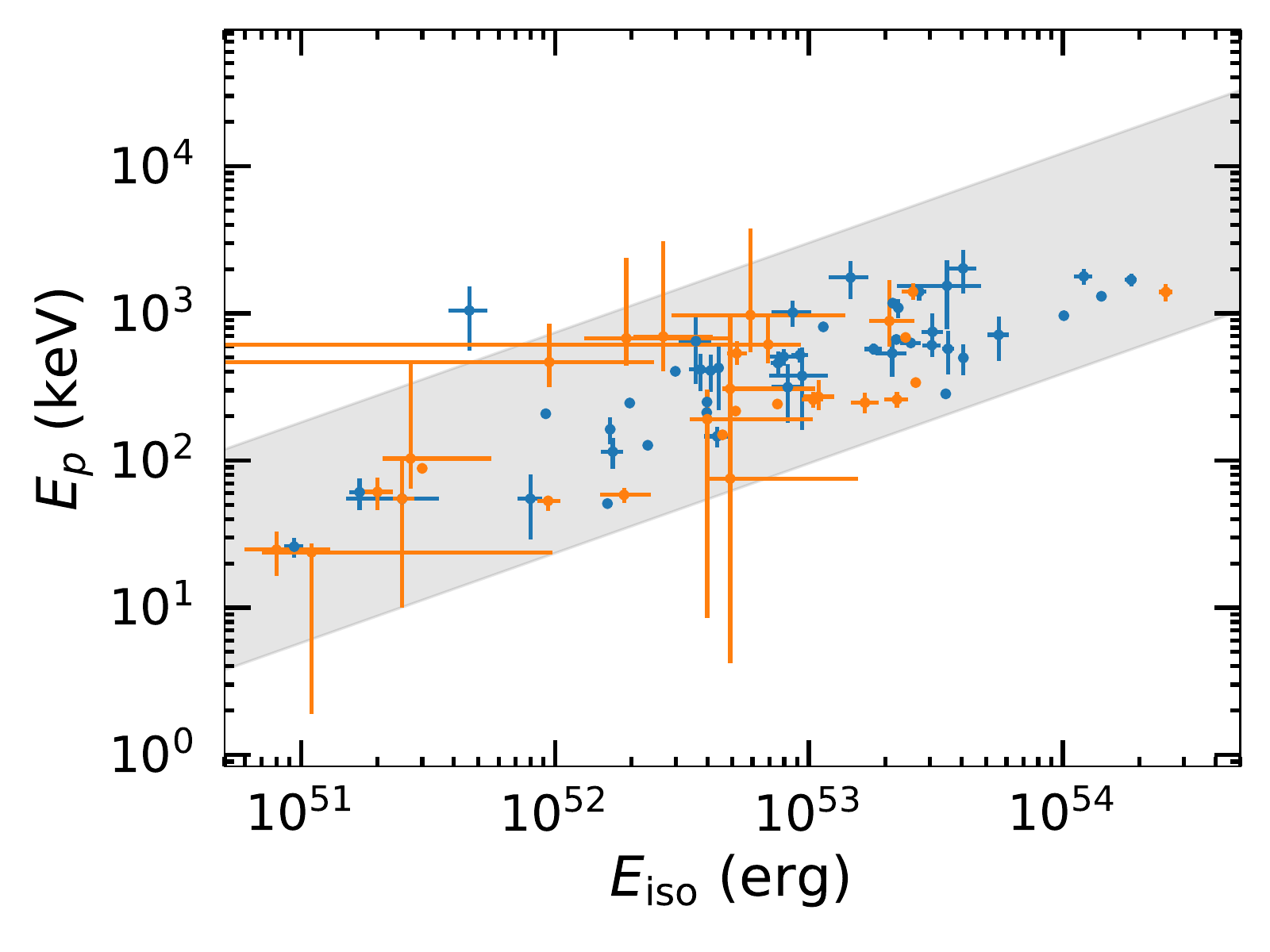}
    \caption{Location in the $E_p-E_{\rm iso} $ plane of the GRB sample selected in our work (orange points, see Tab.~\ref{tab_gen}). The blue points are taken from \cite{nava2012} and the gray region is the corresponding 3$\sigma$ band.}
    \label{Amati}
\end{figure}

The results of our analysis show that in 19 out of 30 GRBs the plateau has an optical-to-X-ray spectrum fully consistent with synchrotron emission from a single population of shock-accelerated electrons ({\it Sample 1}).
The comparison of the temporal properties of the X-ray and optical plateaus further confirms this interpretation. Indeed, the fact that the optical flux densities lie within the allowed range of values extrapolated from X-ray fluxes assuming a single synchrotron spectrum, allows us to infer $\nu_{\rm opt}<\nu_c<\nu_X$. This condition implies a slower evolution of the optical plateau than the X-ray one, in agreement with our findings that $t_{b,o}> t_{b,X}$ in most of the GRBs belonging to {\it Sample 1} (see fig.~\ref{tp-tp}).
In order to test whether the transition of $\nu_c$ across the optical band is simultaneous to a steepening of the optical light curve, we estimate the time $t^*$ for which $\nu_c(t^*)=\nu_{\rm opt}$ and we compare it with $t_{b,opt}$, reported in Tab.~\ref{fit_opt}. The value and the uncertainty of $t^*$ is estimated inverting the relation \ref{rel_nuc}, namely:
$$
t^*=\left(\frac{10^N}{\nu_{\rm opt}/\rm keV}\right)^{1/s} \rm sec
$$
In Tab.~\ref{t*} we compare $t^*$ and $t_{b,opt}$ of all the GRBs of \textit{Sample 1} which have $s>0$. The only case where $t^*$ and $t_{b,opt}$ are not compatible is GRB 090618. In all other cases, we have three possibilities: 1) we have an estimate of both $t^*$ and $t_{b,opt}$ and they are compatible, 2) only a lower limit on $t^*$ is available, but still compatible with $t_{b,opt}$, 3) no break in the optical is observed, possibly indicating that the break may occur later than the available data. While cases 2) and 3) are inconclusive, only for the first case (GRBs 061121, 080413B and 111228A) we can conclude that $t^*\simeq t_{b,opt}$, indicating that the optical temporal break is in agreement with the transition of the cooling frequency across the optical band.
\\
If the condition $\nu_{\rm opt}<\nu_c<\nu_X$ is satisfied, no spectral evolution is expected in the X-rays. We tested the presence of spectral evolution fitting the X-ray photon index temporal trend with a constant. In Tab.~\ref{fit_x} we report the average photon index (the average is taken over the whole available data) and the p-value of the fit. Among 19 cases, 6 GRBs have a p-value $<0.05$
showing significant deviation from a constant trend. For three of them (GRBs 060729, 061121 and 090618) the deviation is given by the hardening of the X-ray spectrum at late times. For GRBs 100621A, 110213A and 140419A there is evidence of spectral softening. For all these six cases, except for GRB 140419A, Tab.~\ref{fit_nuc} shows that the derived $\nu_c$ decreases in time. In this scenario, a softening of the X-ray spectrum is expected as $F_{\nu}\propto \nu^{-(p-1)/2}$ for $\nu<\nu_c$ and $F_{\nu}\propto \nu^{-p/2}$ for $\nu>\nu_c$. Since the synchrotron spectral shape has smooth transitions between a power law segment and the other, a temporal decrease of $\nu_c$ would produce a gradual spectral softening. On the contrary, a decreasing trend of $\nu_c$ is not in agreement with the evidence of spectral hardening. In this regard, we point out that in the derivation of the temporal evolution of $\nu_c$ we assumed that the spectral slopes above and below $\nu_c$ are constant in time. If there is an additional process which induces an intrinsic variation of the spectral slope, the estimation of the temporal trend of $\nu_c$ can be biased.
Hence, we conclude that the results shown for the evolution of the cooling frequency are fully reliable for the GRBs that do not show significant spectral evolution in the X-rays. Among the 12 cases of \textit{Sample 2}, seven of them have a p-value $<0.05$ (GRBs 050730, 080310, 100814A, 110715A, 120404A, 140419A and 150910A), but none of them shows a clear trend which points towards a softening or a hardening of the X-ray spectrum.\\
The consistency of the optical and X-ray data with a single spectrum can be interpreted as an indication that both X-ray and optical photons originate from the same emission region. Moreover, the process responsible for the X-ray plateau should also explain the observed evolution of the cooling frequency. In the standard scenario of a forward shock decelerating through the circumburst medium, the predicted temporal evolution of the characteristic synchrotron frequencies is \citep{Gra2002}:
\begin{equation}
\label{fs_ism}
    \nu_c \propto \epsilon_{B}^{-3 / 2} n_{0}^{-1} E_{52}^{-1 / 2} t_{\text {days }}^{-1 / 2}
\end{equation}
for an ISM with uniform particle density $n_0$, and
\begin{equation}
\label{fs_wind}
    \nu_c \propto  \epsilon_{B}^{-3 / 2} E_{52}^{1 / 2} t_{\mathrm{days}}^{1 / 2}
\end{equation}
for a medium with a wind-like density profile. With $E_{52}$ we indicate the isotropic energy in units of $10^{52}$ erg. The flux density, instead, is expected to decline like $F_{\nu}\propto t^{-a}$ and the predicted values are, in slow cooling regime (e.g., see \citealt{zha2006}):
$$
a=\frac{3}{4}(p-1) \text{ for } \nu<\nu_c, \, a=\frac{3p-2}{4} \text{ for }\nu>\nu_c
$$
for the ISM scenario and
$$
a=\frac{3p-1}{4} \text{ for } \nu<\nu_c, \, a=\frac{3p-2}{4} \text{ for }\nu>\nu_c
$$
for the wind scenario. Since the values of $p$ are likely above 2, the standard picture predicts afterglow light curves not flatter than $t^{-3/4}$. Therefore, even if the observed temporal evolution of $\nu_c$ can be compatible with the temporal behavior specified in eq.~\ref{fs_ism} and \ref{fs_wind}, the observed temporal slope of the X-ray and optical plateaus are incompatible with the standard FS scenario.\\
Modifications to the standard picture have been proposed in the literature \citep{Mis2021}, which invoke the  possible temporal evolution of the shock microphysical parameters, such as $\epsilon_B$. In the specific case, assuming a temporal evolution in the form $\epsilon_B \propto t^{\mu}$ and a circum-burst medium with a density profile $\rho \propto r^{-k}$, the cooling frequency evolves as $\nu_{c} \propto t^{-s}$, with:
$$
s=\frac{4+12 \mu-3 k(1+\mu)}{2(4-k)}.
$$
Moreover, considering the ordering $\nu_{\rm opt}<\nu_c<\nu_X$, the optical and X-ray flux densities are expected to evolve like $F_{\rm opt}\propto t^{a_o}$ and $F_X\propto t^{a_X}$, with
$$
a_o=\frac{1}{2}\left (\mu +\frac{k}{k-4} \right ) + \frac{p-1}{2}\left (\frac{4+12\mu-3k(1+\mu)}{2(k-4)} \right )
$$
and
$$
a_X=\frac{1}{2}\left (\mu +\frac{k}{k-4} \right ) + \frac{p}{2}\left (\frac{4+12\mu-3k(1+\mu)}{2(k-4)} \right ).
$$
In order to check the validity of this scenario, we compare, simultaneously, the expected values of $a_o$, $a_X$ and $s$ with the observed ones. In particular, we search for a combination of $\mu$ and $k$ such that the following relations are simultaneously satisfied:
$$
a_{o,th}\in [a_{o,obs}-\Delta a_o,a_{o,obs}+\Delta a_o ]
$$
$$
a_{X,th}\in [a_{X,obs}-\Delta a_X,a_{X,obs}+\Delta a_X ]
$$
$$
s_{th}\in [s_{obs}-\Delta s,s_{obs}+\Delta s ]
$$
where we indicate with $X_{th}$, $X_{obs}$ and $\Delta X$ the expected value, the observed value and the error of the quantity X, respectively. The intersection of these three conditions defines a region in the $k-\mu$ plane, which is in agreement with observations. We verified for all the GRBs of \textit{Sample 1} that a combination of $k$ and $\mu$ that satisfies the above conditions does not exist.
\\
Another solution invokes the presence of additional energy injected into the forward shock at late times. A single emitting region is compatible with the energy injection scenario, where additional energy is transferred to the external shock due to the late-time activity of the central engine. In this scenario, the dynamical evolution of the blast wave is determined by the rate of energy injection and the efficiency of conversion from injected energy to jet kinetic energy. Hence, the deceleration in the ISM is less severe and the flux drop is shallower. The injection of energy modifies the dynamical evolution of the external shock but it does not have influence on the dominating radiative mechanisms responsible for the dissipation of the particles' energy. Therefore, the spectral energy distribution should be the same of that in the standard scenario of particles dissipating through synchrotron radiation. We notice that, if the energy injection has impact only on the blast wave dynamics, the plateau should be achromatic (e.g. \citealt{fan2006MNRAS.369..197F}). This would imply that the temporal break corresponding to the transition from plateau to post-plateau phase should be the same in optical and X-rays. Though, as pointed out before, this is not the case if a spectral break is between the two bands. \\
Adopting the standard prescription of a injected luminosity in the form:
$$
L_{inj}\propto t^{-q}
$$
the temporal slope of the afterglow light curve at any frequency $\nu_{obs}$ can be predicted depending on the position of $\nu_{obs}$ relative to the synchrotron characteristic frequencies ($\nu_m$ and $\nu_c$). Moreover the temporal behavior of these last are modified consequently and read:
\begin{equation}
   \label{ism_nuc}
\nu_m\propto t^{-(2+q)/2}, \, \nu_c\propto t^{(q-2)/2} 
\end{equation}
for the ISM scenario (jet propagating into a ISM with constant density), and
\begin{equation}
\label{wind_nuc}
    \nu_m\propto t^{-(2+q)/2}, \, \nu_c\propto t^{(2-q)/2}
\end{equation}
for the wind scenario (jet propagating into a medium with density $n\propto r^{-2}$). Therefore, if $q<2$ ($q>2$) the energy injection model predicts a decreasing cooling frequency in the ISM scenario (wind scenario).\\
One of the possible sources of energy responsible for the jet refreshing is the long-lived highly magnetized neutron star (magnetar) left after the production of the GRB \citep{Dai1998,zhang2001,dall2011}. The magnetar looses rotational energy through spin down radiation and the associated released luminosity in the standard scenario of a rotating magnetic dipole depends on the rotational frequency as $L_{sd}\propto \Omega^4$. The temporal evolution of the spin down luminosity is:
$$
L_{\mathrm{sd}}(t)=\frac{L_0}{\left[1+ \frac{t}{\tau} \right]^{2}}
$$
This relation can be further extended including deviations from the standard picture of simple dipole radiation. This modification leads to a spin down luminosity in the form $L_{sd}\propto \Omega^{4-2\alpha}$, where $\alpha$ is related to the braking index $n$ as $n=3-2\alpha$ and $0<\alpha<1$. The corresponding temporal behavior of the spin down luminosity is:
$$
L_{\mathrm{sd}}(t)=\frac{L_0}{\left[1+ (1-\alpha)\frac{t}{\tau} \right]^{\frac{2-\alpha}{1-\alpha}}}
$$
For $t\gg\tau$, the spin down luminosity evolves as $L_{sd}\propto t^{-q}$, with
\begin{equation}
\label{q}
   q=\frac{2-\alpha}{1-\alpha}\geqslant2
\end{equation}
and the forward shock afterglow emission dominates. Indeed, following \cite{dall2011}, the luminosity evolution in the relativistic external shock can be obtained from the balance between radiative losses and energy injection from the spinning down magnetar. Fitting this model on the observed X-rays light curves provides a very good description of the plateau and post-plateau phases, with reasonable values of the magnetic field strength and spin period (e.g., \citealt{dall2011,ber2012,Stratta2018ApJ}). In this scenario, the post-plateau afterglow spectral properties follow the standard forward shock prescriptions, in good agreement with the majority of the GRBs in \textit{Sample 1}. 
Since the majority of the GRBs in \textit{Sample 1} (16/19) show a preference for $\nu_c$ decreasing in time after the plateau phase, even in the magnetar scenario we can apply eqs.~\ref{fs_ism} and \ref{fs_wind} and find agreement with a forward shock propagating in an ISM medium. For the remaining three GRBs, $\nu_c$ increases with time, with a slope compatible with $t^{0.5}$, indicating a preference for wind-like density profile. 

Another interesting scenario has been proposed by \cite{Der2022NatCo..13.5611D}, which claim that the plateau can be still explained in the classical fireball model, provided that the jet is propagating in a wind environment and with a rather low value of bulk Lorentz factor ($\sim$ few tens). This last condition ensures to have a long enough coasting phase and a corresponding light curve whose temporal slope is close to the ones observed during the plateau. Though, in order to have the jet dissipation above the photospheric radius, such low values of Lorentz factor require a not too large isotropic luminosity and a minimum variability time scale of $\sim$ few seconds. Moreover, the transparency requirement is satisfied if rather high values of $\epsilon_B$ ($>0.1$) are assumed. Therefore, also this scenario requires further investigation. 
The HLE from a structured jet, as well, provides a viable explanation for the presence of a plateau phase in the X-ray light curves of GRBs. In the approach of \cite{Oganesyan2020ApJ}, the photons emitted during the prompt emission at large angles with respect to the jet core arrive to the observer at late times and less Doppler boosted. The duration and temporal slope of the plateau phase depend on the jet structure, radius and Lorentz factor of the dissipation site. In the approximation of an instantaneous prompt emission and in the limit of an on-axis observer, there is a biunivocal relation between the arrival time $t_{obs}$ of the photon and $\theta$, where $\theta$ is the polar angle between the jet axis and the patch of the shock front from which the photon departed. Namely, the flux observed at time $t_{obs}$ corresponds to the contribution from a jet ring at a polar angle $\theta(t_{obs})$. This implies that, if the energy spectrum in the shock comoving frame is the same along the jet ring, then also the observed spectrum at each time should reflect the same spectral shape of the prompt emission. Though, since the Doppler boosting decreases at higher latitudes, the observed spectral peak should decrease with time as well. Therefore, also in the HLE scenario, we expect an optical and X-ray emission compatible with a single synchrotron component, whose characteristic frequencies ($\nu_m$ and $\nu_c$) decrease in time due to the Doppler effect. The specific rate of temporal decrease of the characteristic frequencies depends on the jet structure. We note here that HLE would be dominant over the afterglow from forward shock emission throughout the light curve, still being compatible with a single synchrotron component.

Another approach based on the HLE from a structured jet is the one adopted by \cite{Ben2020b,Ben2020a}, where the photons responsible for the plateau emission are produced in the deceleration phase of the forward shock. In particular, in \cite{ben2022}, the authors show also the predictions of the temporal evolution of the cooling frequency. As shown in Fig.~1 and in Tab.~1 of \cite{ben2022}, the model predicts
$$
-1<s=-\frac{3k-4}{2}<2, \text{ for } 0<k<2
$$
for the pre-deceleration and post deceleration phase, while
$$
-1/2<s=-\frac{3k-4}{8-2k}<1/2, \text{ for } 0<k<2
$$
for the angular structure dominated phase. As before, $\rho\propto r^{-k}$. These ranges of $s$ are fully consistent with the ones reported in Tab.~\ref{fit_nuc}. In this class of models, the plateau appears when the jet is observed slightly off-axis, namely at $\theta_{obs}\gtrsim \theta_j$, where $\theta_j$ is the opening angle of the jet core. Even if the condition $\theta_{obs}\gtrsim \theta_j$ is plausible, looking at the distribution in the $E_p-E_{\rm iso}$ plane of the analysed GRBs in Fig.~\ref{Amati}, we can instead exclude that $\theta_{obs}\gg \theta_j$, since we would expect, in that case, an isotropic energy and peak energy of the GRBs, on average, smaller than the ones of the whole population of GRBs.
Among the GRBs in \textit{Sample 1}, the case of GRB 100621A shows several peculiarities \citep{gre2013}. The optical light curve clearly shows a bump around $5\times 10^3-2 \times 10^4$ s, while the X-ray light curve is well fitted by a broken power law. Moreover, the X-ray spectrum shows an evident softening at late times. Such a softening imposes the introduction of an additional spectral break between optical and X-rays, in order to explain the broad band emission with a single synchrotron component. If this spectral break is identified with $\nu_m$, the observations would imply a $\nu_m$ which increases with time. This increase is hardly explained in the framework of standard forward shock theory, even including an energy injection term (see eqs.~\ref{ism_nuc} and \ref{wind_nuc}). Hence, even if GRB 100621A satisfies the conditions to be in \textit{Sample 1}, the shape of the optical light curve and the X-ray spectral softening are hardly explained in the context of standard synchrotron emission from a single emission zone. 

For the GRBs belonging to \textit{Sample 2}, optical  and X-ray data are incompatible with a single synchrotron spectrum in at least one temporal bin. In at least two events (GRB 100814A and GRB 150910A), the optical light curve has a shape which differs substantially from the X-ray one. In a few cases, though, the two light curves behave in a similar way, yet the optical emission appears to be a factor of a few above the maximum extrapolated from the X-ray flux (e.g., GRB 060614, GRB 081007 and GRB 100418A). 
When the optical light curve shows substantial differences with the X-ray one, 
a single synchrotron-emitting region cannot explain the observed broad-band spectrum. One viable explanation is that the HLE associated to the prompt phase and external shock emission from a structured jet can simultaneously contribute to the broad-band spectrum. The HLE can dominate in the X-rays, while the optical would be dominated by external shock. If this last option is valid, this means that the two components cannot have the same spectral energy distribution, otherwise one of them would dominate in both the bands, which is not the case.
Moreover, if the component that dominates in the optical is characterised by a non-thermal spectrum, we should expect an optical spectral index softer than the X-ray one. Indeed, in order to have an excess flux in optical, the spectral component dominating in optical should have a steeper spectral slope. If in optical $F_{\nu}\sim \nu^{-\beta_{\rm opt}}$ and in X-rays $F_{\nu}\sim \nu^{-\beta_{\rm X}}$, this would translate in $\beta_{\rm opt}>\beta_{X}$. Though, comparing the optical spectral slopes with the X-ray spectral index for \textit{Sample 2}, we verified that they have compatible values. A possible solution would be that the optical flux is dominated by a component with a non-thermal spectrum and a cut-off at energies $\nu_{\rm opt}<\nu_{\rm cut-off}<\nu_X$, or, alternatively, by a thermal component with a characteristic temperature $kT\sim h\nu_{\rm opt}$.  \\

\section{Conclusions}
\label{conclusion}
The origin of the plateau phase in the X-ray light curve of GRBs is still matter of debate. In this work we analyzed a complete sample of GRBs with simultaneous optical and X-ray data during the plateau phase, in order to shed light on its physical origin. We performed a time-resolved spectral analysis in the X-rays and we compared X-ray and optical data to verify if they are compatible with a single synchrotron spectrum. While the majority of the cases show compatibility with a single component synchrotron origin of the multi-band emission, we collect evidence that some GRBs ($\sim 1/3$ of the entire sample) are incompatible with the standard forward shock emission from a single dissipation zone.\\
For the sample of GRBs compatible with a single zone emission, we derive the temporal evolution of the cooling frequency and we compare it with the predictions from several models. We show that the majority of the GRBs shows a cooling frequency which decreases in time. For the GRBs which show a plateau both in X-ray and in optical (19 over 30), this leads to a duration of the optical plateau larger than the X-ray one. Moreover, we verify that the transition of the cooling frequency across the optical band is compatible with a simultaneous steepening of the optical light curve. We find that both an energy injection model and the scenario of HLE from a structured jet predict a temporal decay slope in optical/X-ray and trend of the cooling frequency compatible with the observations. This is due to the fact that both scenarios assume classic synchrotron emission. However, while in energy injection this would come from the afterglow (external shock), the HLE emission model would imply that everything comes from the prompt emission region, while the external shock would always remain subdominant. A model in which the plateau is produced by the external shock viewed off-axis, like the one proposed by \cite{ben2022}, would be degenerate with energy injection in the framework of our study. Therefore, further analysis is necessary to test which scenario is more favoured. \\
Concerning the second sample of GRBs not compatible with a single synchrotron spectrum, the optical emission lies above the extrapolation inferred from the X-ray analysis. Neither the energy injection model nor the HLE model alone can account for this behavior. Such a result necessarily requires the interplay of two processes. A confirmation of this "mixed" scenario requires further investigation, through the detailed modeling of the different emission components.

\section*{Acknowledgements}
We acknowledge financial contribution from the agreement ASI-INAF n.2017-14-H.O. GS acknowledges the support by the State of Hesse within the Research Cluster ELEMENTS (Project ID 500$/$10.006). 
DAK acknowledges support from Spanish National Research Project RTI2018-098104-J-I00 (GRBPhot).
MB and GO acknowledge financial support from the AHEAD2020 project (grant agreement n. 871158). 
AR acknowledges support from the INAF project Premiale Supporto Arizona $\&$ Italia. SD acknowledges funding from the European Union’s Horizon2020 research and innovation programme under the Marie Skłodowska-Curie (grant agreement No.754496).

\bibliographystyle{aa} 
\bibliography{mybib} 

\clearpage

\begin{table*}[]
    \centering
    \caption{General information of the full sample.}
    \begin{tabular}{l c c c c c }
GRB     & z             & $t_{90}$ (s) &  E$_{\rm iso}(10^{52}$ erg)& E$_p$(keV) & Ref.\\ 
\hline
\hline
050319  &	3.24	&	$152	\pm 	11 		$ & $3.98		_{-0.59		}^{+6.39 	}	  	$ & $		190.8	_{-182.3 	}^{+114.5	}$ &  (1)  \\ \hline
050416A &	0.65	&	$6.7	\pm 	3.4		$ & $0.08		_{-0.02		}^{+0.05 	}	  	$ & $		24.8 	_{-8.3	 	}^{+8.3		}$ &  (1)  \\ \hline
050730	& 	3.97 	& 	$155   \pm 	19 			$ & $5.89		_{-3.02		}^{+8.07 	}	  	$ & $		973.8	_{-432.3 	}^{+2797	}$ &  (1)  \\ \hline
050824  &	0.83	&	$25	\pm 	5.6			$ & $0.11		_{-0.04		}^{+0.87 	}	  	$ & $		23.8	_{-21.9 	}^{+3.7		}$ &  (1)  \\ \hline
051109A &	2.35	&	$37	\pm 	6			$ & $0.95		_{-6.46		}^{+1.50 	}	  	$ & $		466.8	_{-150.6 	}^{+388.1	}$ &  (1)  \\ \hline
060526  &	3.21	&	$298	\pm 	23		$ & $4.90		_{-0.35		}^{+5.72 	}	  	$ & $		307.4	_{-303.2 	}^{+635.9	}$ &  (1)  \\ \hline
060605  &	3.8		&	$80	\pm 	7			$ & $1.91		_{-0.61		}^{+3.11 	}	  	$ & $		677.8	_{-238.7 	}^{+1714	}$ &  (1)  \\ \hline
060614  & 	0.125 	& 	$109   \pm 	3 			$ & $0.25  		_{-0.02		}^{+0.03	}	 	 $ & $ 		55.0 	_{-45.0		}^{+45.0	}$ &  (2)  \\ \hline
060729  &	0.54	&	$113	\pm 	22		$ & $0.27		_{-0.06		}^{+0.29 	}	  	$ & $		103.2	_{-38.5 	}^{+352.7	}$ &  (1)  \\ \hline
061121  &	1.31	&	$81	\pm 	5			$ & $25.70		_{-2.48		}^{+1.33 	}	  	$ & $		1402	_{-166.6 	}^{+208.3	}$ &  (1)  \\ \hline
080310 	& 	2.43 	& 	$363   \pm 	17 			$ & $4.90		_{-0.99		}^{+10.71 	}  		$ & $		75.4	_{-30.8 	}^{+72.0	}$ &  (1)  \\ \hline
080413B &	1.1		&	$8	\pm 	2			$ & $6.92		_{-6.89		}^{+2.41 	}	  	$ & $		614.5	_{-154.5 	}^{+350.2	}$ &  (1)  \\ \hline
080605  &	1.64	&	$18	\pm 	1			$ & $24.04	 	_{-0.28 	}^{+0.28	}		 $ & $	  	686.3 	_{-26.4 	}^{+23.8	}$ &  (3)  \\ \hline
081007 	& 	0.53 	& 	$   9.7 \pm 4.9 	    $ & $0.20		_{-0.03		}^{+0.03 	} 	  	$ & $		61.2 	_{-15.3 	}^{+15.3 	}$ &  (3)  \\ \hline
081029 	& 	3.85 	& 	$275   \pm 	   49 		$ & $20.75 	  	_{-3.45		}^{+5.25	} 	  	$ & $		887.2	_{-290.9	}^{+804.8	}$ &  (3)  \\ \hline
090618  &	0.54	&	$113	\pm 	1		$ & $26.36 	  	_{-0.36		}^{+0.37	} 	  	$ & $		338.8 	_{-12.3		}^{+12.3	}$ &  (3)  \\ \hline
091018  &   0.97    &	$4.4   \pm 	0.6 		$ & $0.94 	 	_{-0.09 	}^{+0.11	}		 $ & $	  	53.2 	_{-7.9 		}^{+3.9 	}$ &  (3)  \\ \hline
091029  &	2.75	&	$39	\pm 	5			$ & $16.63  	_{-1.98		}^{+2.16	}	 	$ & $	  	247.6 	_{-37.5		}^{+41.3	}$ &  (3)  \\ \hline
100219A &   4.67    &   $27    \pm 	   9 		$ & $2.67  	 	_{-0.64		}^{+1.50	}	 	 $ & $ 		696.2	_{-293.1 	}^{+2393.1	}$ &  (3)  \\ \hline
100418A &   0.62 	& 	$7.9   \pm 	1.1 		$ & $22.23  	_{-2.37 	}^{+2.37	}	 	$ & $	  	259.6 	_{-30.7		}^{+33.9	}$ &  (3)  \\ \hline
100621A &	0.54	&	$64	\pm 	2			$ & $4.57		_{-0.17 	}^{+0.18 	}	 	 $ & $ 		149.6  	_{-10.8  	}^{+12.3  	}$ &  (3)  \\ \hline
100814A &	1.44	&	$	177\pm 11			$ & $7.52 	 	_{-0.19 	}^{+0.19	} 	  	$ & $		242.1 	_{-17.0 	}^{+20.9	}$ &  (3)  \\ \hline
110213A &	1.46	&	$48	\pm 	16			$ & $5.15 	 	_{-0.20 	}^{+0.22	} 	  	$ & $		216.9 	_{-12.8 	}^{+12.8	}$ &  (3)  \\ \hline
110715A &   0.82 	& 	$13    \pm 	   4 		$ & $10.4 		_{-1 		}^{+1		}	 	 $ & $ 		259		_{-31	    }^{+34 		}$ &  (3)  \\ \hline
111228A &	0.71	&	$101	\pm 	5		$ & $1.87 	 	_{-0.36 	}^{+0.52	} 	  	$ & $		58.4 	_{-6.9 		}^{+6.9		}$ &  (3)  \\ \hline
120404A &   2.87 	& 	$39 	\pm 	4  		$ & $10.91 	 	_{-1.39 	}^{+1.70 	}		 $ & $	  	271.4 	_{-50.4  	}^{+81.4	}$ &  (3)  \\ \hline
130702A &	0.15	&	$\sim59		$	&	 $6.6 	 	_{-0.4		}^{+0.4		} 	  	$ & - &  (4) \\ \hline
140419A &	3.96	&	$80	\pm 	4			$ & $254.68	 	_{-14.80 	}^{+16.34 	}	 	 $ & $ 		1397.6  _{-188.3 	}^{+188.3	}$ &  (3)  \\ \hline
150910A &	1.36	&	$112	\pm 	37		$ & $5.20 	 	_{-0.45 	}^{+0.49 	}	 	 $ & $ 		535.4 	_{-87.3		}^{+113.2	}$ &  (3) \\ \hline
180728A &	0.12	&	$8.7	\pm 	0.3		$ & $0.30 	 	_{-0.0002 	}^{+0.0002 	}	 	 $ & $ 		88.5  	_{-1.6 		}^{+1.6 	}$ &  (3) \\ \hline 	
  \end{tabular}
  
  \tablefoot{(1) \cite{Kann2010ApJ}, (2) \cite{Kann2011ApJ}, (3) Kann et al. (in preparation), (4) \cite{Vol2017}. }

    \label{tab_gen}
\end{table*}

\begin{table*}[]
    \centering
    \caption{Optical general information}
    \begin{tabular}{l c c c c c }
GRB & $\beta_{opt}$ &$A_V$(mags) & $t_1$(s) & $t_2$(s) & Model \\ 
\hline
\hline
050319  &	$0.74 \pm 0.42 $ & $0.05 \pm 0.09 $ &381   &  400550  & SMC    	\\ \hline
050416A &	$0.92 \pm 0.3  $ & $0.21 \pm 0.14 $ &657   &  144815  & SMC    	\\ \hline
050730	&	$0.52 \pm 0.05 $ & $0.1  \pm 0.015$ &1555  &  12563   & SMC    	\\ \hline
050824  &	$0.45 \pm 0.18 $ & $0.14 \pm 0.13 $ & 634.7  &  34478   & SMC  	\\ \hline
051109A &	$0.42    	   $ & $0.09 \pm 0.03 $ &167   &  44747   & SMC    	\\ \hline
060526  &	$0.65 \pm 0.06 $ & $0       $ &3080  &  462574        & N/A    	\\ \hline
060605  &	$0.6           $ & $0       $ &86    &  23377         & N/A    	\\ \hline
060614  &	$0.41 \pm 0.09 $ & $0.28 \pm 0.07 $ &4733  &  246090  & SMC    	\\ \hline
060729  &	$0.67\pm 0.07  $ & $0       $ &18042 &  662500        & N/A    	\\ \hline
061121  &	$0.6           $ & $0       $ &305   &  72360         & N/A 	\\ \hline
080310 	&	$0.42 \pm 0.12 $ & $0.19 \pm 0.05 $ &153   &  252216  & SMC    	\\ \hline
080413B &	$0.74 \pm 0.04 $ & $0       $ &96    &  780506        & N/A    	\\ \hline
080605  &	$0.58\pm 0.35$   & $0.51\pm 0.19$ &414   &  124475  & SMC    	\\ \hline
081007 	&  $0.27 \pm 0.11$	 & $0.82 \pm 0.09$  &  94 & 328979   & SMC   		\\ \hline
081029 	&  $0.33\pm 0.05$   & $0.24\pm 0.02$ &150   &  108578  & SMC		\\ \hline
090618  &	$0.71 \pm 0.02 $ & $0       $ &405   &  454723        & N/A    	\\ \hline
091018  &	$0.61\pm 0.02$   & $0       $ &301   &  534067        & N/A    	\\ \hline
091029  &	$0.429\pm 0.026$ & $0       $ &311   &  188708        & N/A    	\\ \hline
100219A &  $0.66 \pm 0.14 $ & $0.13 \pm 0.05 $ &31708 &  398304  & SMC		\\ \hline
100418A &	$1.06\pm 0.02$   & $0.09\pm 0.04$ &27742 &  476546  & SMC    	\\ \hline
100621A &	$0.78\pm 0.09$   & $3.72\pm 0.10$ &237   &  10974   & SMC    	\\ \hline
100814A &  $0.41  \pm 0.04$	 & $0.16\pm	0.02$ & 526	& 960827    & SMC     		\\ \hline
110213A &	$0.9  \pm 0.07 $ & $0.132\pm 0.003$ &193   &  5546    & SMC    	\\ \hline
110715A &  $0.63\pm 0.28$   & $0.47\pm 0.22$ &217211&  736849  & SMC		\\ \hline
111228A &	$0.69 \pm 0.07 $ & $0.16 \pm 0.04 $ &349   &  663118  & SMC    	\\ \hline
120404A &	$1.02    $ 	     & $0.07\pm 0.02$ &730   &  19824         & MW  \\ \hline
130702A &	$0.71\pm 0.02$ & $0       $ &101088&  335296        & N/A    	 \\ \hline
140419A &	$0.76\pm 0.08$ & $0       $ &280   &  10780         & N/A    	\\ \hline
150910A &	$0.53\pm 0.14$ & $0.17\pm 0.05$ &728   &  120966  & SMC    		 \\ \hline
180728A &	$0.67\pm 0.05$ & $0       $ &2072  &  180050        & N/A    		\\ \hline		
  \end{tabular}
  
  \tablefoot{For GRBs 060605 and 061121 we assumed $\beta_{opt}=0.6$ as in \cite{Kann2006ApJ}.
    $t_1$(s) and $t_2$(s) define the interval within which the optical modelling has been obtained.
    For GRB 100219A , the values reported are from \cite{Thoene2013a}.}

    \label{tab_opt}
\end{table*}


\begin{table*}[]
    \centering
    \begin{tabular}{c c c c c c c c c}
name & N ($\times 100$ mJy) & a & b & $t_b/10^4 s
$ & $\chi^2/$dof & $\log_{10}(\frac{L_{\rm X}}{\rm erg/s})$ & ph. index. & p-value\\ \hline
\hline

\multicolumn{9}{c}{\emph{Sample 1}}\\ \hline
\hline
050824 & 	 $ 0.03 \pm 0.02 $ & 	 $ 0.04 \pm 0.32 $ & 	 $ 0.98 \pm 0.17 $ & 	 $ 3.75 \pm 3.75 $ & 	53.0/41  	& 	 $ 45.66 _{-0.12}^{+0.10}$ & $ 1.96\pm0.07$ & $ 0.078    $ \\ \hline 
050319 & 	 $ 0.07 \pm 0.01 $ & 	 $ 0.45 \pm 0.04 $ & 	 $ 1.90 \pm 0.18 $ & 	 $ 5.36 \pm 1.15 $ & 	65.7/82  	& 	 $ 48.09 _{-0.11}^{+0.09}$ & $ 1.92\pm0.03$ & $ 0.763    $ \\ \hline 
050416A & 	 $ 0.43 \pm 0.30 $ & 	 $ 0.32 \pm 0.17 $ & 	 $ 0.96 \pm 0.05 $ & 	 $ 0.10 \pm 0.10 $ & 	90.8/96  	& 	 $ 46.79 _{-0.11}^{+0.09}$ & $ 1.90\pm0.04$ & $ 0.765    $ \\ \hline 
051109A & 	 $ 0.94 \pm 0.84 $ & 	 $ 0.52 \pm 0.39 $ & 	 $ 1.34 \pm 0.08 $ & 	 $ 0.47 \pm 0.46 $ & 	151.7/151   & 	 $ 48.25 _{-0.11}^{+0.09}$ & $ 2.12\pm0.03$ & $ 0.412    $ \\ \hline 
060526 & 	 $ 0.02 \pm 0.01 $ & 	 $ 0.56 \pm 0.06 $ & 	 $ 2.25 \pm 0.23 $ & 	 $ 5.29 \pm 1.45 $ & 	92.4/37     & 	 $ 48.13 _{-0.11}^{+0.09}$ & $ 1.98\pm0.07$ & $ 0.508    $ \\ \hline 
060605 & 	 $ 0.14 \pm 0.03 $ & 	 $ 0.39 \pm 0.10 $ & 	 $ 2.45 \pm 0.15 $ & 	 $ 1.06 \pm 0.17 $ & 	48.7/63     & 	 $ 48.23 _{-0.11}^{+0.09}$ & $ 2.00\pm0.04$ & $ 0.386    $ \\ \hline 
060729 & 	 $ 0.26 \pm 0.01 $ & 	 $ 0.03 \pm 0.02 $ & 	 $ 1.53 \pm 0.02 $ & 	 $ 6.83 \pm 0.39 $ & 	867.3/675  & 	 $ 46.40 _{-0.11}^{+0.09}$ & $ 1.93\pm0.01$ & $ 0.002(^*)    $ \\ \hline 
061121 & 	 $ 1.23 \pm 0.16 $ & 	 $ 0.34 \pm 0.03 $ & 	 $ 1.55 \pm 0.03 $ & 	 $ 0.77 \pm 0.10 $ & 	323.1/276  & 	 $ 48.39 _{-0.09}^{+0.07}$ & $ 1.93\pm0.02$ & $ 0.001 (^*)   $ \\ \hline 
080413B & 	 $ 1.36 \pm 0.71 $ & 	 $ 0.51 \pm 0.09 $ & 	 $ 1.22 \pm 0.04 $ & 	 $ 0.15 \pm 0.09 $ & 	300.8/225  & 	 $ 48.09 _{-0.10}^{+0.08}$ & $ 1.86\pm0.02$ & $ 0.226    $ \\ \hline 
080605 & 	 $ 10.81 \pm 2.58 $ & 	 $ 0.40 \pm 0.08 $ & 	 $ 1.50 \pm 0.03 $ & 	 $ 0.06 \pm 0.01 $ & 	383.2/311  & 	 $ 49.31 _{-0.07}^{+0.06}$ & $ 1.71\pm0.02$ & $ 0.194    $ \\ \hline 
090618 & 	 $ 4.65 \pm 0.54 $ & 	 $ 0.33 \pm 0.06 $ & 	 $ 1.51 \pm 0.02 $ & 	 $ 0.41 \pm 0.05 $ & 	812.8/775  & 	 $ 47.66 _{-0.08}^{+0.07}$ & $ 1.91\pm0.01$ & $ <10^{-3}(^*) $ \\ \hline 
091018 & 	 $ 5.63 \pm 0.79 $ & 	 $ -0.17 \pm 0.26 $ & 	 $ 1.25 \pm 0.03 $ & 	 $ 0.03 \pm 0.01 $ & 	155.0/115  & 	 $ 48.14 _{-0.10}^{+0.08}$ & $ 1.92\pm0.03$ & $ 0.059    $ \\ \hline 
091029 & 	 $ 0.17 \pm 0.03 $ & 	 $ -0.06 \pm 0.10 $ & 	 $ 1.22 \pm 0.04 $ & 	 $ 0.83 \pm 0.19 $ & 	127.8/120  & 	 $ 47.78 _{-0.11}^{+0.09}$ & $ 1.92\pm0.04$ & $ 0.558    $ \\ \hline 
100621A & 	 $ 0.11 \pm 0.04 $ & 	 $ 0.68 \pm 0.03 $ & 	 $ 1.69 \pm 0.09 $ & 	 $ 8.05 \pm 2.48 $ & 	249.5/178  & 	 $ 46.70 _{-0.11}^{+0.09}$ & $ 2.37\pm0.03$ & $ <10^{-3}(^*) $ \\ \hline 
110213A & 	 $ 5.03 \pm 0.37 $ & 	 $ -0.19 \pm 0.06 $ & 	 $ 1.90 \pm 0.03 $ & 	 $ 0.31 \pm 0.02 $ & 	245.0/229  & 	 $ 48.57 _{-0.11}^{+0.09}$ & $ 2.04\pm0.02$ & $ 0.020  (^*)  $ \\ \hline 
111228A & 	 $ 0.57 \pm 0.12 $ & 	 $ 0.19 \pm 0.07 $ & 	 $ 1.28 \pm 0.04 $ & 	 $ 0.75 \pm 0.19 $ & 	195.4/150  & 	 $ 47.00 _{-0.11}^{+0.09}$ & $ 1.97\pm0.03$ & $ 0.589    $ \\ \hline 
130702A & 	 $ 0.26 \pm 0.10 $ & 	 $ 0.10 \pm 0.61 $ & 	 $ 1.35 \pm 0.10 $ & 	 $ 8.93 \pm 4.66 $ & 	210.2/226  & 	 $ 44.81 _{-0.11}^{+0.09}$ & $ 1.81\pm0.03$ & $ 0.343    $ \\ \hline 
140419A & 	 $ 1.21 \pm 1.03 $ & 	 $ 0.60 \pm 0.20 $ & 	 $ 1.55 \pm 0.11 $ & 	 $ 0.60 \pm 0.46 $ & 	201.6/198  & 	 $ 49.38 _{-0.11}^{+0.09}$ &  $1.88 	 \pm0.03$ &	 $0.017(^*)$ \\ \hline 
180728A & 	 $ 1.78 \pm 1.54 $ & 	 $ 0.69 \pm 0.17 $ & 	 $ 1.44 \pm 0.07 $ & 	 $ 2.12 \pm 1.71 $ & 	583.5/500  & 	 $ 45.76 _{-0.11}^{+0.09}$ & $ 1.76\pm0.02$ & $ 0.055    $ \\ \hline 

\multicolumn{9}{c}{\emph{Sample 2}}\\ \hline
\hline
050730 & 	 $ 2.64 \pm 0.33 $ & 	 $ -0.02 \pm 0.22 $ & 	 $ 2.62 \pm 0.05 $ & 	 $ 0.76 \pm  0.06 $ & 	481.8/334  & 	 $ 49.54_{-0.11}^{+0.09}$ & $ 1.55\pm0.02 $ & $<10^{-3} (^*)$ \\ \hline 
060614 & 	 $ 0.09 \pm 0.01 $ & 	 $ -0.06 \pm 0.06 $ & 	 $ 2.16 \pm 0.06 $ & 	 $ 5.21 \pm  0.40 $ & 	119.8/153  & 	 $ 44.49_{-0.11}^{+0.09}$ & $ 1.77\pm0.03 $ & $0.098 	$ \\ \hline 
080310 & 	 $ 0.13 \pm 0.03 $ & 	 $ 0.32 \pm 0.09 $ & 	 $ 1.81 \pm 0.09 $ & 	 $ 1.42 \pm  0.32 $ & 	148.8/77   & 	 $ 48.07_{-0.11}^{+0.09}$ & $ 1.92\pm0.05 $ & $<10^{-3} (^*)$ \\ \hline 
081007 & 	 $ 0.01 \pm 0.01 $ & 	 $ 0.68 \pm 0.05 $ & 	 $ 1.56 \pm 0.17 $ & 	 $ 11.34 \pm  7.17 $ & 	76.1/64    & 	 $ 46.32_{-0.11}^{+0.09}$ & $ 1.87\pm0.05 $ & $0.453 	$ \\ \hline 
081029 & 	 $ 0.08 \pm 0.01 $ & 	 $ 0.18 \pm 0.10 $ & 	 $ 3.00 \pm 0.22 $ & 	 $ 1.80 \pm  0.21 $ & 	95.3/76    & 	 $ 48.08_{-0.11}^{+0.09}$ & $ 1.86\pm0.05 $ & $0.297 	$ \\ \hline 
100219A & 	 $ 0.04 \pm 0.01 $ & 	 $ 0.53 \pm 0.06 $ & 	 $ 3.00 \pm 0.29 $ & 	 $ 2.77 \pm  0.56 $ & 	42.3/23    & 	 $ 48.55_{-0.11}^{+0.09}$ & $ 1.58\pm0.06 $ & $0.973 	$ \\ \hline 
100418A & 	 $ 0.02 \pm 0.00 $ & 	 $ -0.45 \pm 0.15 $ & 	 $ 1.63 \pm 0.11 $ & 	 $ 7.18 \pm  1.78 $ & 	27.9/26    & 	 $ 45.15_{-0.13}^{+0.10}$ & $ 1.76\pm0.09 $ & $0.441 	$ \\ \hline 
100814A & 	 $ 0.05 \pm 0.00 $ & 	 $ 0.50 \pm 0.02 $ & 	 $ 2.38 \pm 0.09 $ & 	 $ 21.53 \pm  1.47 $ & 	731.6/293  & 	 $ 47.25_{-0.11}^{+0.09}$ & $ 1.84\pm0.02 $ & $0.049 (^*)	$ \\ \hline 
110715A & 	 $ 24.76 \pm 1.15 $ & 	 $ -0.80 \pm 0.32 $ & 	 $ 1.00 \pm 0.01 $ & 	 $ 0.01 \pm  0.00 $ & 	452.1/247  & 	 $ 48.57_{-0.07}^{+0.06}$ & $ 1.85\pm0.02 $ & $<10^{-3} (^*)$ \\ \hline 
120404A & 	 $ 0.42 \pm 0.17 $ & 	 $ -0.16 \pm 0.38 $ & 	 $ 1.97 \pm 0.14 $ & 	 $ 0.28 \pm  0.09 $ & 	29.7/36   & 	 $ 48.31_{-0.11}^{+0.09}$ & $ 1.78\pm0.06 $ & $0.001(^*)	$ \\ \hline 
150910A & 	 $ 1.92 \pm 0.15 $ & 	 $ 0.34 \pm 0.02 $ & 	 $ 2.49 \pm 0.06 $ & 	 $ 0.67 \pm  0.04 $ & 	487.1/328  & 	 $ 48.72_{-0.07}^{+0.06}$ & $ 1.52\pm0.02 $ & $<10^{-3} (^*)$ \\ \hline 
\end{tabular}
    \caption{Results relative to the temporal fit of the X-ray light curve. The first four columns are the best fit parameters. $t_b$ is in the observer frame. The fifth column is the reduced $\chi^2$ and $L_X$ is the average X-ray luminosity during the plateau. Further details are specified in the text. Errors are given at the 1 sigma level of confidence. The last two columns report the average X-ray photon index and the corresponding p-value of the fit with a constant of the photon index as a function of time. Cases denoted with $(^*)$ have a p-value $<0.05$, indicating that the photon index temporal trend significantly deviates from a constant.}
    \label{fit_x}
\end{table*}

\begin{table*}[]
    \centering
    \begin{tabular}{ c c c c c c c }

name & N ($\times 100$ mJy) & a & b & $t_b/ 10^4 s$ & $\chi^2/$dof & $\log_{10}(L_{\rm opt}$(erg/s)) \\ \hline
\hline
\multicolumn{7}{ c }{\emph{Sample 1}}\\ \hline
\hline
050824 & 	 $ 4.2 \pm 0.1 $ & 	 $ 0.66 \pm 0.01 $ & - & - & 	 58.9/27 &  - \\ \hline 
050319 & 	 $ 0.9 \pm 0.1 $ & 	 $ 0.53 \pm 0.02 $ & 	 $ 2.85 \pm 0.30 $ & 	 $ 41.57 \pm 3.40 $ & 	 $ 39.9/37 $ & 	 $ 46.26 _{-0.05}^{+0.06} $  \\ \hline 
050416A & 	 $ 11.9 \pm 2.1 $ & 	 $ -0.31 \pm 0.32 $ & 	 $ 0.92 \pm 0.07 $ & 	 $ 0.05 \pm 0.02 $ & 	 $ 40.1/28 $ & 	 $ 44.36 _{-0.05}^{+0.05} $  \\ \hline 
051109A & 	 $ 17.2 \pm 0.3 $ & 	 $ 0.81 \pm 0.02 $ & - & - & 	 25.0/21 &  - \\ \hline 
060526 & 	 $ 2.5 \pm 0.2 $ & 	 $ 0.60 \pm 0.01 $ & 	 $ 2.64 \pm 0.10 $ & 	 $ 15.78 \pm 0.85 $ & 	 $ 199.0/43 $ & 	 $ 46.56 _{-0.05}^{+0.05} $  \\ \hline 
060605 & 	 $ 12.0 \pm 1.6 $ & 	 $ 0.80 \pm 0.03 $ & 	 $ 3.00 \pm 0.17 $ & 	 $ 2.45 \pm 0.18 $ & 	 $ 165.5/29 $ & 	 $ 47.42 _{-0.04}^{+0.04} $  \\ \hline 
060729 & 	 $ 31.0 \pm 2.1 $ & 	 $ 0.02 \pm 0.03 $ & 	 $ 1.60 \pm 0.03 $ & 	 $ 8.16 \pm 0.72 $ & 	 $ 305.0/41 $ & 	 $ 44.99 _{-0.05}^{+0.06} $  \\ \hline 
061121 & 	 $ 4.9 \pm 4.2 $ & 	 $ 0.53 \pm 0.08 $ & 	 $ 1.26 \pm 0.14 $ & 	 $ 4.09 \pm 3.93 $ & 	 $ 58.2/30 $ & 	 $ 45.98 _{-0.09}^{+0.12} $  \\ \hline 
080413B & 	 $ 1.5 \pm 0.1 $ & 	 $ 0.43 \pm 0.01 $ & 	 $ 3.00 \pm 0.27 $ & 	 $ 40.12 \pm 2.46 $ & 	 $ 213.7/25 $ & 	 $ 45.42 _{-0.04}^{+0.04} $  \\ \hline 
080605 & 	 $ 8.0 \pm 0.1 $ & 	 $ 0.64 \pm 0.01 $ & - & - & 	 172.4/39 &  - \\ \hline 
090618 & 	 $ 44.3 \pm 10.9 $ & 	 $ 0.51 \pm 0.05 $ & 	 $ 1.45 \pm 0.04 $ & 	 $ 1.72 \pm 0.43 $ & 	 $ 43.2/43 $ & 	 $ 45.26 _{-0.03}^{+0.03} $  \\ \hline 
091018 & 	 $ 1.2 \pm 0.1 $ & 	 $ 0.91 \pm 0.01 $ & 	 $ 2.84 \pm 0.32 $ & 	 $ 12.51 \pm 0.88 $ & 	 $ 42.5/36 $ & 	 $ 45.97 _{-0.03}^{+0.03} $  \\ \hline 
091029 & 	 $ 3.8 \pm 0.8 $ & 	 $ 0.39 \pm 0.03 $ & 	 $ 1.63 \pm 0.13 $ & 	 $ 5.87 \pm 1.43 $ & 	 $ 48.6/37 $ & 	 $ 46.08 _{-0.04}^{+0.05} $  \\ \hline 
100621A & 	 $ 32.2 \pm 2.7 $ & 	 $ 0.22 \pm 0.02 $ & 	 $ 2.41 \pm 0.18 $ & 	 $ 4.00 \pm 0.40 $ & 	 $ 2560.6/29 $ & 	 $ 45.32 _{-0.02}^{+0.02} $  \\ \hline 
110213A & 	 $ 290.4 \pm 5.6 $ & 	 $ 0.20 \pm 0.01 $ & 	 $ 2.01 \pm 0.02 $ & 	 $ 1.52 \pm 0.04 $ & 	 $ 2645.9/41 $ & 	 $ 47.09 _{-0.02}^{+0.02} $  \\ \hline 
111228A & 	 $ 29.4 \pm 2.6 $ & 	 $ 0.25 \pm 0.03 $ & 	 $ 1.54 \pm 0.04 $ & 	 $ 1.91 \pm 0.20 $ & 	 $ 72.5/39 $ & 	 $ 45.28 _{-0.04}^{+0.05} $  \\ \hline 
130702A & 	 $ 320.1 \pm 24.6 $ & 	 $ 1.31 \pm 0.02 $ & - & - & 	 41.8/40 &  - \\ \hline 
140419A & 	 $ 15.1 \pm 0.2 $ & 	 $ 1.06 \pm 0.01 $ & - & - 	 119.6/34 &  - \\ \hline
180728A & 	 $ 123.7 \pm 1.7 $ & 	 $ 1.08 \pm 0.01 $ & - & - & 	 92.5/39 &  - \\ \hline

\multicolumn{7}{ c }{\emph{Sample 2}}\\ \hline
\hline
050730 & 	 $ 74.9 \pm 13.0 $ & 	 $ -0.22 \pm 0.28 $ & 	 $ 1.58 \pm 0.08 $ & 	 $ 0.74 \pm 0.19 $ & 	 $ 236.9/18 $ & 	 $ 47.13 _{-0.02}^{+0.02} $  \\ \hline 
060614 & 	 $ 8.6 \pm 0.6 $ & 	 $ 0.03 \pm 0.06 $ & 	 $ 2.47 \pm 0.08 $ & 	 $ 7.55 \pm 0.46 $ & 	 $ 48.4/22 $ & 	 $ 42.81 _{-0.04}^{+0.05} $  \\ \hline 
080310 & 	 $ 175.7 \pm 5.0 $ & 	 $ -0.60 \pm 0.10 $ & 	 $ 1.16 \pm 0.01 $ & 	 $ 0.18 \pm 0.01 $ & 	 $ 795.5/43 $ & 	 $ 46.93 _{-0.03}^{+0.03} $  \\ \hline 
081007 & 	 $ 8.9 \pm 0.1 $ & 	 $ 0.73 \pm 0.01 $ & - & - & 	 272.4/38 &  - \\ \hline 
081029 & 	 $ 197.6 \pm 3.0 $ & 	 $ -1.31 \pm 0.06 $ & 	 $ 1.90 \pm 0.02 $ & 	 $ 0.86 \pm 0.02 $ & 	 $ 316.7/30 $ & 	 $ 47.36 _{-0.02}^{+0.02} $  \\ \hline 
100219A & 	 $ 6.9 \pm 0.2 $ & 	 $ 0.99 \pm 0.02 $ & - & - & 	 73.1/21 &  - \\ \hline 
100418A & 	 $ 21.6 \pm 0.8 $ & 	 $ -0.32 \pm 0.03 $ & 	 $ 1.66 \pm 0.04 $ & 	 $ 5.72 \pm 0.33 $ & 	 $ 200.0/31 $ & 	 $ 44.58 _{-0.03}^{+0.04} $  \\ \hline 
100814A & 	 $ 8.2 \pm 0.2 $ & 	 $ 0.09 \pm 0.01 $ & 	 $ 3.90 \pm 0.15 $ & 	 $ 39.14 \pm 0.70 $ & 	 $ 3100.6/47 $ & 	 $ 45.68 _{-0.02}^{+0.02} $  \\ \hline 
110715A & 	 $ 9.6 \pm 1.3 $ & 	 $ 0.52 \pm 0.01 $ & 	 $ 2.87 \pm 0.30 $ & 	 $ 37.49 \pm 3.19 $ & 	 $ 123.4/30 $ & 	 $ 46.12 _{-0.05}^{+0.06} $  \\ \hline 
120404A & 	 $ 147.0 \pm 3.0 $ & 	 $ -1.65 \pm 0.11 $ & 	 $ 1.60 \pm 0.04 $ & 	 $ 0.26 \pm 0.01 $ & 	 $ 22.3/38 $ & 	 $ 46.79 _{-0.05}^{+0.06} $  \\ \hline 
150910A & 	 $ 315.1 \pm 11.6 $ & 	 $ -4.00 \pm 0.34 $ & 	 $ 1.26 \pm 0.01 $ & 	 $ 0.11 \pm 0.00 $ & 	 $ 75.3/36 $ & 	 $ 46.20 _{-0.10}^{+0.12} $  \\ \hline 

    \end{tabular}
    \caption{Results relative to the temporal and spectral optical analysis. The first four columns are the best fit parameters. $t_b$ is in the observer frame. The fifth column is the reduced $\chi^2$ and $L_{\rm opt}$ is the average optical luminosity during the plateau, if present. The reported parameters are specified in the text. Errors are given at 1 sigma level of confidence. When the fit with a broken power law does not converge, we fit with a single power law and therefore only the value of the slope $a$ is reported.}
    \label{fit_opt}
\end{table*}

\begin{table*}[]
    \centering
    \begin{tabular}{ c c c }
name & $\log_{10}(t^*)$ & $\log_{10}(t_{b,opt})$ \\ \hline
\hline
050416A	&	>3.64						 	&	$2.70_{-0.22}^{+	0.15}$	\\ \hline
051109A	&	$6.77	_{-2.92	}^{+6.16 	}$ 	&	- 	 						\\ \hline
060729	&	>3.65						  	&	$4.20_{-0.26}^{+	0.16}$	\\ \hline
061121	&	$6.27	_{-1.75	}^{+2.69 	}$  	&	$4.61_{-1.41}^{+	0.29}$	\\ \hline
080413B	&	$4.98	_{-1.38	}^{+2.04 	}$  	&	$5.60_{-0.03}^{+	0.03}$	\\ \hline
080605	&	$5.41	_{-1.95	}^{+3.75 	}$ 	&	-							\\ \hline
090618	&	$5.26	_{-0.64	}^{+0.74 	}$ 	&	$4.23_{-0.32}^{+	0.18}$	\\ \hline
091029	&	>1.29							&	$4.77_{-0.12}^{+	0.09}$	\\ \hline
110213A	&	>3.09							&	$4.18_{-0.01}^{+	0.01}$	\\ \hline
111228A	&	$5.58	_{-1.97	}^{+3.23 	}$ 	&	$4.28_{-0.05}^{+	0.04}$	\\ \hline
130702A	&	>1.30							& 	-							\\ \hline	
180728A	&	$5.54	_{-2.42	}^{+4.60 	}$ 	&	-						\\ \hline		

	    \end{tabular}
    \caption{Comparison between $t^*$ and $t_{b,opt}$. $t*$ is the time at which $\nu_c$ crosses the optical band. }
    \label{t*}
\end{table*}

\begin{figure*}[t]
     \centering
     \begin{subfigure}{0.47\textwidth}
         \centering
         \includegraphics[width=\textwidth]{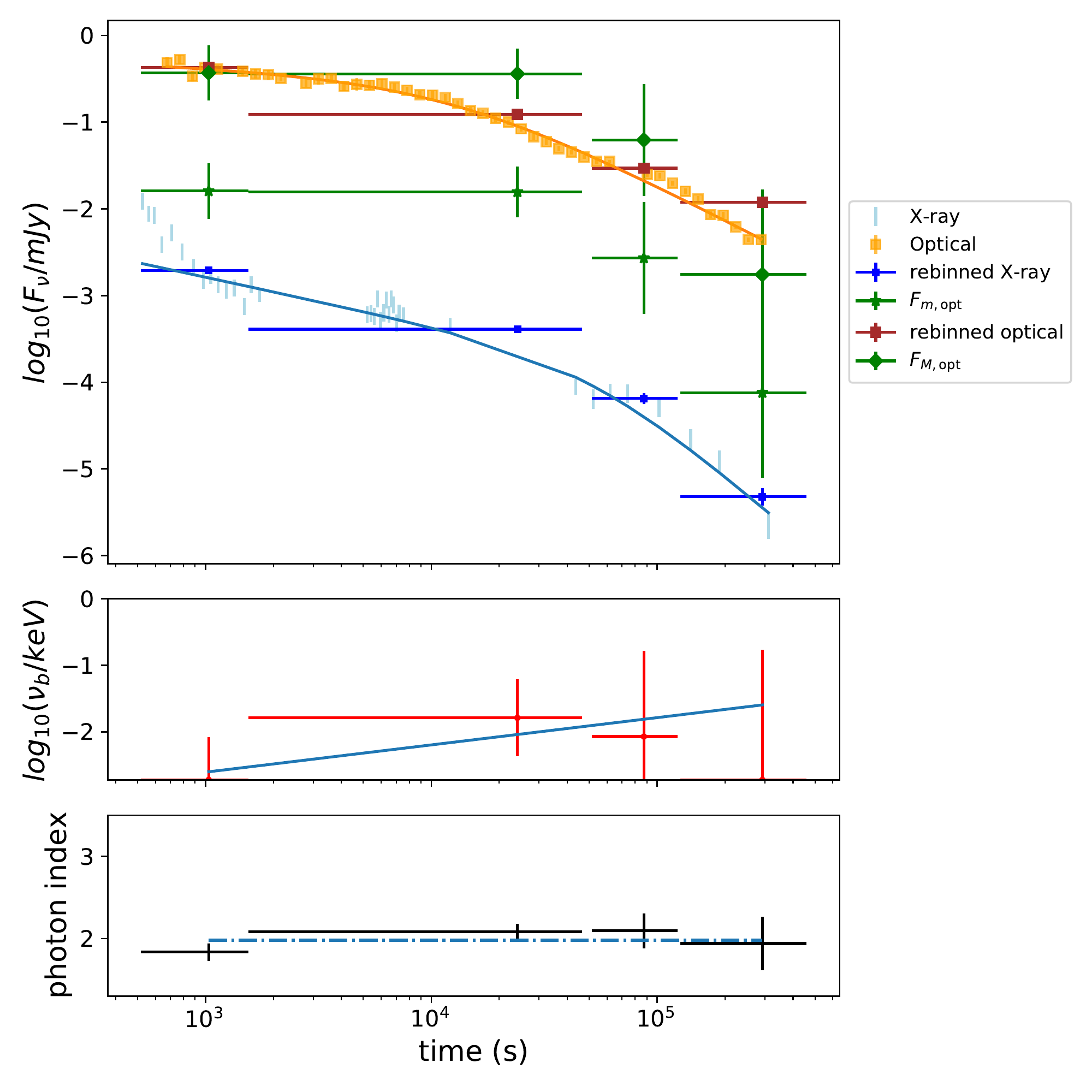}
         \caption{060526}
         \label{fig:y equals x}
     \end{subfigure}
     \hfill
     \begin{subfigure}{0.47\textwidth}
         \centering
         \includegraphics[width=\textwidth]{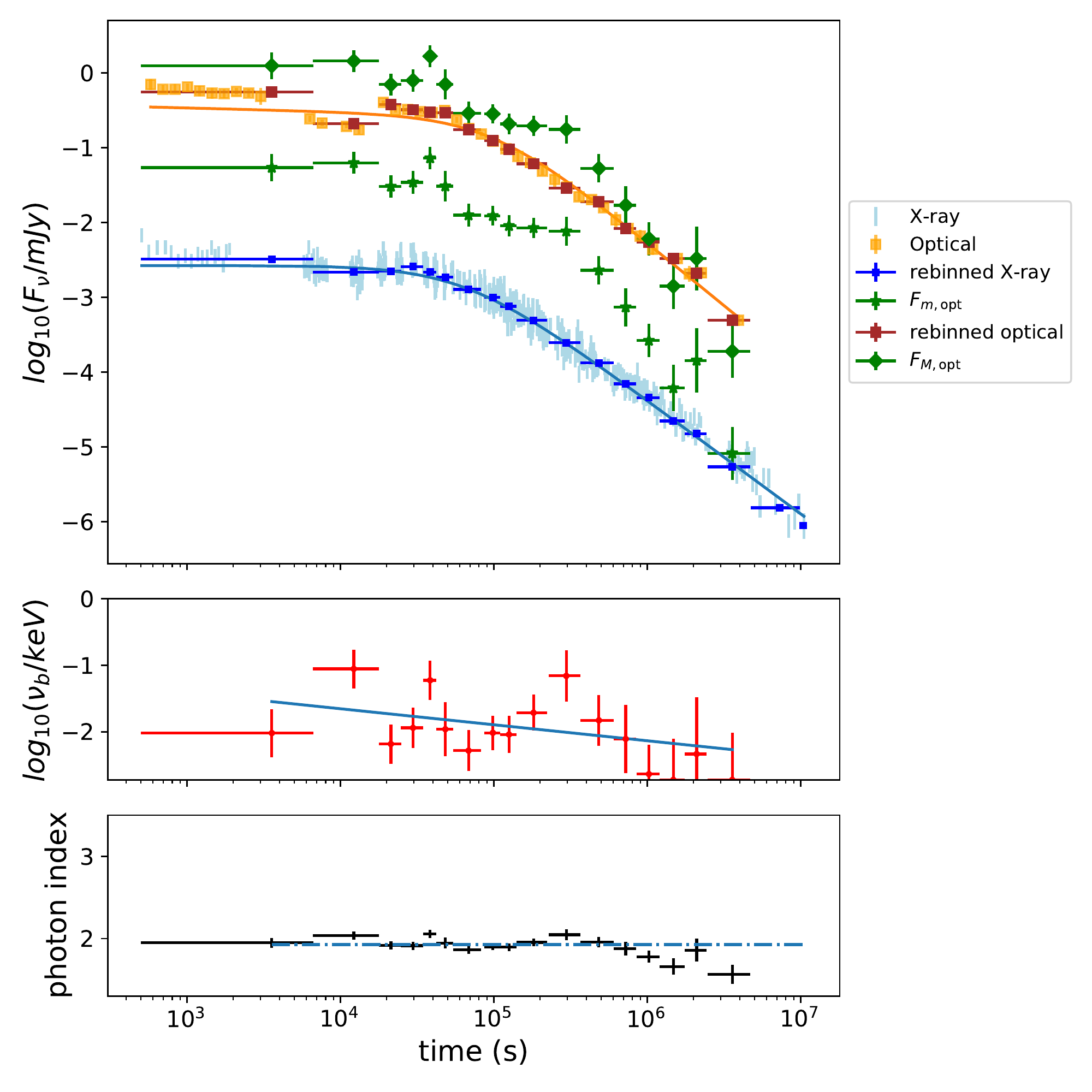}
         \caption{060729}
         \label{fig:three sin x}
     \end{subfigure}
     \\
     \begin{subfigure}{0.47\textwidth}
         \centering
         \includegraphics[width=\textwidth]{opt_X_050824.pdf}
         \caption{050824}
         \label{fig:five over x}
     \end{subfigure}
     \hfill
     \begin{subfigure}{0.47\textwidth}
         \centering
         \includegraphics[width=\textwidth]{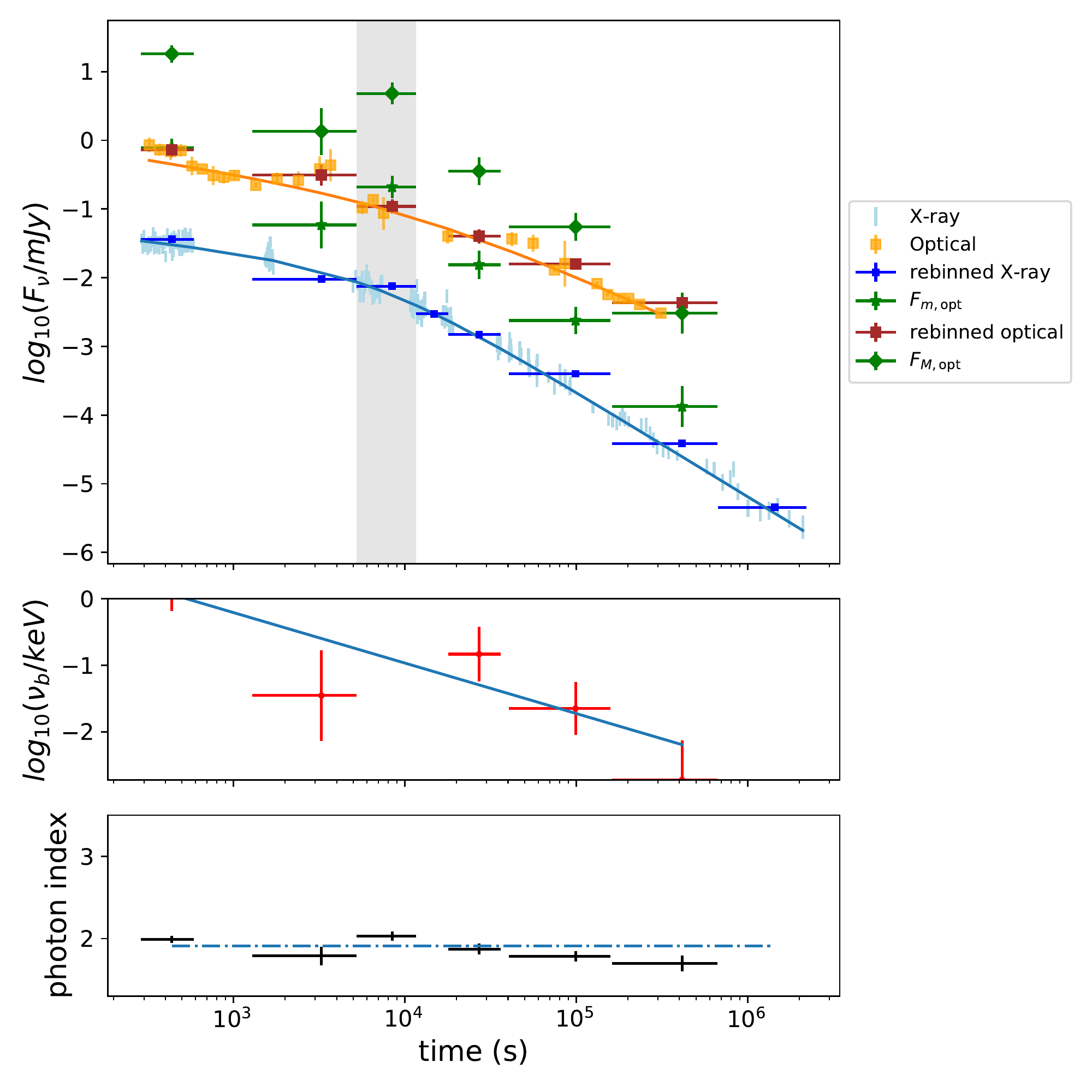}
         \caption{061121}
         \label{fig:five over x}
     \end{subfigure}
        \caption{\emph{Sample 1} - continued}
\end{figure*}

\begin{figure*}[t]
     \centering
     \begin{subfigure}{0.47\textwidth}
         \centering
         \includegraphics[width=\textwidth]{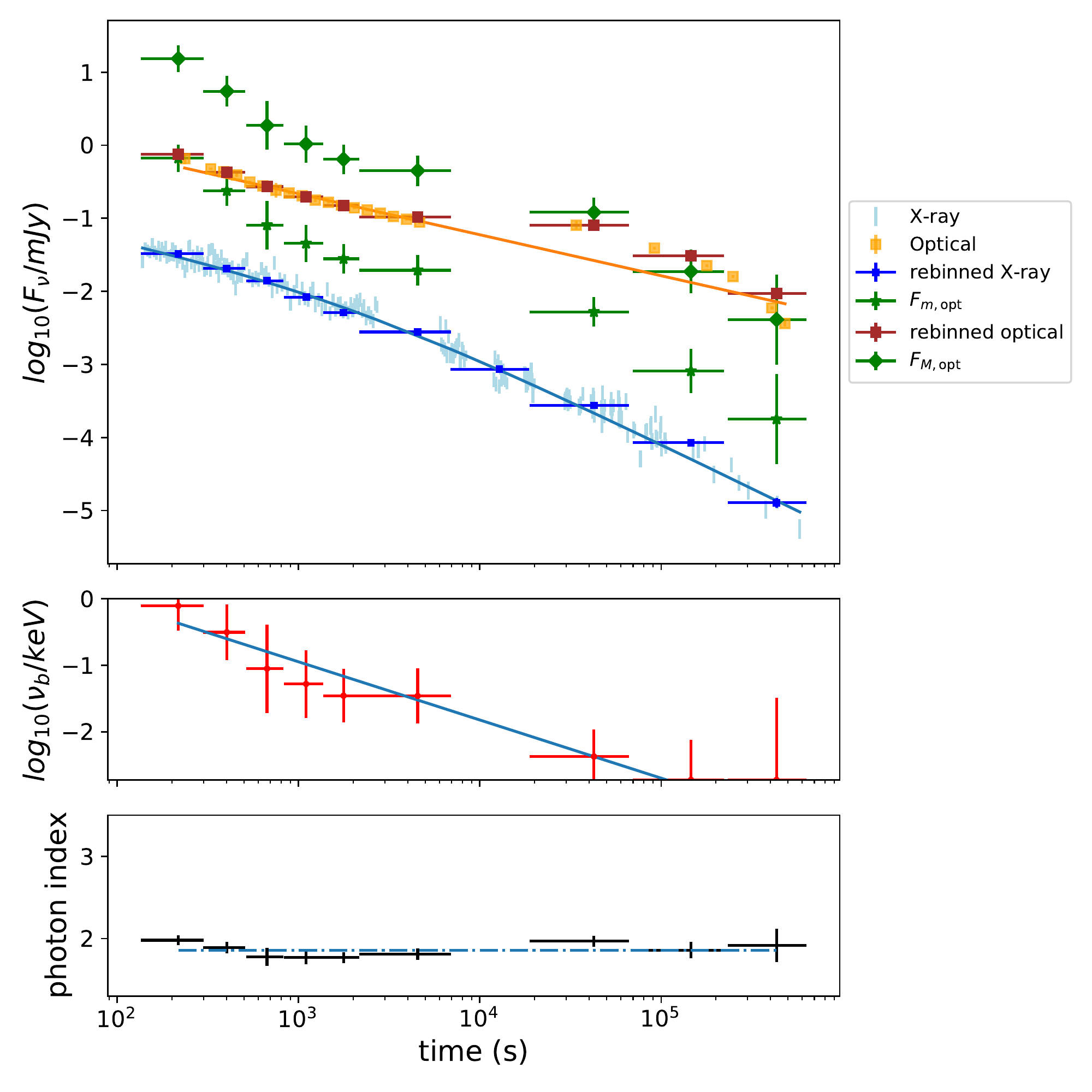}
         \caption{080413B}
         \label{fig_080413B}
     \end{subfigure}
     \hfill
     \begin{subfigure}{0.47\textwidth}
         \centering
         \includegraphics[width=\textwidth]{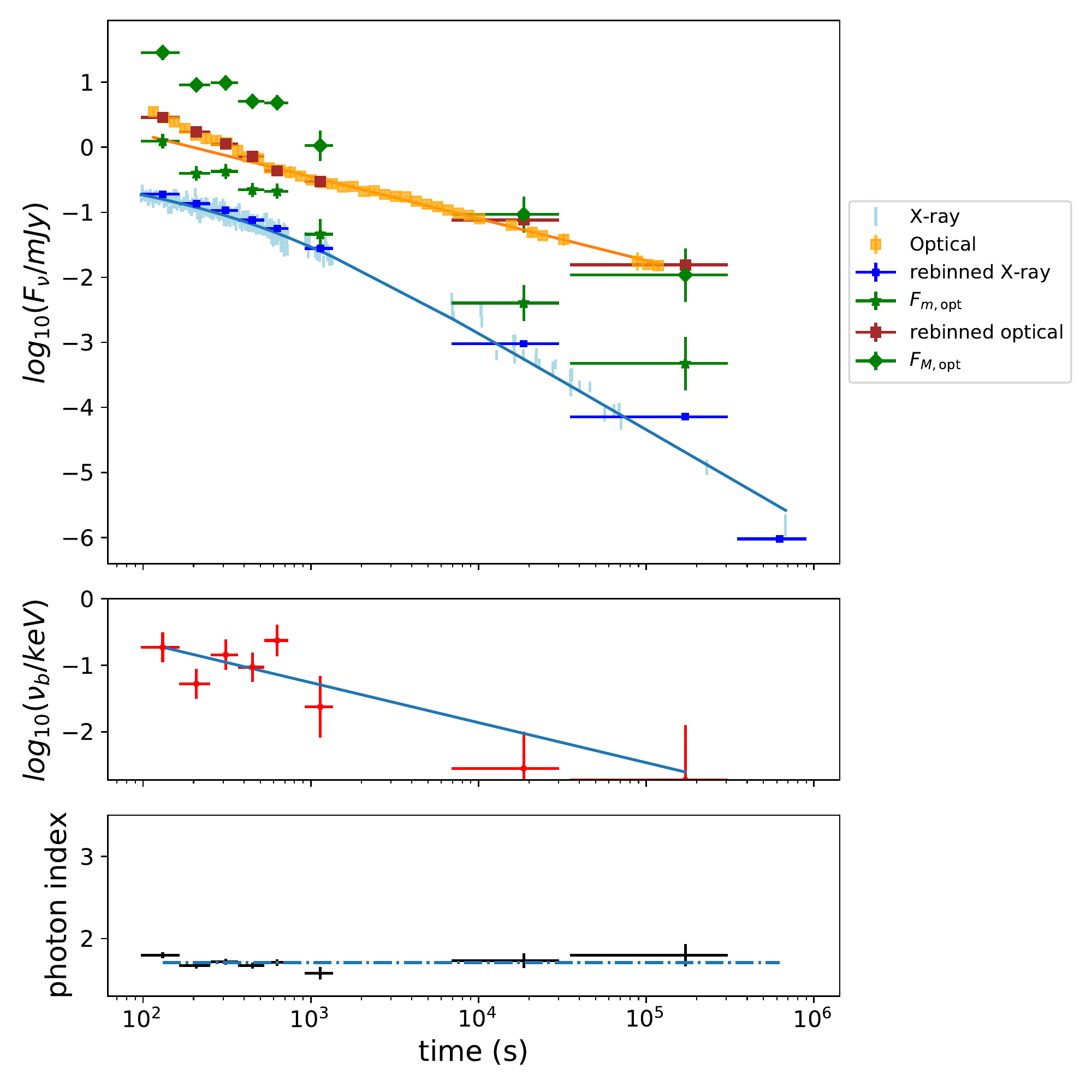}
         \caption{080605}
         \label{fig:three sin x}
     \end{subfigure}
     \\
     \begin{subfigure}{0.47\textwidth}
         \centering
         \includegraphics[width=\textwidth]{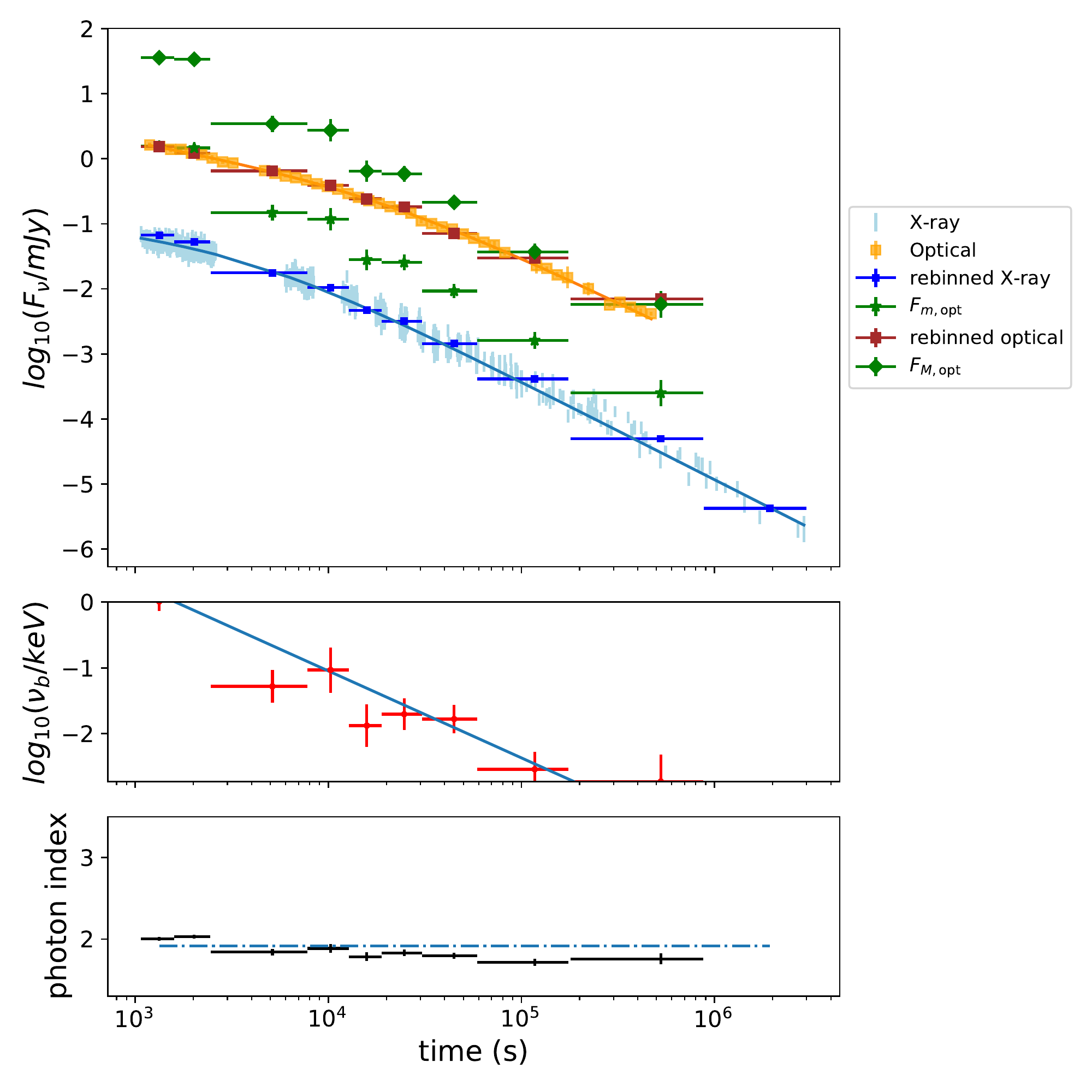}
         \caption{090618}
         \label{fig:five over x}
     \end{subfigure}
     \hfill
     \begin{subfigure}{0.47\textwidth}
         \centering
         \includegraphics[width=\textwidth]{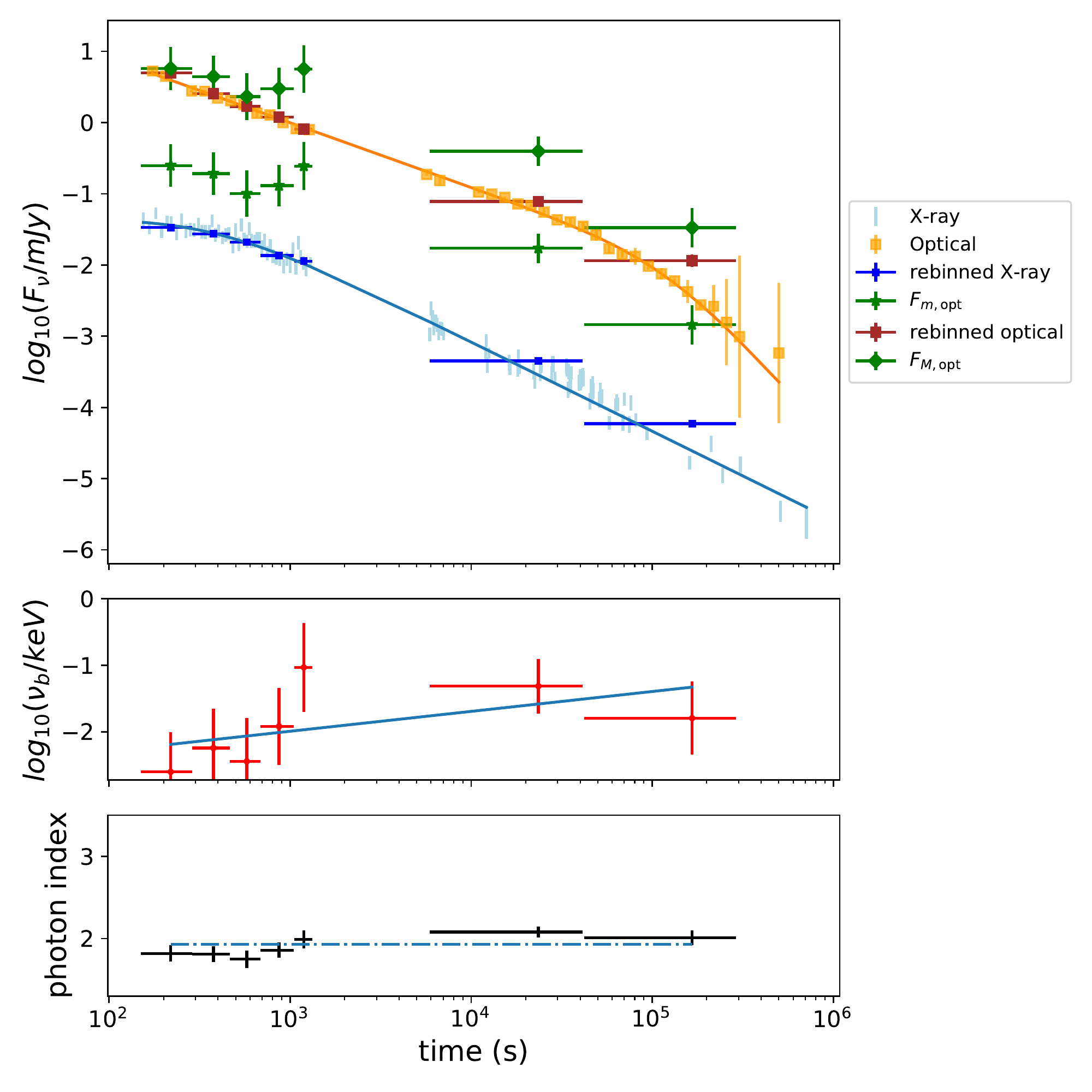}
         \caption{091018}
         \label{fig:five over x}
     \end{subfigure}
        \caption{\emph{Sample 1} - continued}
\end{figure*}

\begin{figure*}[t]
     \centering
     \begin{subfigure}{0.47\textwidth}
         \centering
         \includegraphics[width=\textwidth]{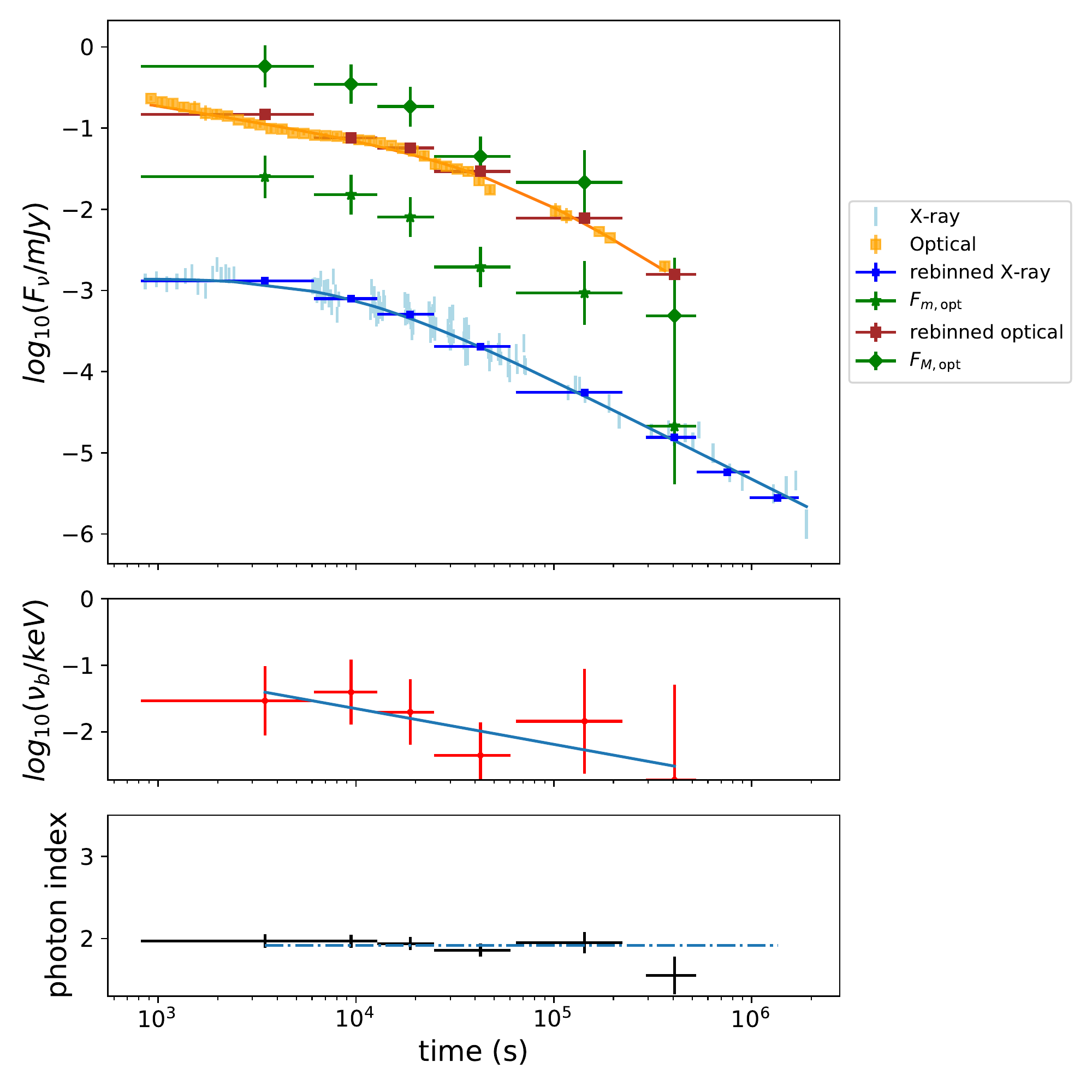}
         \caption{091029}
         \label{fig:y equals x}
     \end{subfigure}
     \hfill
     \begin{subfigure}{0.47\textwidth}
         \centering
         \includegraphics[width=\textwidth]{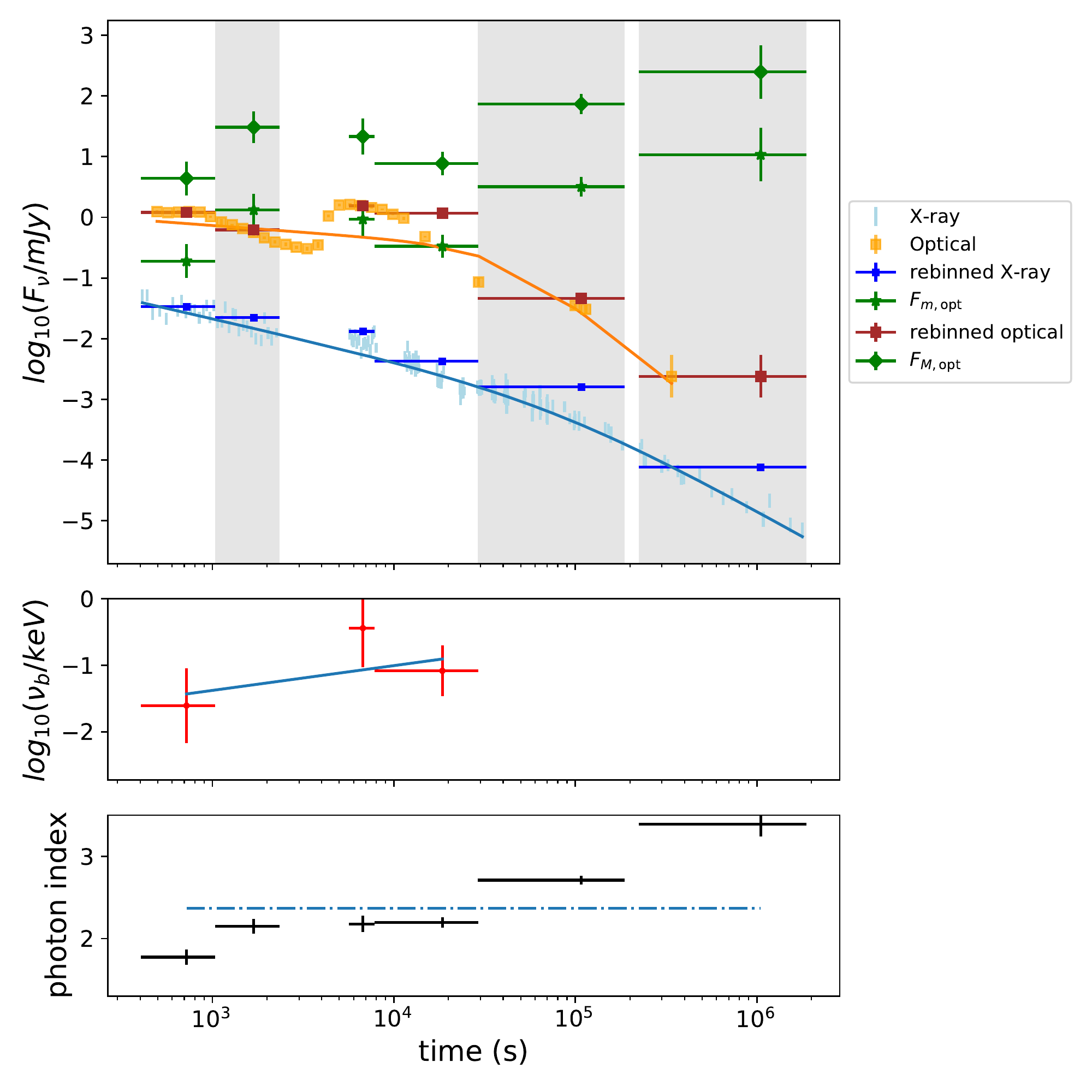}
         \caption{100621A}
         \label{fig:three sin x}
     \end{subfigure}
     \\
     \begin{subfigure}{0.47\textwidth}
         \centering
         \includegraphics[width=\textwidth]{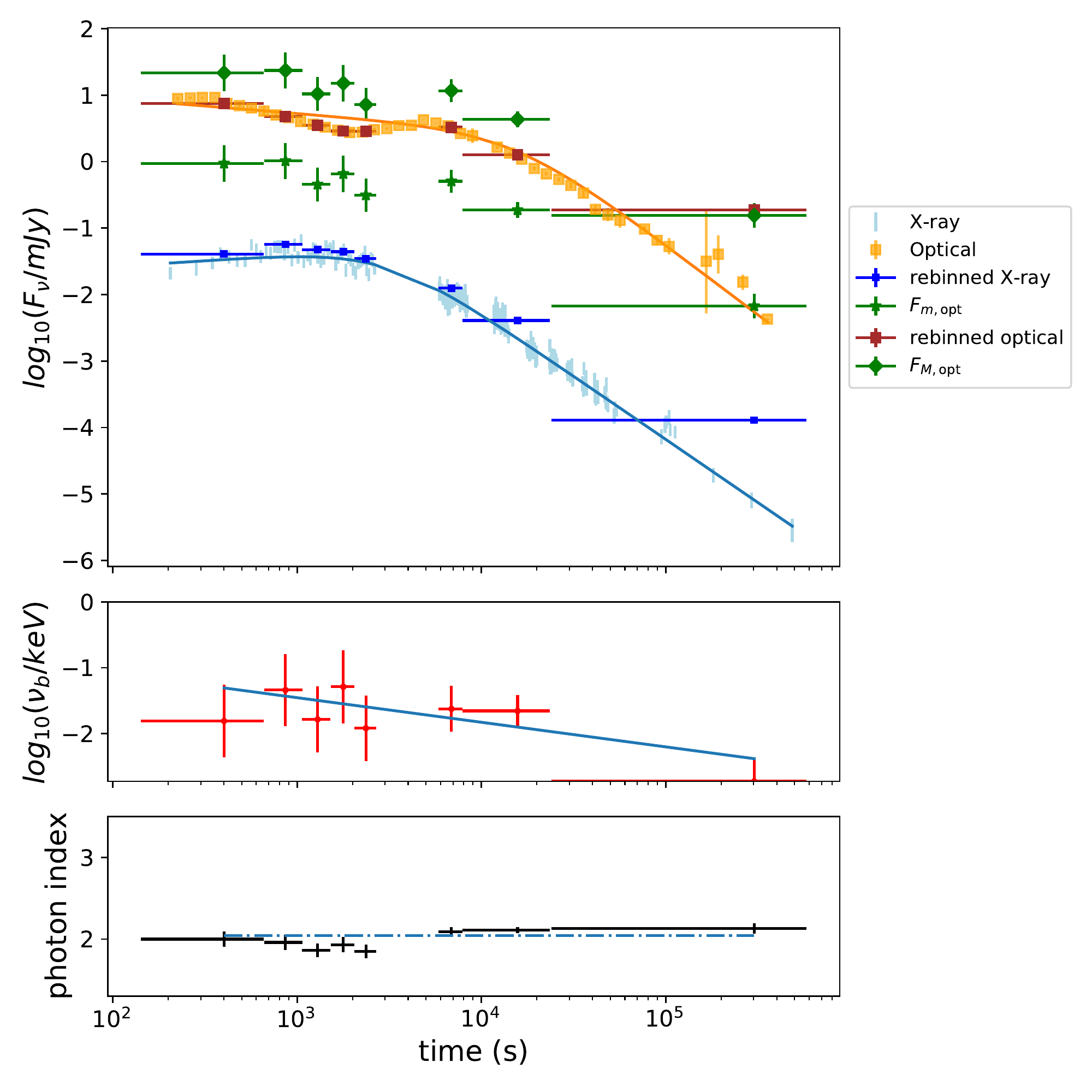}
         \caption{110213A}
         \label{fig:five over x}
     \end{subfigure}
     \hfill
     \begin{subfigure}{0.47\textwidth}
         \centering
         \includegraphics[width=\textwidth]{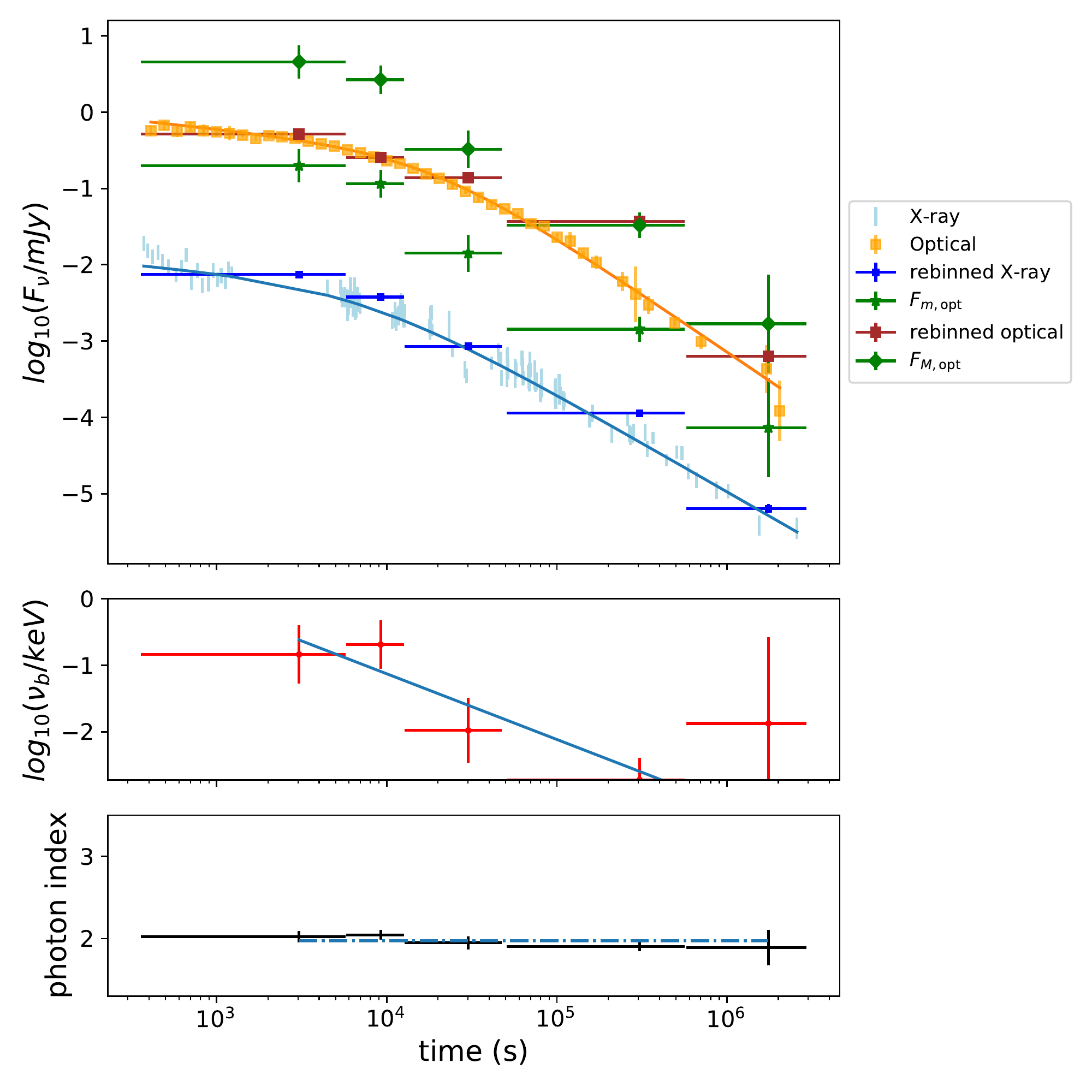}
         \caption{111228A}
         \label{fig:five over x}
     \end{subfigure}
        \caption{\emph{Sample 1} - continued}
        
\end{figure*}

\begin{figure*}[t]
     \centering
     \begin{subfigure}{0.47\textwidth}
         \centering
         \includegraphics[width=\textwidth]{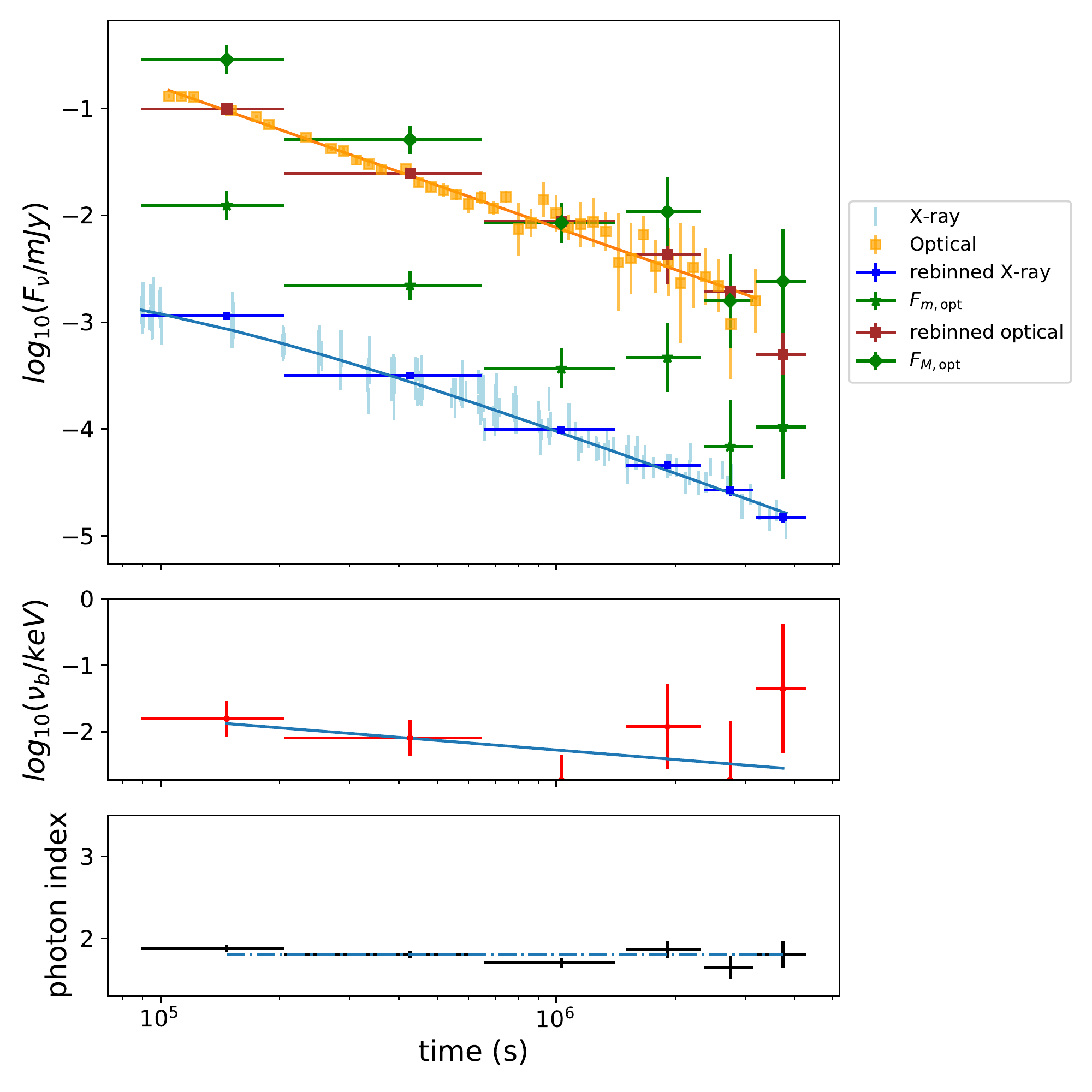}
         \caption{130702A}
         \label{fig:y equals x}
     \end{subfigure}
     \hfill
     \begin{subfigure}{0.47\textwidth}
         \centering
         \includegraphics[width=\textwidth]{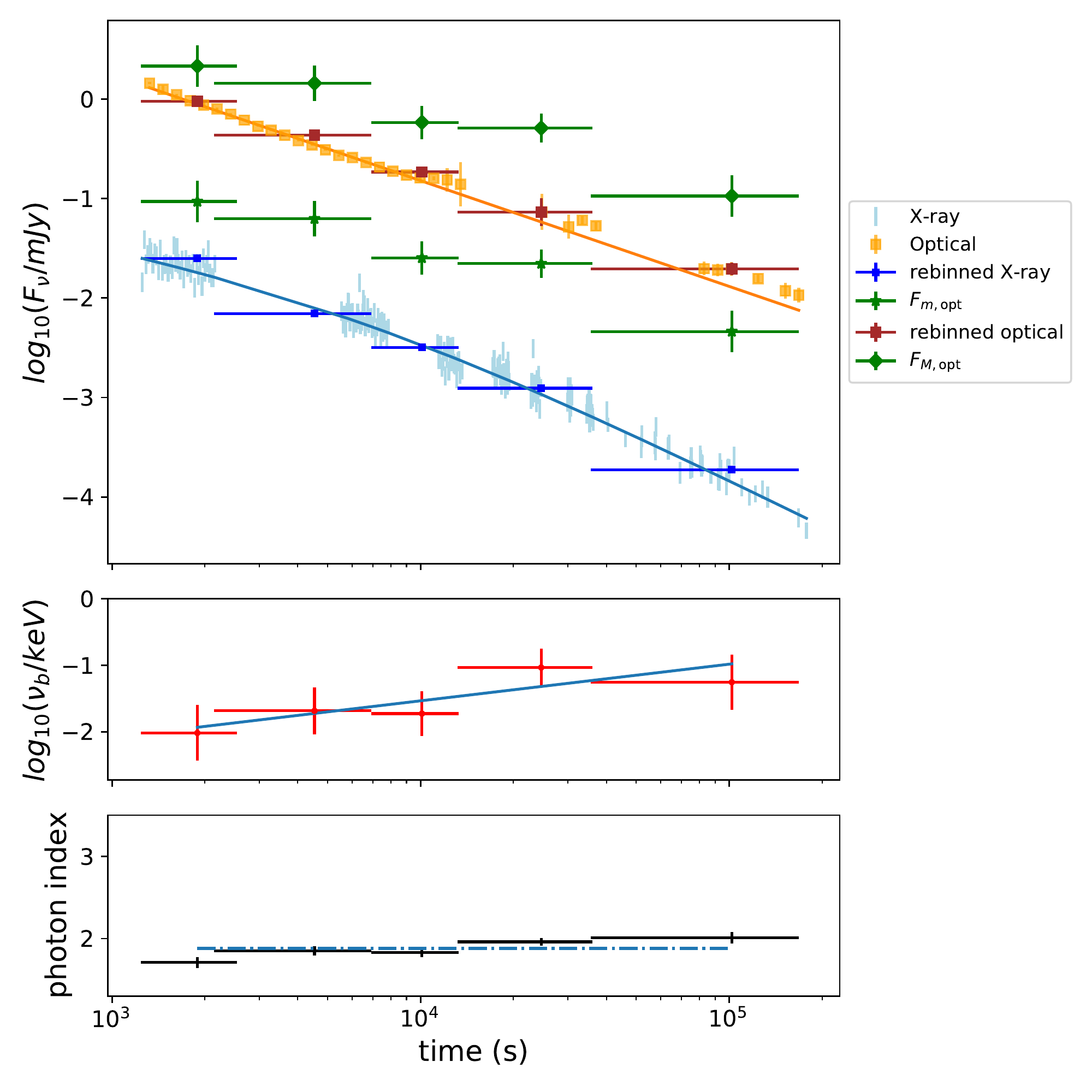}
         \caption{140419A}
         \label{fig:three sin x}
     \end{subfigure}\\
     
     \begin{subfigure}{0.47\textwidth}
         \centering
         \includegraphics[width=\textwidth]{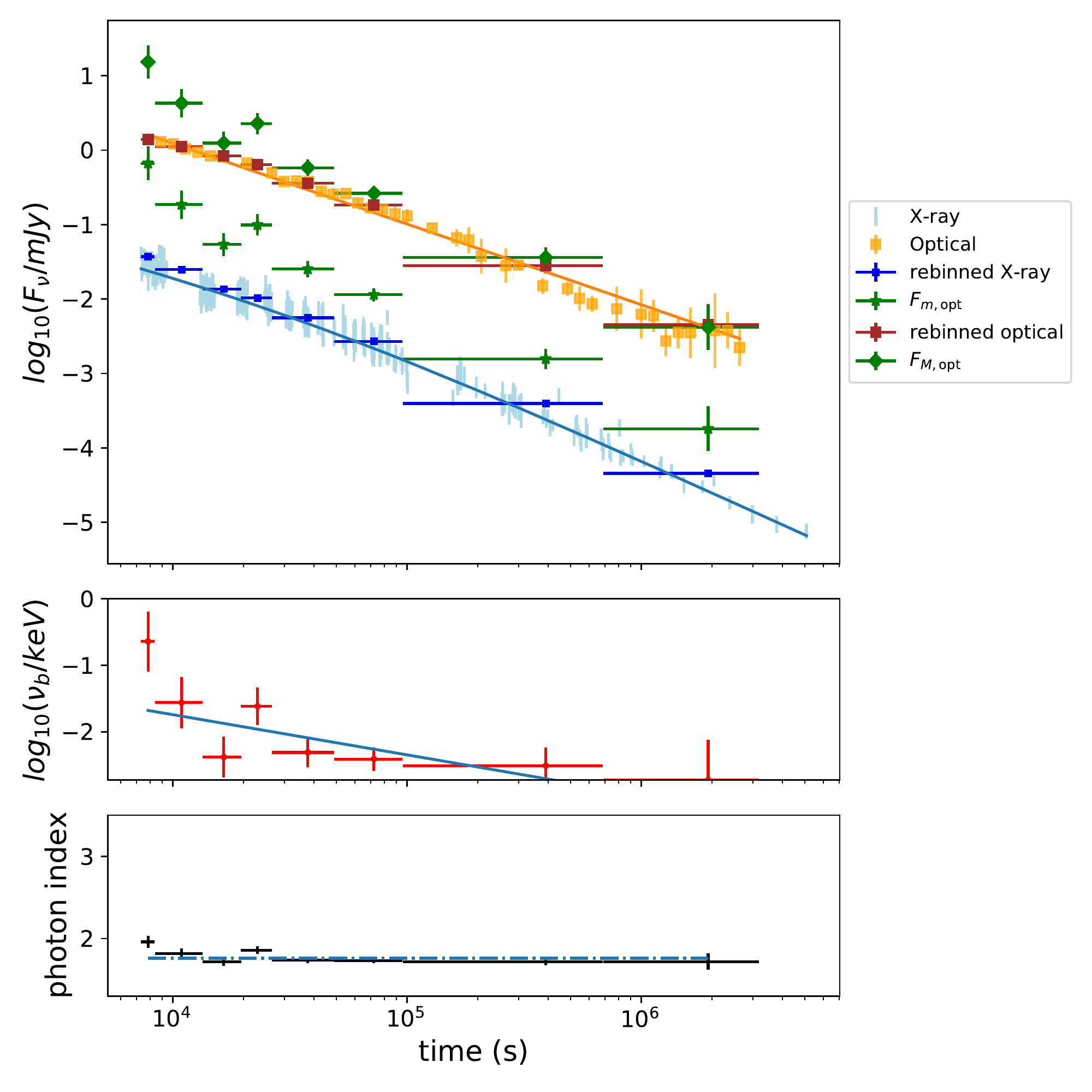}
         \caption{180728A}
         \label{fig:three sin x}
     \end{subfigure}
      \caption{\emph{Sample 1} - continued}
    
\end{figure*}

\begin{figure*}[t]
     \centering
     \begin{subfigure}{0.47\textwidth}
         \centering
         \includegraphics[width=\textwidth]{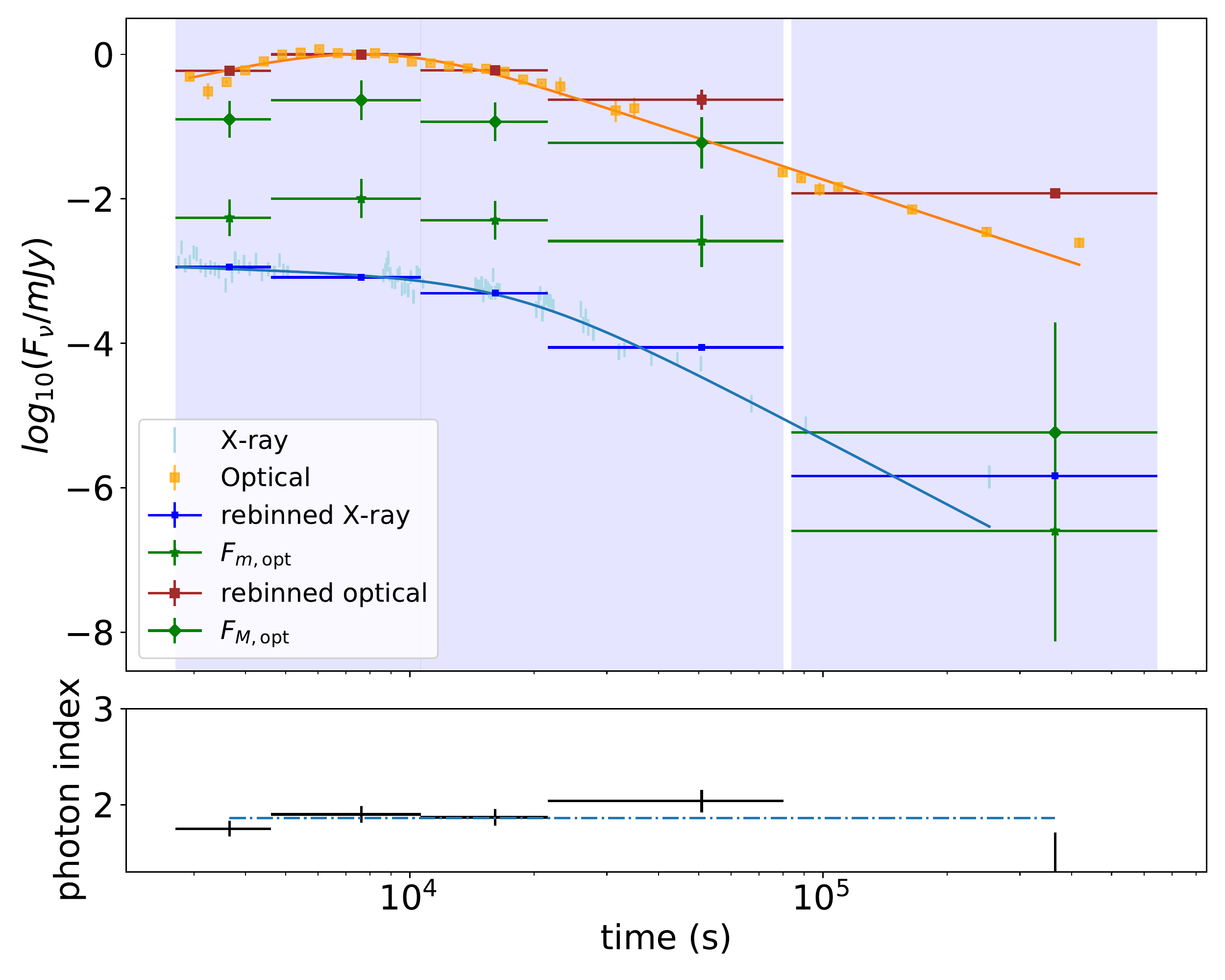}
         \caption{081029}
         \label{fig:y equals x}
     \end{subfigure}
     \hfill
     \begin{subfigure}{0.47\textwidth}
         \centering
         \includegraphics[width=\textwidth]{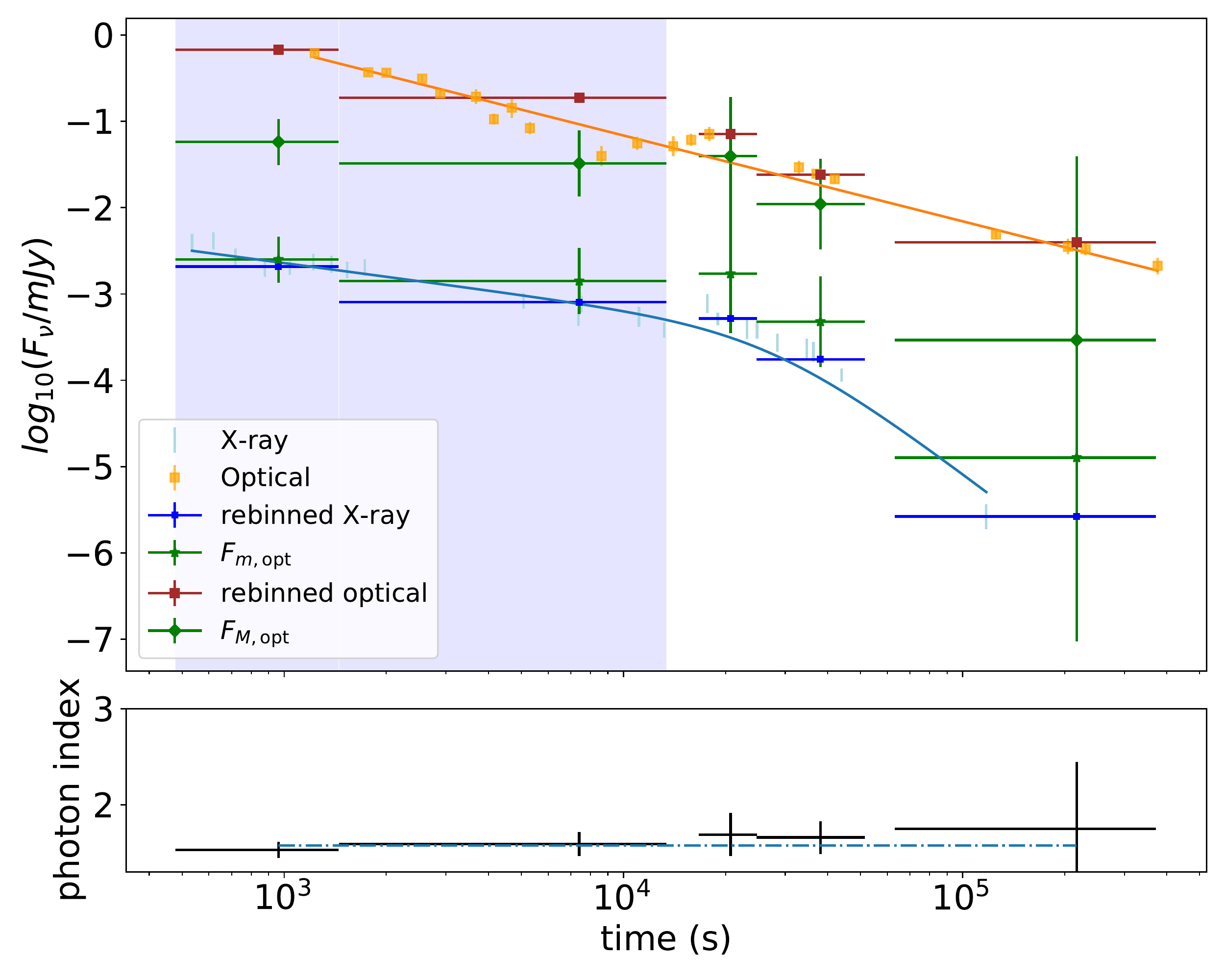}
         \caption{100219A}
         \label{fig:three sin x}
     \end{subfigure}
     \\
     \begin{subfigure}{0.47\textwidth}
         \centering
         \includegraphics[width=\textwidth]{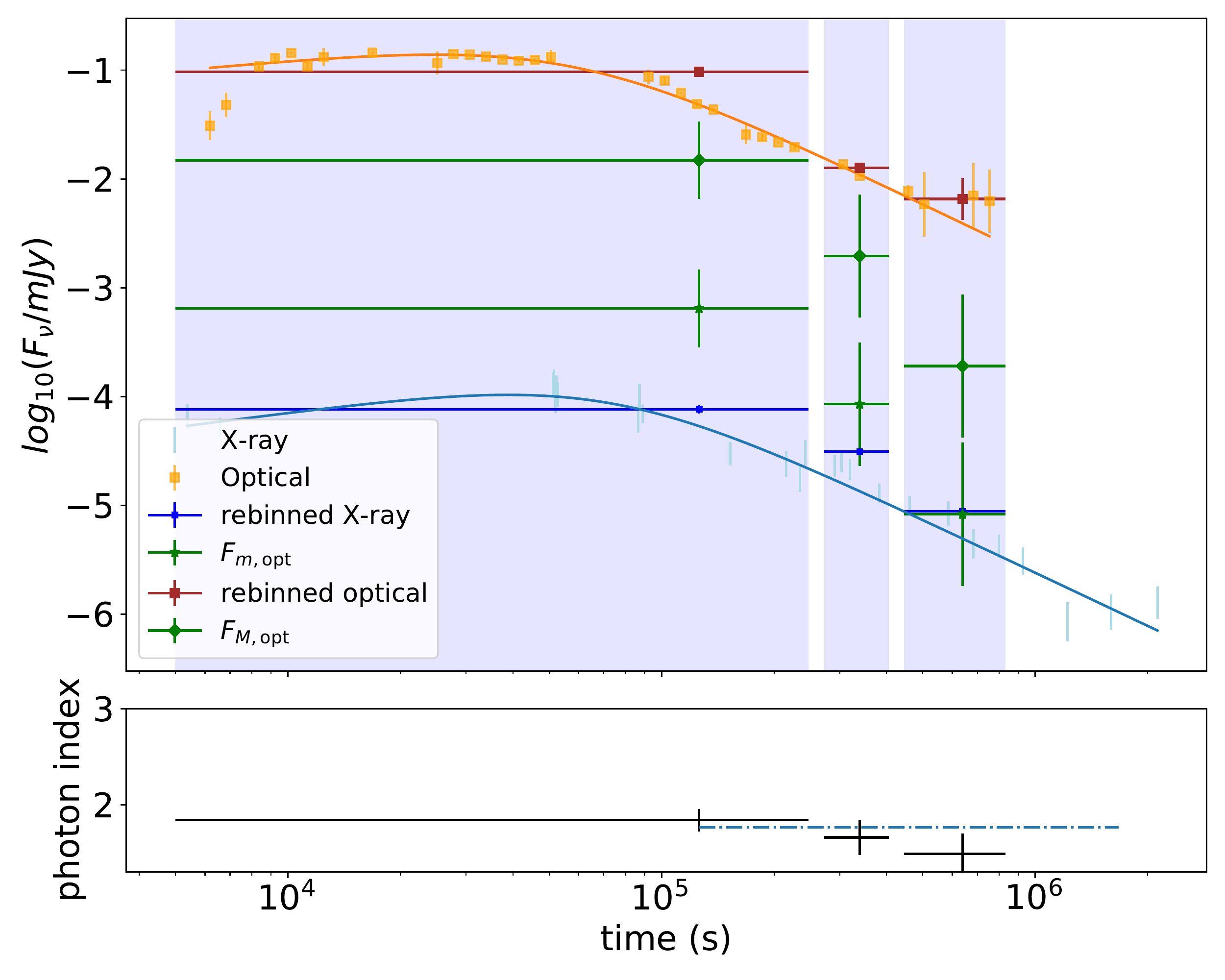}
         \caption{100418A}
         \label{fig:five over x}
     \end{subfigure}
     \hfill
     \begin{subfigure}{0.47\textwidth}
         \centering
         \includegraphics[width=\textwidth]{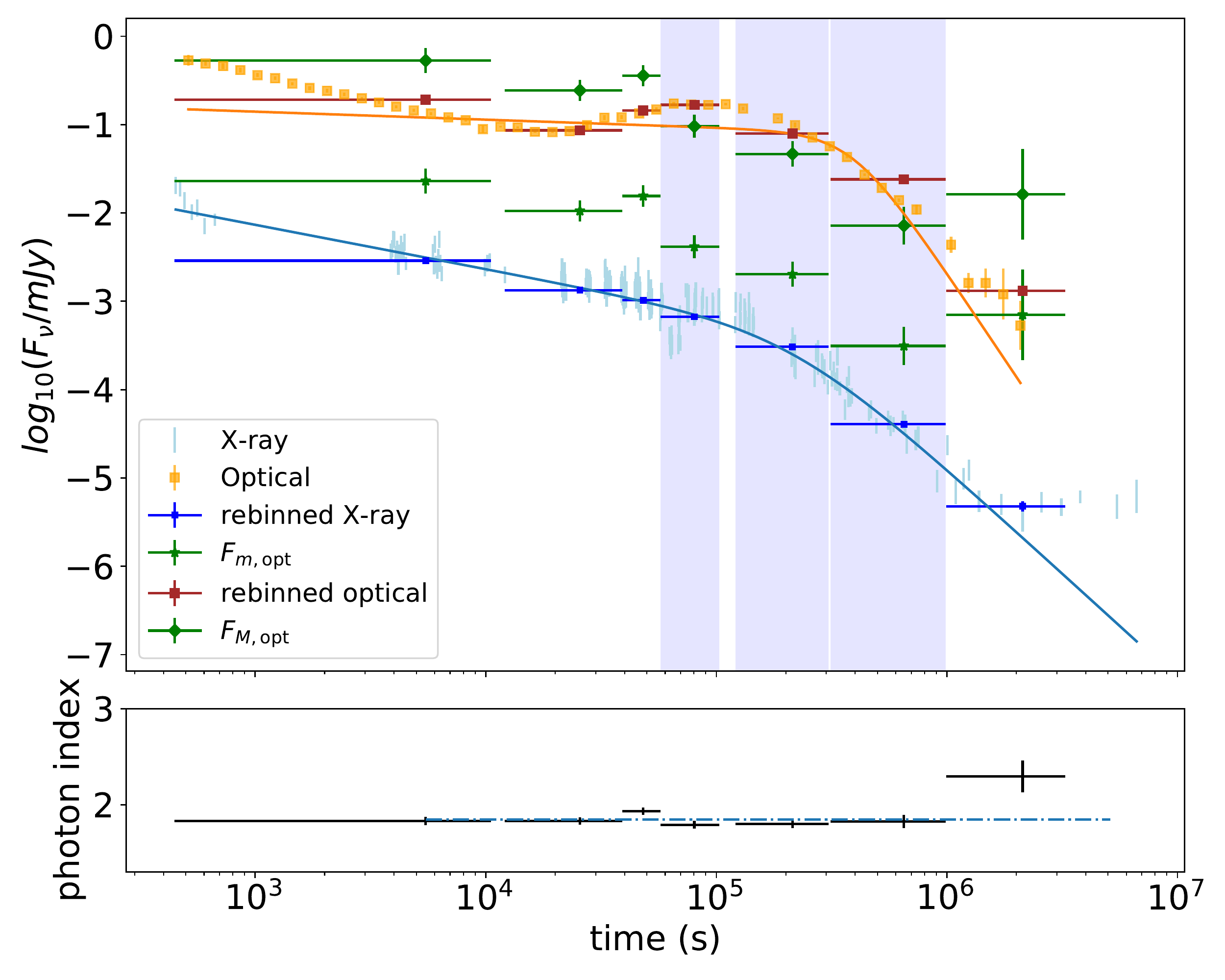}
         \caption{100814A}
         \label{fig:five over x}
     \end{subfigure}
        \caption{\emph{Sample 2} - continued}
        
\end{figure*}

\begin{figure*}[t]
     \centering
     \begin{subfigure}{0.47\textwidth}
         \centering
         \includegraphics[width=\textwidth]{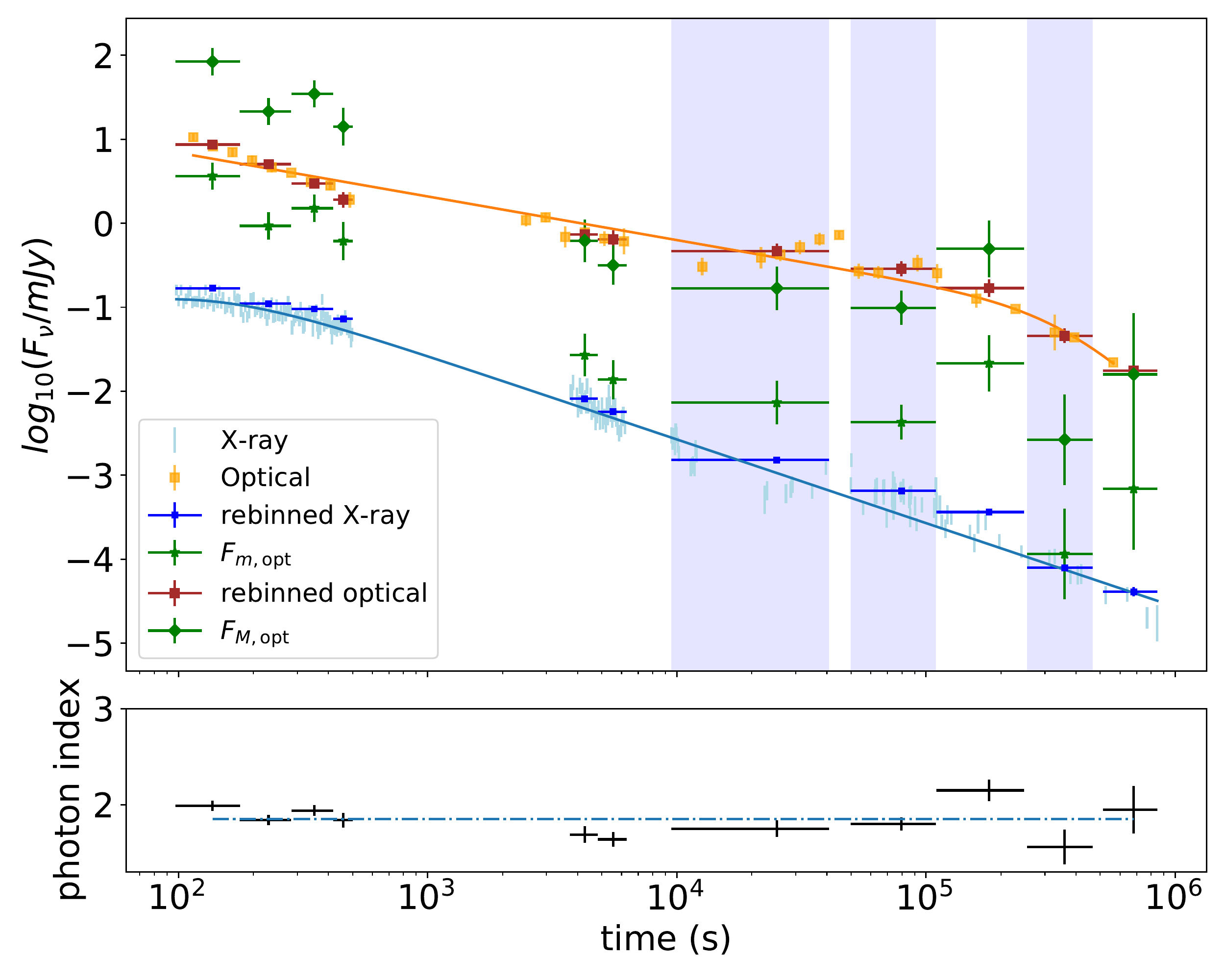}
         \caption{110715A}
         \label{mmm}
     \end{subfigure}
     \hfill
     \begin{subfigure}{0.47\textwidth}
         \centering
         \includegraphics[width=\textwidth]{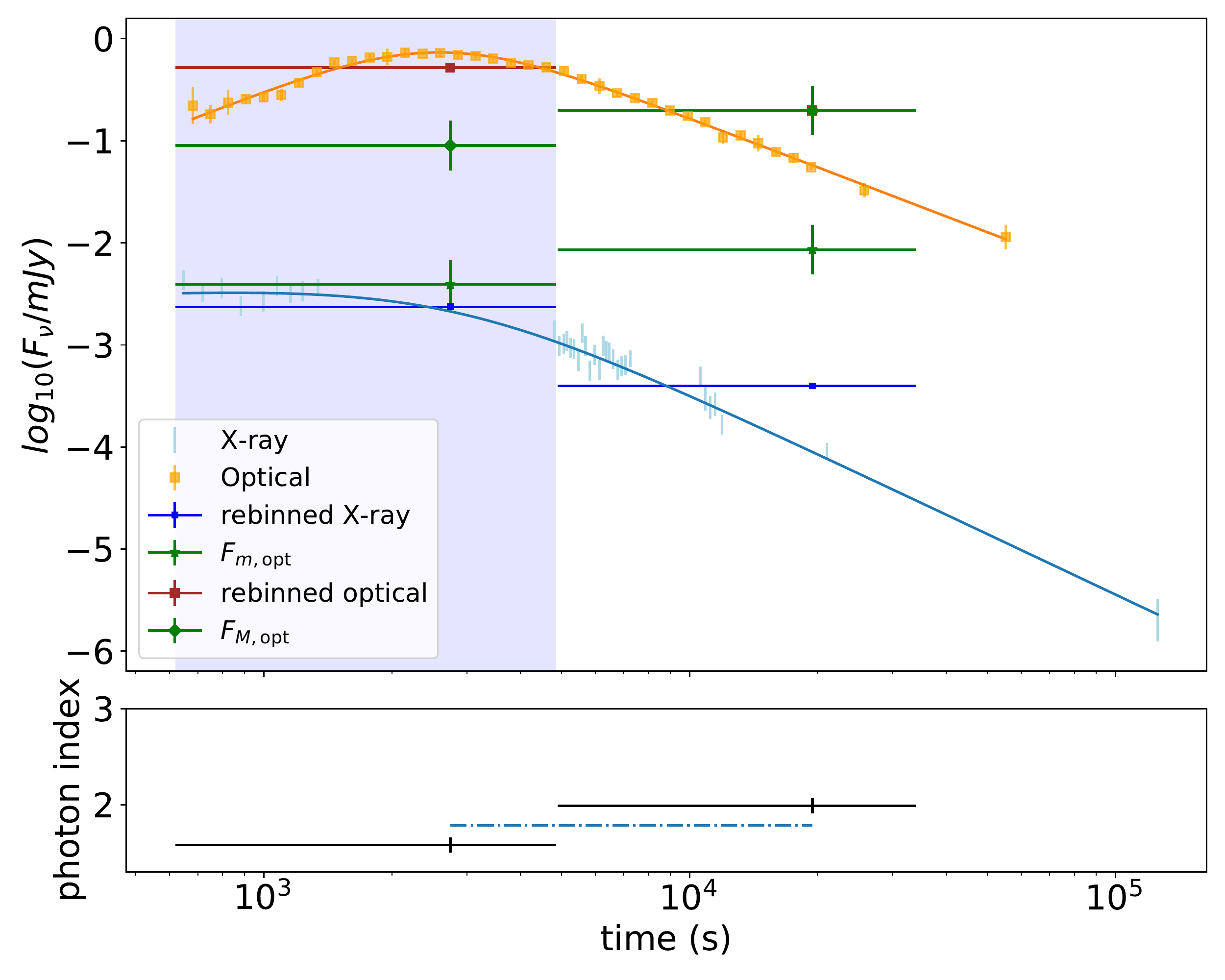}
         \caption{120404A}
         \label{fig:three sin x}
     \end{subfigure}
     \\
     \begin{subfigure}{0.47\textwidth}
         \centering
         \includegraphics[width=\textwidth]{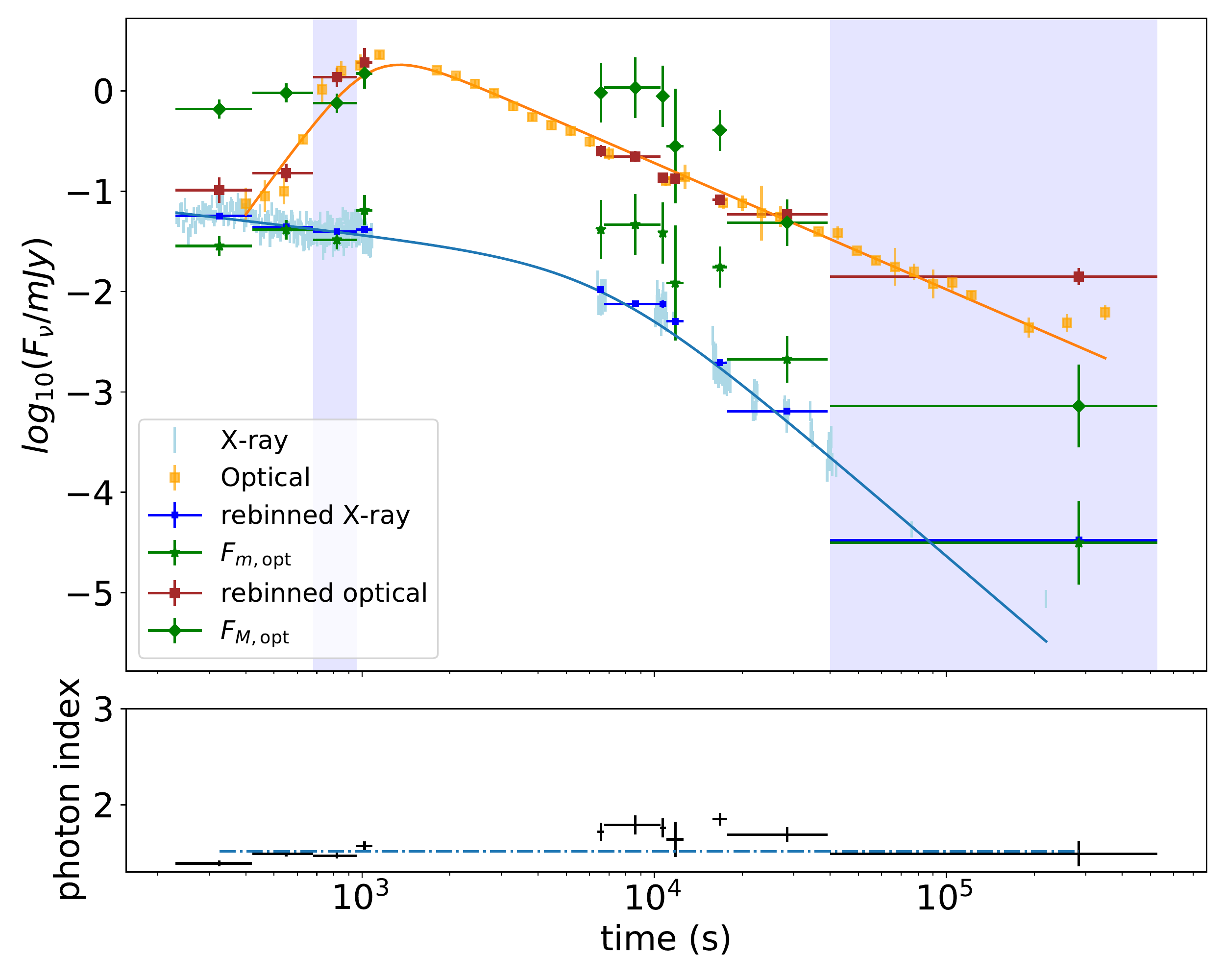}
         \caption{150910A}
         \label{fig:five over x}
     \end{subfigure}
        \caption{\emph{Sample 2} - continued}
        
\end{figure*}

\end{document}